%% file: paper.tex
\documentclass[acmsmall,screen,nonacm]{acmart}
\acmConference[PL'18]{ACM SIGPLAN Conference on Programming Languages}{January 01--03, 2018}{New York, NY, USA}
\acmYear{2018}
\acmISBN{} %
\acmDOI{} %
\startPage{1}

\setcopyright{none}

\usepackage[utf8]{inputenc}
\usepackage{amsmath}
\usepackage{refcount}
\usepackage{amsfonts}
\usepackage{listings}
\usepackage{stmaryrd}
\usepackage{wrapfig}
\usepackage{color}
\usepackage{proof}
\usepackage{thmtools,thm-restate}
\usepackage{tabularx}
\usepackage{caption}
\usepackage{subcaption}
\usepackage{stackengine}
\usepackage{pifont}
\usepackage[clock]{ifsym}
\usepackage{tikz}
\usetikzlibrary{positioning}
\usepackage{program}
\usepackage{listings,lstautogobble}
\usepackage{soul}
\usepackage{algorithm}
\usepackage{algpseudocode}
\usepackage{verbatim}
\usepackage{multirow}
\usepackage{xspace}
\usepackage{pifont}
\usepackage[normalem]{ulem}
\usepackage{adjustbox}
\usepackage{colortbl}
\usepackage{booktabs}
\usepackage{float}

\usepackage{tikz}
\usetikzlibrary{arrows.meta, positioning, fit}

\usepackage{definitions}

\begin{document}

\title{Named Commutative Blocks for Parallelization}
\ifARXIV
\subtitle{Extended Version}
\fi

\author{Parisa Fathololumi}
\orcid{0000-0002-8602-2862}
\affiliation{\institution{Stevens Institute of Technology} \country{USA}}
\email{pfathol1@stevens.edu}

\author{Adam Chen}
\orcid{0000-0002-5159-6532}
\affiliation{\institution{Stevens Institute of Technology} \country{USA}}
\email{achen19@stevens.edu}

\author{Maurice Herlihy}
\orcid{0000-0002-3059-8926}
\affiliation{\institution{Brown University} \country{USA}}
\email{mph@cs.brown.edu}

\author{Eric Koskinen}
\orcid{0000-0001-7363-634X}
\affiliation{\institution{Stevens Institute of Technology} \country{USA}}
\email{eric.koskinen@stevens.edu}

\begin{abstract}
This paper proposes a new way to exploit code commutativity to parallelize sequential programs. While prior work provided the ability to identify and exploit commutative function calls or adjacent code blocks, this paper describes the first method that dynamically exploits statically-verified commutativity conditions
of arbitrary code regions. In this method, programmers give parameterized names to code blocks anywhere in the program, and provide commutativity expressions over those parameters and the global state. Novel algorithms exploit this commutativity and dependence information to automatically construct a program-specific parallelization scheme, yielding asynchronous tasks executed through a commutativity-aware scheduler.
We furthermore show that commutativity conditions can be verified automatically, reducing reliance on manual effort and increasing confidence in correctness. We implement our approach in a new compiler and runtime system called \tool{} that takes in a sequential program and produces a parallel version that dynamically exploits the commutativity conditions.
We experimentally demonstrate that \tool{} (i) strictly generalizes prior commutativity-based parallelization techniques, (ii) enables parallelization opportunities beyond the reach of existing methods, and (iii) automatically verifies commutativity conditions.
\ifARXIV

\medskip
\noindent
\emph{Note: This is the extended version of \citet{DBLP:conf/popl/FathololumiCHK27} which appeared in POPL 2027.}
\fi
\end{abstract}


\maketitle

\newcommand\TODORevFeedback[1]{\textcolor{red}{TODO: ``#1''}}

\section{Introduction}
\label{sec:intro}
Concurrency is important to the performance of programs.
Unfortunately, reasoning about and creating effective concurrent programs is difficult.
Researchers and practitioners have thus been developing techniques to make writing correct concurrent programs easier.

One avenue is through automated (or semi-automated) parallelization: transforming a programmer's sequential code into a concurrent program.
Some classic works aimed to discover dependence-free regions of a program to identify opportunities for parallelism~\cite{DBLP:books/mk/AllenK2001}. 
Often these works are based on a 
program dependence graph (PDG)~\cite{DBLP:journals/toplas/FerranteOW87}.
DOALL parallelization~\cite{DBLP:books/mk/AllenK2001} and more recent
papers on decoupled software pipelining (DSWP)~\cite{DBLP:conf/IEEEpact/RanganVVA04,ps-dswp} are parallelization techniques that exploit dependence information.

Meanwhile, \emph{commutativity} has been widely studied as an ingredient for concurrency.
Code commutativity is when two regions of code executed in either order lead to the same final state. 
Commutativity enables concurrency because different interleavings become equivalent. This benefit of commutativity has a long history including
database design~\cite{DBLP:journals/tc/Weihl88}, early parallelizing compilers~\cite{DBLP:journals/toplas/RinardD97}, data parallelism~\cite{galois}, transactional memory~\cite{DBLP:conf/ppopp/NiMAHHMSS07,DBLP:conf/ppopp/HerlihyK08,DBLP:conf/podc/BronsonCCO10,DBLP:journals/tpds/HassanPPR17,DBLP:conf/aplas/DickersonKGH19} the scalable commutativity rule~\cite{scalablecommutativityrule} and replicated data types (CRDTs)~\cite{DBLP:conf/forte/ZellerBP14,DBLP:journals/pacmpl/GomesKMB17,DBLP:conf/cav/NagarJ19,DBLP:journals/pacmpl/LiuPRK0V20}.

Some works allow users to write annotations for code that \emph{always} commutes. 
\citet{DBLP:conf/oopsla/BocchinoADAHKOSSV09} allowed programmers to mark class methods as being always commutative and 
\citet{bridges-micro07} provided a commutative function annotation specifying that all calls to that function commute.
CommSet~\cite{commset} combines dependence-based parallelization with commutativity annotation, and parallelization applies when it is statically determinable that they always commute.

Often code regions only sometimes commute, thus requiring commutativity \emph{conditions} to be written and then evaluated at runtime. Jade~\cite{jade} provided a declarative language for specifying how tasks could concurrently access objects, including the conditions under which those specifications applied.
Galois~\cite{galois} uses commutativity conditions over data structures such as graphs. 

As commutativity conditions have become more prevalent, another line of research has been devoted to \emph{verify} whether those conditions are correct. \citet{kim-rinard-pldi11} focused on manual commutativity verification of ADT specifications. This then led to automated commutativity synthesis/verification techniques such as the {\sc Servois2} tool~\cite{DBLP:conf/tacas/BansalKT18,DBLP:conf/atva/ChenFNPBK23}. Others focused on (unsound) learning of commutativity conditions~\cite{DBLP:conf/pldi/DimitrovRVK14}.
However, these works do not perform parallelization or even execution.

Despite the known connection between commutativity and parallelization, 
almost no prior automatic parallelization is able to 
\emph{exploit verified commutativity conditions at runtime}.
An exception is the recent Veracity~\cite{veracity} language, which uses commutativity for conditional parallel execution of code blocks. However the parallelism is limited to small windows: code blocks must be \emph{adjacent} (i.e. sequentially composed), which also restricts it from being applied to the appealing setting of parallelizing loop iterations. 

\emph{Goal.} We thus seek to enable new opportunities for automated, dependence-based  parallelization  
\begin{wrapfigure}[8]{r}[1pt]{6.5cm}
	\vspace{-10pt}
	\begin{lstlisting}[style=vcy,language=VCY,basicstyle=\ttfamily\scriptsize]
	commutativity {
	  |$\nblockA{f}{x}$|, |$\nblockA{g}{i,y}$|: ( y == 0 |$\mid\mid$| x != i );
	  |$\nblockA{g}{i,y}$|, |$\nblockA{g}{i$'$,y$'$}$|: ( i != i$'$ );
	}
	int main(int[] arr, int size, int x) {
	  |$\nblockA{f}{x}$|:{ arr[x] = arr[x] / 2; }
	  int y = x;  
	  for(int i=0; i < size; i++) {
		  |$\nblockA{g}{i,y}$|:{ arr[i] = arr[i] + y; } } }
\end{lstlisting}
\end{wrapfigure}
through concurrent execution of arbitrary (\ie{} nonadjacent) regions of code whose commutativity \emph{conditions are verified statically, and then exploited dynamically}.
To this end, we begin with a small language extension allowing programmers to identify regions of code and specify the conditions under which they commute. This requires an ability to parametrically name those commutative blocks, and then provide 
commutativity specifications over those names/parameters. Above is a simple example.
In this example, there is a block labeled $\nblockA{f}{x}$ at the beginning of \code{main}, and one labeled $\nblockA{g}{i,y}$ in the body of the loop. 
Typical dependence-based parallelization procedures analyze programs to determine data dependences, and parallelize to exploit independence.
Normally these $\nblock{f}$ and $\nblock{g}$ blocks could not be executed concurrently because they both read and write array \code{arr}, possibly even in the same slot.

\emph{This paper.}
We begin by introducing \emph{named commutative blocks} (NCBs) into the programming language, allowing the programmer to provide a label for various code fragments such as $\nblockA{f}{x}$ above. Next, the programmer can specify the conditions under which two named commutative blocks (sequentially) commute, e.g., 
$\nblockA{f}{x}$ and $\nblockA{g}{i,y}$ commute whenever 
\texttt{y==0} or \texttt{x!=i}.
These commutativity conditions describe a conditional independence that we will use as part of a dependence-based parallelization procedure to allow instances of $\nblockA{f}{x}$ and $\nblockA{g}{i,y}$ to execute concurrently whenever that condition holds at runtime (also using lock synthesis~\cite{DBLP:conf/popl/EmmiFJM07,DBLP:conf/pldi/CheremCG08,DBLP:conf/pldi/FlanaganQ03,DBLP:conf/popl/VechevYY10,DBLP:conf/ppopp/Golan-GuetaRSY15}  for pair-wise atomicity).
Conditional commutativity can be exploited not only between instances of two different NCBs but also two instances of the same NCB, such as two iterations of a loop, \eg{} $\nblockA{g}{i,y}$ and $\nblockA{g}{i$'$,y$'$}$ when \texttt{i!=i$'$}.
(We will see later examples in which inter-iteration conditions don't always hold.)

\medskip
\noindent
{\bf Challenges \& Contributions.} This paper addresses (i) how NCB commutativity conditions can be automatically verified and (ii) novel dependence-based parallelization procedures and runtimes that can dynamically exploit those commutativity conditions. In summary, our contributions are:

$\bullet$ (Sec.~\ref{sec:ncb}) \emph{Language extension.} 
We first formalize the NCB language extension, 
allowing programmers to specify the commutativity of nonadjacent code fragments. At runtime those specifications will be evaluated to true/false to decide whether it is safe for the fragments to proceed in parallel.

$\bullet$ (Sec.~\ref{sec:verification}) \emph{Verification.}
If a commutativity condition is unsound, then non-commutative NCBs may be executed concurrently, leading to executions that are not equivalent to some sequential execution (unserializable). Moreover, writing valid commutativity conditions can often be difficult for users. We thus describe a method for verifying commutativity conditions through a translation that accounts for the fact that NCBs may share some common outer scope (e.g. global variables) but also have fragment-local scope. Also, multiple instances of a single NCB fragment (e.g. the body of a loop) may be executed concurrently thus requiring a notion of self-commutativity.

$\bullet$ (Sec.~\ref{sec:analysis}) \emph{Program analysis and task
identification.} We next address how an NCB program can be parallelized. 
First, we extend the standard notion of a program dependence graph (PDG)~\cite{DBLP:journals/toplas/FerranteOW87,DBLP:books/mk/AllenK2001} to one that includes commutativity conditions on edges between NCB nodes. Next, the analysis constructs a parallelization scheme over the entire program body (including NCBs) deciding which code fragments should be grouped together, i.e., which can be executed asynchronously and which must be synchronous.

We next construct a parallel program as a collection of pieces called ``tasks'' (including a designated initial task), along with their corresponding memory read/write \emph{data dependencies}. These tasks are also equipped with NCB commutativity conditions that dictate when task instances may proceed in parallel despite otherwise having a potential data dependency. To ensure that instances of those tasks are pair-wise atomic, we employ off-the-shelf lock synthesis~\cite{DBLP:conf/popl/EmmiFJM07}.

$\bullet$ (Sec.~\ref{sec:execution_model}) \emph{Execution.} An execution of a task is a ``job,'' which may spawn additional concurrent jobs corresponding to other tasks. These jobs can exhibit data dependencies among one another. We therefore describe a runtime system that schedules jobs while respecting these dependencies, and exploiting commutativity when possible. 

$\bullet$ (Sec.~\ref{sec:correctness}) \emph{Correctness.} We prove that every execution of the concurrent execution model is equivalent to a sequential execution of the original program (Theorem~\ref{thm:correctness}). This has been mechanized in Rocq and relies  on the following assumptions: soundness and stability of commutativity conditions (ensured by Sec.~\ref{sec:verification}), soundness of the dependence analysis, pair-wise atomicity of jobs (ensured by lock synthesis), and termination of every job.

$\bullet$ (Sec.~\ref{sec:implementation} and~\ref{sec:eval}) \emph{Implementation, verification, scaling.}
We implemented our work in a language called \name{}, including verification of commutativity conditions, the parallelization analysis, and the runtime scheduler. \name{} is implemented in OCaml and operates as an interpreter using the multicore support of OCaml $\geq$5. The implementation also includes a simple program analysis for lock synthesis, applied as a post-parallelization stage to the job-generating program to ensure pairwise atomicity of jobs.

We evaluate against both existing and new benchmarks:
We first adapt existing benchmarks and show that \name{} largely
subsumes the parallelization opportunities in the closest related
works PS-DSWP~\cite{ps-dswp}, CommSet~\cite{commset}, and
Veracity~\cite{veracity}, obtaining similar scaling behavior. We then
introduce six new benchmarks including I/O, banking, smart contract
voting~\cite{voting}, and ERC20 transactions~\cite{erc20}.
These new benchmarks demonstrate that named commutative blocks and our dynamic commutativity runtime unlock new forms of parallelization not achievable through prior work. We show that the NCB commutativity conditions can be verified (Table~\ref{table:benchmarks}) and that speedups approach the theoretical maximum for each benchmark (Table~\ref{table:execution}).

\emph{Limitations.}
This paper introduces the NCB language extension and, for now, assumes the programmer decides where to place NCBs. In future work, profiling and/or resource analysis information could be used. This is not straight-forward because, as we summarize in Sec.~\ref{sec:eval}, there are tradeoffs that pertain to sizes and numbers of NCBs.
\ARXIV{Also, while our work verifies commutativity conditions, in principle it can also synthesize them (due to our reliance on {\sc Servois2}). We found that synthesis timed out, so we leave that to future work.}
In addition to arrays, \name{} supports dictionaries and file I/O (create, open, read, write, etc.). Programs with heap operations can be executed, but verification of heaps with aliasing is the subject of ongoing orthogonal work.
Furthermore, since the bulk of the novelty of NCBs is at the language level, \Tool{} is an interpreter but in the future we hope to explore further opportunities through a compiler backend and add inter-procedural support.

\POPL{
\medskip%
\noindent%
\emph{Note: Additional details can be found in the extended version~\cite{technicalreport}.}}

\section{Programming with named commutative blocks}
\label{sec:overview}

We start with some simple examples of how a programmer can describe the commutativity of code regions through the use of named blocks and commutativity specifications. We will then discuss the more technical challenges and contributions involved in realizing this programming model.
\begin{wrapfigure}[6]{r}[1pt]{6.8cm}
\vspace{-5pt}
\begin{lstlisting}[style=vcy,language=VCY,basicstyle=\ttfamily\scriptsize]
commutativity { |$\nblockA{f}{i}$|, |$\nblockA{f}{j}$|: ( i!=j ); }
int main(int[] arr, int size) {
  for(int i=0; i < size; i++) {
    |$\nblockA{f}{i}$|:{ arr[i] = (arr[i] + 1)/2; } }
}
\end{lstlisting}
\end{wrapfigure}
In the body of \code{main} on the right, a loop is traversing array \code{arr} and at each iteration \code{i}, incrementing/dividing element \code{arr[i]}. The body of the loop is given a block label $\nblock{f}$. This should be thought of as simply a label to name the block, with a small addition of some variable names included. This $\nblockA{f}{i}$ notation is not to be interpreted as a function application, hence we avoid normal parens.

Since each iteration \code{i} of the loop reads and writes \emph{only} \code{arr[i]}, each iteration can be executed in parallel with all other iterations, as would be done by a classical parallelizing compiler. More abstractly, we can think of this as that each iteration of the loop \emph{commutes} with every other iteration of the loop, in this example, simply because the iterations mutate disjoint array cells. The above example expresses this information as a \code{commutativity} annotation at the top. The variable names given with the block label are the variables from the local state which may be used 
\begin{wrapfigure}[7]{l}[1pt]{6.8cm}
\begin{lstlisting}[style=vcy,language=VCY,basicstyle=\ttfamily\scriptsize]
commutativity { 
   |$\nblock{f}$|, |$\nblockA{g}{i}$|: ( x<=y |$\mid\mid$| x != i ); }
int main(int[] arr, int size, x, y) {
  |$\nblock{f}$|:{ if (x>y) arr[x] = y; }
  x = y;
  for(int i=0; i < size; i++) {
    |$\nblockA{g}{i}$|:{ arr[i] = (arr[i] + 1); }}}
\end{lstlisting}
\end{wrapfigure}
as formal parameters to the commutativity condition. In this case, the commutativity condition expresses that an iteration of the body $\nblock{f}$ with a given value of \code{i} commutes with another iteration of the body $\nblock{f}$ with a given value \code{j}, as long as \code{i != j}, that is, they are different iterations.
Commutativity specifications can include boolean combinations of (pure) arithmetic program expressions including constants, variables, array expressions, etc. (formalized in Def.~\ref{def:commutativity-spec-syntax}).
The next example on the left illustrates how named-blocks with commutativity specifications can be combined with dependence-based parallelization~\cite{DBLP:books/mk/AllenK2001,ps-dswp,commset}.
Such techniques typically begin by constructing a program dependence graph (PDG)~\cite{DBLP:journals/toplas/FerranteOW87}, determining for every program statement $s$, every other statement $s'$ upon which $s$ depends either due to data or due to control. Ignoring the named fragment annotations for the moment, a PDG for the above example is as follows:
\vspace{-10pt}
\begin{center}
\scalebox{0.9}{
\begin{tikzpicture}[node distance={14mm}, main/.style = {draw, rectangle}, ctl/.style={dashed, dash pattern=on 1.5pt off 1.5pt}]
\node[main](e){Entry};
\node[main](f)[right=12mm of e]{$\nblock{f}$};
\node[main](xy)[right=12mm of f]{\code{x:=y}};
\node[main](for)[right=12mm of xy]{For};
\node[main](g)[right=12mm of for]{$\nblock{g}$};
\draw[->,ctl] (e) edge[bend right] (f);
\draw[->,ctl] (e) to [out=-20,in=200] (xy);
\draw[->,ctl] (e) to [out=17,in=165] (for);
\draw[->,ctl] (for) edge[bend left]  (g);
\draw[->,ctl] (for.north) edge[loop,distance=7mm] (for.north);
        
\draw[<-] (f.west) -- node[above, pos=0.5, inner sep=1pt] {\code{x,y}} (e.east);
\draw[->] (f.east) -- node[above, pos=0.5, inner sep=1pt] {\code{x,y}} (xy.west);
\draw[->] (f) to [out=-15,in=195] node[above, inner sep=1.2pt] {\code{arr}} (g);
\end{tikzpicture}}
\end{center}
\vspace{-10pt}
For this example, since $\nblockA{g}{i}$ has no dependence on the \code{x:=y} block, it can be executed in parallel. But what of $\nblockA{g}{i}$'s possible dependence on $\nblock{f}$? A typical dependence-based parallelization would require these two blocks to execute sequentially because they both may access the same location of \code{arr}. However, we can improve over PDG dependences with conditional commutativity. As shown in the \code{commutativity} specification above, $\nblock{f}$ can be executed in parallel with any loop iteration $\nblockA{g}{i}$ provided that \code{x<=y} (in which case $\nblock{f}$ does not write to any slot in \code{arr}) or \code{x!=i} (in which case $\nblockA{g}{i}$ writes to a slot not written by $\nblock{f}$), and we are able to use this information to execute them in parallel.

\emph{Intuition: Nonadjacent commutative fragments.}
In many existing settings, commutativity pertains to code fragments \code{s$_1$;s$_2$} that are 
\emph{adjacent} in a program. By contrast, named blocks need not be adjacent. $\nblock{f}$ and $\nblock{g}$ 
appear at different program locations, with other code (\code{x:=y} and the loop header \code{i<size;i++}) in between.
Nonetheless, the programmer can still think about whether two nonadjacent blocks commute (without having to think about a complicated concurrent execution model).  
Much as one can understand that \code{y:=x+1} depends on an earlier \code{x:=42}, without thinking about how a compiler uses that dependence, one can think about the commutativity of $\nblock{f}$ and $\nblock{g}$ as if the two bodies ran back-to-back in isolation, reading their inputs (such as \code{i} for $\nblock{g}$) and possibly mutating their common outer scope (such as \code{arr}).
Commutativity is thus a three-way notion, akin to read/write dependences, which involve the two statements and their common variables.

This intuition is formalized as a partitioning of the state, along with statically checkable well-formedness constraints (Sec.~\ref{sec:ncb}).
Specifically, the state is partitioned into three portions:
(1) the \emph{shared} portion $\sigma_{gl}$, here \code{arr}, \code{x}, and \code{y}, which belong to the common outer scope of both fragments;
(2) $\nblock{f}$'s inputs $\sigma_\nblock{f}$, here empty, since $\nblock{f}$ has no parameters; and
(3) $\nblock{g}$'s inputs $\sigma_\nblock{g}$, here \code{i}, which is local to the loop and differs across iterations, and is thus named as a parameter of $\nblock{g}$.
A useful analogy is transactions: $\sigma_{gl}$ is an object's current state, and two transactions, each with its own inputs ($\sigma_\nblock{f}$ and $\sigma_\nblock{g}$), are being considered for execution. A variable written by one fragment and read by the other, such as \code{arr}, is always in the shared portion. The well-formedness constraints ensure that,
like a transaction's arguments, NCB parameters are \emph{read-only}: neither fragment assigns \code{i}, so its value at $\nblockA{g}{i}$'s entry is also the value $\nblockA{g}{i}$ uses throughout, and the scheduler can safely hand each instance a private copy. Parameters serve only to name per-instance values in a condition, so that two instances that refer to the same variable (here, the \code{i} of two different iterations) can be told apart. Everything a fragment writes, and that outlives it, is in the shared portion.

Parallelization of this program (discussed below) attempts to find ways for the various portions of the program (including block $\nblock{f}$, the \code{x:=y} block, and block $\nblock{g}$) to execute simultaneously. However, this is only as permitted by dependences and commutativity conditions. Here, $\nblock{f}$ must complete before \code{x:=y} executes, since \code{x:=y} writes \code{x} which $\nblock{f}$ reads. Consequently, whenever the condition of $\nblock{f}$ and $\nblockA{g}{i}$ is evaluated while $\nblock{f}$ is pending, \code{x} and \code{y} still hold their values at $\nblock{f}$'s entry. The iterations $\nblockA{g}{i}$, in contrast, do not depend on \code{x:=y}, and \emph{may execute concurrently with} $\nblock{f}$ whenever the commutativity condition holds, their pairwise atomicity is ensured by lock synthesis (Sec.~\ref{sec:analysis}).
Intervening code such as \code{x:=y} becomes a separate task (with its own dependence edges) so commutativity of blocks $\nblock{f}$ and $\nblock{g}$ can be specified without concern for \code{x:=y}.

\begin{wrapfigure}[10]{r}[1pt]{6.5cm}
\vspace{-10pt}
\begin{lstlisting}[style=vcy,language=VCY,basicstyle=\ttfamily\scriptsize]
commutativity {
  |$\nblockA{f}{i}$|, |$\nblockA{g}{j}$|: ( j % 2 != 0 );
  |$\nblockA{g}{i}$|, |$\nblockA{g}{j}$|: ( i != j );
  |$\nblockA{f}{i}$|, |$\nblockA{f}{j}$|: ( i != j ); }
int main(int[] arr, int size) {
  for(int i=0; i < size; i++) {
  	|$\nblockA{f}{i}$|:{
    	if (i % 2 == 0) arr[i] = 0;
    	else arr[i] = arr[i] + 1; } }
  for(int i=0; i < size; i++) { 
  	|$\nblockA{g}{i}$|:{ arr[i] = arr[i] - 1; } }
}
\end{lstlisting}
\end{wrapfigure}
In the example to the right we attempt to parallelize all iterations of all loops altogether. The first loop increments all odd array slots, and writes 0 to all even slots. The second loop decrements all array slots. The second loop's subtraction commutes when the index matches up with the first loop's additions, but not when it matches up with the first loop's assignments of 0. This is expressed in the first commutativity condition specifying that $\nblock{f}$ iterations commute with $\nblock{g}$ iterations of even index.
All $\nblock{g}$ and $\nblock{f}$ iterations commute amongst themselves, since each iteration has a unique index, so the condition will always be satisfied.

{\bf Running example.} While the above examples illustrate some of the paper goals, we use the next Example~\ref{ex:motivation} (below to the right) to highlight the challenges addressed in this paper. The program 
iterates through array \code{arr}. At each iteration \code{i}, first in one code block labeled $\nblock{f}$, a value \code{x} is calculated and its square is added to \code{arr[i]}. Then, in code block labeled $\nblock{g}$, another computation is performed and, if \code{arr[i]} is positive, \code{arr[i]} will be decremented and \code{calc2} performed; otherwise \code{calc3} will be performed. In either case the result of the calculations will be written to \code{out[i]}. 

If the \code{calc} computations in this example are expensive, it would be appealing to perform them in parallel. At a high level there are two possible strategies. First, since there are no loop-carried dependences, the loop iterations all commute; in fact all $\nblock{f}$ iterations commute with each other, all $\nblock{g}$ iterations commute with each other, and $\nblock{f}$ iterations commute
\begin{wrapfigure}[15]{r}[1pt]{7.3cm}
\vspace*{-\baselineskip}
\begin{example}[Motivating example]\label{ex:motivation}$\;$
\begin{lstlisting}[style=vcy,language=VCY,basicstyle=\ttfamily\scriptsize]
commutativity { 
  |$\nblockA{f}{i}, \nblockA{g}{j}$|: i!=j |$\mid\mid$| (i==j && arr[i]>0);
  |$\nblockA{f}{i}, \nblockA{f}{j}$|: i!=j;
  |$\nblockA{g}{i}, \nblockA{g}{j}$|: i!=j; }
int main(int[] arr, int size) {
 int[] out = new int[size]; int i = 0;
 while(i < size){
   |$\nblockA{f}{i}$|:{ // will only increase arr[i]
     int x = calc(i); 
     arr[i] = arr[i] + (x*x);
   }
   |$\nblockA{g}{i}$|:{ // only observes if arr[i]>0 
      int t = compute(i);
      if (arr[i] > 0 && t < 0) {
         arr[i] = arr[i] - 1; 
         out[i] = calc2(t);
      } else {
         out[i] = calc3(t);
   } }
   i = i + 1;
} }
\end{lstlisting}
\vspace{-20pt}
\end{example}
\end{wrapfigure}
with $\nblock{g}$ block iterations when \code{i!=j}. These facts are captured by the \code{commutativity} specifications that all have an \code{i!=j} component. 

A next goal is to parallelize the \code{calc} computation with either \code{calc2} or \code{calc3} within \emph{the same} iteration of the loop. Both $\nblock{f}$ and $\nblock{g}$ read and write \code{arr[i]}, so there is a data dependence. Nonetheless, for every iteration \code{i}, if the initial value of \code{arr[i]} is positive, then $\nblock{f}$ will not change this, and so the outcome of the test \code{arr[i]>0} inside $\nblock{g}$ will not change.  Moreover, the increment of \code{arr[i]} in $\nblock{f}$ commutes with its decrement in $\nblock{g}$. Therefore, we also have the disjunct \code{(i==j && arr[i]>0)} in the commutativity specification for  $\nblock{f}$ with  $\nblock{g}$.

\subsection{Challenges \& Contributions}

The core challenges are (i) whether NCB commutativity conditions are correct,  (ii) how to perform a data-dependence analysis that incorporates NCBs and their commutativity, and (iii) how to exploit that commutativity at runtime to allow commutative code fragments to execute in parallel when they otherwise would be blocked due to data-dependence.
We now discuss how we address these using Ex.~\ref{ex:motivation}. 
\ARXIV{The following diagram illustrates how such programs with named commutativity blocks are analyzed for dependences and commutativity, transformed into a concurrent ``job-generating'' program that spawns jobs, and executed by a commutativity-aware runtime scheduler.
\begin{center}
\includegraphics[width=5.2in]{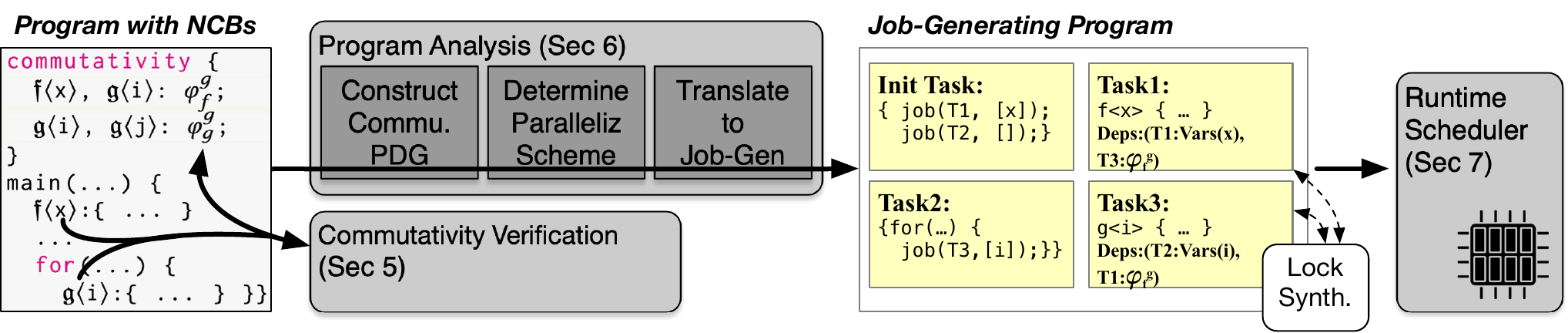}
\end{center}
}

\emph{Challenge 1: Correctness of commutativity specifications.} 
Commutativity conditions are sometimes difficult for a programmer to write and reason about, and incorrect conditions can lead to unsound execution. To complicate verification, fragments may share a common outer variable scope such as the \lstinline|out| array in Ex.~\ref{ex:motivation}, but also have fragment-specific instances of variables such as the \lstinline|i| that pertains to NCB $\nblockA{f}{i}$ as compared to the \lstinline|i| that pertains to NCB $\nblockA{g}{i}$. Moreover, code fragments may need to commute with \emph{themselves}, such as seen in a loop body. 
To address verification, we thus describe a translation $Tr(\nblockA{f}{i},\nblockA{g}{i})$ (Sec.~\ref{sec:verification}) that reduces NCB commutativity verification to pre/post specifications of methods operating on an abstract data type (ADT). The {\sc Servois2} commutativity verifier~\cite{DBLP:conf/atva/ChenFNPBK23}  can then be applied to the resulting ADT. Our translation accounts for fragment-local state, and we prove that verification of the translated property $Tr(\varphi_\nblock{f}^\nblock{g})$ implies that  $\varphi_{\nblock{f}}^{\nblock{g}}$ is a sound commutativity condition for $\nblockA{f}{i}$ and $\nblockA{g}{i}$.

\emph{Challenge 2: Dependence analysis and parallelization of NCBs.}
Classical dependence analyses~\cite{DBLP:books/mk/AllenK2001} do not take into consideration program fragments with conditional commutativity, nor how to determine a parallelization scheme, nor a format to represent the resulting scheme so that commutativity can be exploited at runtime. We next address this in three steps.

(Step 2.1) Existing analyses construct a so-called program dependence graph (PDG)~\cite{DBLP:journals/toplas/FerranteOW87}, with a node for each program statement $s$, and an edge $s'\xrightarrow{x_1,...}s$ whenever statement $s$ depends on variables modified by statement $s'$. Additionally, an edge $s\dashrightarrow{}s'$ is introduced when $s$ is a control statement that determines the execution of $s'$.
Various techniques can then parallelize programs in a way that respects these dependences~\cite{DBLP:books/mk/AllenK2001}.
Support for named commutative blocks (NCBs) requires that we extend PDGs with commutativity information and develop new algorithms for parallelizing programs in accordance with these extended PDGs. 
To this end, we first incorporate conditional commutativity information about NCBs into an extended PDG that we call a \emph{conditionally commutative PDG} (or PDG$_C$) in two ways:
(i) nodes can be for an entire named commutative block $\nblock{f}$ rather than further delving into the body of the block, and
(ii) we add a new kind of edge 
containing the commutativity condition $\varphi_{\nblock{f}}^{\nblock{g}}$ between NCBs $\nblock{f}$ and $\nblock{g}$. For Ex.~\ref{ex:motivation}, the PDG$_C$ includes the edges:\\
$\begin{array}{lll}
	[\codemath{i=0}] \xrightarrow{\codemath{i}} [\codemath{while(i<size)}], &
	[\codemath{while(i<size)}] \xrightarrow{\codemath{i}} \nblockA{f}{i}, &
	\multicolumn{1}{l}{
	\nblockA{f}{i} \xrightarrow{\text{commute}\;\codemath{(i!=j $\mid\mid$ (i==j\&\& arr[i]>0))}} \nblockA{g}{j}
	}
\end{array}$

\smallskip
(Step 2.2)
We next describe algorithms that discover a program-specific parallelization scheme and (Step 2.3) transform the input program with NCBs into a collection of inter-operating tasks called a \emph{job-generating program}. Each task in this program graph consists of some subtree of the AST,
\begin{wrapfigure}[11]{r}[1pt]{2.8cm}
        \begin{tikzpicture}[
            node distance=1.7cm and 1.7cm,
            every node/.style={font=\small},
            tasknode/.style={draw, minimum size=1.2cm, font=\small},
            subtask/.style={Stealth-, thick, draw=black},
            commute/.style={Stealth-Stealth, thick, draw=black},
            scale=0.8, transform shape
        ]
		\node[tasknode] (s1) {\begin{minipage}[t]{1.0in}  \lstinline|while(i < size){ $\;\;\newJob$(2,[i]); $\;\;\newJob$(3,[i]);|  \newline \lstinline|$\;\;$i = i + 1; }| \ignore{\newline Deps: [Init:Vars(\code{i})]} \end{minipage}};
		\node[rotate=90] (t1) [left=.3cm of s1.north west] {{\bf Task1}};
        \node[tasknode] (s2) [below=.3cm of s1] {\begin{minipage}[t]{1.0in} \lstinline|$\nblockA{f}{i}\block{...}$| \newline Deps: \{T1:\code{i}, T3:$\varphi_{\nblock{f}}^{\nblock{g}}$\}\end{minipage}};
		\node[rotate=90] (t2) [left=.3cm of s2.north west] {{\bf Task2}};
        \node[tasknode] (s3) [below=.3cm of s2] {\begin{minipage}[t]{1.0in}\lstinline|$\nblockA{g}{j}\block{...}$| \newline Deps: \{T1:\code{i}, T2:$\varphi_{\nblock{f}}^{\nblock{g}}$\} \end{minipage}};
		\node[rotate=90] (t3) [left=.3cm of s3.north west] {{\bf Task3}};

        \end{tikzpicture}
\end{wrapfigure}
but with child subtrees possibly replaced with asynchronous calls to other tasks. For example, when applied to Ex.~\ref{ex:motivation}, our algorithm generates the collection of tasks to the right.
This task program organizes units of work along with the data and/or commutativity dependences between those jobs, indicated via dependences, such as the commutativity specifications in {\bf Task2} with respect to {\bf Task3}. 
These tasks are given as input to the execution model, discussed next, which has the responsibility of ensuring jobs correctly await their dependences but are allowed to execute in parallel when runtime commutativity conditions hold.

The $\newJob{}$ statements in {\bf Task1} are special instructions that tell the runtime to create a new job (instance of a task with runtime values). This is the main way new jobs (and thus parallel units of work) are created and scheduled. 
Sec.~\ref{sec:analysis} describes the algorithms that
construct the PDG$_C$, a parallelization scheme that decides which tasks can be asynchronous.
The process merges PDG nodes with consideration to loop-carried dependences and commutativity of NCBs, and 
reconstructs AST subtrees to assemble each task.
Jobs (whether from NCBs or other parallelized blocks) may touch shared
memory. {To ensure correctness, a job and a job that it overtook (one that precedes it in program order, but whose dependence it eliminated because their commutativity condition held) must execute atomically with respect to each other: their steps may run concurrently, but only in ways equivalent to running one entirely before the other (details \POPL{in~\cite{technicalreport}}\ARXIV{in Sec.~\ref{sec:correctness}}). Other jobs that run concurrently need no synchronization of this kind, as the correctness proof shows that they are independent.} This can be enforced by standard protocols such as two-phase locking or
optimistic concurrency control~\cite{DBLP:conf/pldi/FlanaganQ03}. Our
implementation {addresses} it via a simple program analysis for lock synthesis~\cite{DBLP:conf/popl/EmmiFJM07},
run as a dedicated post-parallelization stage: given the job-generating
program output by the parallelization analysis, this stage infers and inserts
{locks around conflicting accesses to shared variables, which suffice for most pairs of blocks in our benchmarks; \name{} can also enforce the requirement directly (Sec.~\ref{sec:correctness}).}

\emph{Challenge 3: Execution model for NCB programs.} The task graph above is given as input to the runtime. The principal runtime challenge is to execute jobs according to a schedule that respects dependences, does not deadlock, utilizes commutativity, and ensures the concurrent execution is equivalent to a sequential execution of the original program. While a job executes, it may create a new job (via \newJob), which may have data or commutative dependences. To complicate matters, the job may also depend on \emph{all other future jobs of the same kind (i.e. same task)} or it may even require waiting on a parent's asynchronous execution of other AST siblings. 

We present a formal execution model (Sec.~\ref{sec:execution_model}) and implementation (Sec.~\ref{sec:implementation}) that organizes waiting safely, yet allows jobs without dependences or those that commute with existing jobs to proceed concurrently. 
\POPL{This execution model bears some similarity to that of Jade~\cite{jade}, and improves on the state of the art (e.g.~\cite{ps-dswp,jade,commset}) by exploiting verified commutativity conditions.}
\ARXIV{During execution every job goes through a few kinds of waiting ``phases,'' ensuring not only that it waits for every dependent job, but also any job that could later be spawned that it must be ordered after. This execution model improves on the state of the art. PS-DSWP~\cite{ps-dswp} lacks commutativity and lacks the ability to specify regions of code to be treated as a single unit, CommSet~\cite{commset} could not dynamically exploit conditional commutativity, and Veracity~\cite{veracity} was restricted to adjacent pairs of commutative code.}

\emph{Challenge 4: Performance and Verification.}
Finally, our evaluation shows that \name{} (i) largely subsumes existing works~\cite{ps-dswp,commset,veracity}, (ii) enables new forms of parallelization not possible with any existing work, and (iii) is able to verify commutativity conditions across all of the benchmarks.

\section{Preliminaries}
\label{sec:prelim}

\emph{States.}
We would like to capture a notion of scoping for variables, \eg{} for expressing a global state on top of a local state.
So, we introduce structure to our notion of states to capture scoping.
A state $\sigma \in \Sigma$ is either a partial mapping $f_\sigma : \mathit{Varnames} \sqcup \mathit{MemLocs} \rightharpoonup V$ (\ie~the disjoint union of variable names and memory locations to program values) or the sum of two states $\sigma_0 \oplus \sigma_1$.
A lookup is defined as such: if the state is a partial map, it is merely the value of the partial map; if the state is $\sigma_0 \oplus \sigma_1$, then it is defined recursively $(\sigma_0 \oplus \sigma_1) := \mathit{if} \sigma_0(x) \, \mathit{is} \: \mathit{defined} \: \mathit{then} \, \sigma_0(x) \, \mathit{else} \, \sigma_1(x)$.
For a state $\sigma$ and a set $X$ of variables and memory locations, the \emph{restriction} $\sigma{\upharpoonright}_X$ is the partial map that agrees with $\sigma$ on $X$ and is undefined elsewhere: $(\sigma{\upharpoonright}_X)(x) = \sigma(x)$ if $x \in X$.
Updates are defined similarly: if the state is a partial map, $\sigma[x \mapsto \mathit{val}] := \lambda y. \mathit{if} \: x = y \: \mathit{then} \: \mathit{val} \: \mathit{else} \: f_\sigma(y)$, and if the state is a sum, $(\sigma_0 \oplus \sigma_1)[x \mapsto \mathit{val}] := \mathit{if} \sigma_0(x) \, \mathit{is} \: \mathit{undefined} \: \mathit{and} \: \sigma_1(x) \, \mathit{is} \: \mathit{defined} \: \mathit{then} \: \sigma_0 \oplus (\sigma_1[x \mapsto \mathit{val}]) \: \mathit{else} \: (\sigma_0[x \mapsto \mathit{val}]) \oplus \sigma_1$.

{\bf Programming language.}
The syntax is that of a standard imperative language with \texttt{if}, \texttt{for}, and \texttt{while}.
There are constants $c$, references, mutable variables and declarations. We use $t$ to refer to types, $c$ to constants, $v$ to variables, and $e$ to expressions. 
\POPL{The full grammar is given \inApx{apx:redexes}.}
\ifARXIV
The full grammar is given in Fig.~\ref{fig:apx-grammar}.
\begin{figure*}[p!thb]
	\begin{center}
	$\begin{array}{llllll}
		\mathit{id} &::=& \texttt{[a-z\_][a-zA-Z\_0-9]*} \\
		c &::=& \mathit{int\text{-}literal} \mid \mathit{string\text{-}literal} \mid \mathit{bool\text{-}literal} \mid () \\
		e &::=& c \mid e_0 \; \mathit{bop} \: e_1 \mid \mathit{uop} \: e \mid e_0 ? e_1 : e_2 \mid e_0[e_1] \mid (e) \\
		n &::=& \texttt{new} \: t[e] \mid \texttt{new hashtable}[t_0, t_1] \\
		\mathit{v\text{-}decl} &::=& t \: \mathit{id} = e &
		\mathit{v\text{-}decls} &::=& \mathit{v\text{-}decl} \mid \mathit{v\text{-}decl} , \mathit{v\text{-}decls} \\
		\mathit{lval} &::=& \mathit{id} \mid e_0[e_1] &
		\mathit{blk\text{-}lbl} &::=& \mathit{id} \mid \mathit{id_{lbl} \langle id_0, ... , id_n \rangle} \\
		\mathit{blk} & ::= & \{ \mathit{stmt}* \} \\
		\mathit{stmt} &::=& 
			\multicolumn{4}{l}{
				\mathit{v\text{-}decl}; \mid \mathit{lval} = e; \mid \texttt{if(} e \texttt{)} \mathit{blk} \mid \texttt{if(} e \texttt{)} \mathit{blk} \: \texttt{else} \: \mathit{blk} \mid \texttt{while(} e \texttt{)} \: \mathit{blk}
			}\\
		&& \multicolumn{4}{l}{ 
			\texttt{for(} \mathit{v\text{-}decls}; e_0; s_0 \texttt{)} \: \mathit{blk} \mid \mathit{blk} \mid \mathit{blk\text{-}lbl} : \mathit{blk}
		}
	\end{array}$
	\end{center}
	\caption{Complete grammar, including $t$ types, $c$ constants, $v$ variables, $e$  expressions, and more.}\label{fig:apx-grammar}
\end{figure*}
\fi

\emph{Semantics.}
\noindent
We define redexes in the usual way, and define reduction rules on redexes.
\POPL{The full list of redexes is given \inApx{apx:redexes}; examples include:
\[\begin{array}{lll}
	r &::=& 
	v \mid c_0[c_1] \mid \texttt{new }t[c]  \mid \texttt{new hashtable}[t_0,t_1] \mid c_0 \text{ bop } c_1 \\
	&& \mid lval = c \mid \texttt{while}(e) \block{s} \mid \sskip; s  \mid \texttt{if(true)}\block{s_0}\texttt{else}\block{s_1} 
\end{array}\]}
\ARXIV{The redexes are defined as follows:}
\ifARXIV
	\begin{center}
	$\begin{array}{lll}
		r &::=& 
		v \mid \text{deref } p \mid c_0[c_1] \mid \texttt{new }t[c] 
		\mid \texttt{new hashtable}[t_0,t_1] \mid \text{uop } c \mid c_0 \text{ bop } c_1 \\
		&&\mid \texttt{true} ? e_0 : e_1 \mid \texttt{false} ? e_0 : e_1 \mid c.\text{fieldname} 
		\mid lval = c \mid t\text{ }v = c \mid \texttt{while}(e) \block{s} \mid \sskip; s  \\
		&& \mid \texttt{for(} \mathit{v\text{-}decls} \texttt{; } e \texttt{; } s_0 \texttt{)} \block{s_1} \mid \texttt{if(true)}\block{s_0}\texttt{else}\block{s_1}
		\mid \texttt{if(false)}\block{s_0}\texttt{else}\block{s_1} \\
		&& \mid \block{s} \mid \nblockA{f}{$v_0, ... , v_n$}\block{s}
	\end{array}$
	\end{center}
\fi
Redex reduction rules are straightforward and denoted
$\angles{r, \sigma} \leadsto \angles{r', \sigma'}$.
\POPL{A few examples are in Fig.~\ref{fig:reductions} and the rest are \inApx{apx:redexes}.}
\ARXIV{The redex reduction rules are given in Fig.~\ref{fig:apx-redex}.}
The reductions indicate that variables and memory accesses are looked up atomically from the program state.

\ifARXIV
\begin{figure*}[t]
	\begin{center}
		$
		\begin{array}{rcl}
			\angles{v, \sigma} & \leadsto & \angles{\sigma(v), \sigma} \\ 
			\angles{\text{deref } p, \sigma} & \leadsto & \angles{\sigma(p), \sigma} \\ 
			\angles{c_0[c_1], \sigma} & \leadsto & \angles{\sem{c_0[c_1] : ref}\sigma, \sigma} \\
			\angles{\texttt{new }t[c], \sigma} & \leadsto & \angles{p, \sigma[p \mapsto \texttt{new }t[c] : \text{ref}]} \\
			\angles{\texttt{new hashtable}[t_0,t_1], \sigma} & \leadsto & \angles{p, \sigma[p \mapsto \texttt{new hashtable}[t_0,t_1] : \text{ref}]} \\
			\angles{\text{uop } c, \sigma} & \leadsto & \angles{\sem{\text{uop }c}\sigma, \sigma} \\
			\angles{c_0 \text{ bop } c_1, \sigma} & \leadsto & \angles{\sem{c_0 \text{ bop } c_1}\sigma, \sigma} \\
			\angles{\texttt{true} ? e_0 : e_1, \sigma} & \leadsto & \angles{e_0, \sigma} \\
			\angles{\texttt{false} ? e_0 : e_1, \sigma} & \leadsto & \angles{e_1, \sigma} \\
			\\
			\angles{lval = c, \sigma} & \leadsto & \angles{\sskip, \sigma[lval \mapsto c]} \\
			\angles{t\text{ }v = c, \sigma} & \leadsto & \angles{\sskip, \sigma[v \mapsto c]} \\
			\angles{\texttt{if(true)}\block{s_0}\texttt{else}\block{s_1}, \sigma} & \leadsto & \angles{s_0, \sigma} \\
			\angles{\texttt{if(false)}\block{s_0}\texttt{else}\block{s_1}, \sigma} & \leadsto & \angles{s_1, \sigma} \\
			\angles{\texttt{for}(\mathit{v\text{-}decls};e;s_0) \block{s_1}, \sigma} & \leadsto & \angles{v\text{-}decls; \texttt{while}(e)\block{s_1; s_0}, \sigma} \\
			\angles{\texttt{while}(e) \block{s}, \sigma} & \leadsto & \angles{\texttt{if}(e)\block{s; \texttt{while}(e)\block{s}}\texttt{else}\block{\sskip}, \sigma} \\
			\angles{\block{s}, \sigma} & \leadsto & \angles{s, \sigma} \\
			\angles{\nblockA{f}{$v_0, ..., v_n$}:\block{s}, \sigma} & \leadsto & \angles{s, \sigma} \\
			\angles{\sskip; s, \sigma} & \leadsto & \angles{s, \sigma} \\
		\end{array}
		$
	\end{center}
	\caption{Redex reductions in the extended language.}\label{fig:apx-redex}
\end{figure*}
To obtain the small-step semantics from the redexes, we decompose a program into a context and redex, and reduce the redex:
\[
\infer[\text{SS}]{\angles{H[r], \sigma} \leadsto \angles{H[r'], \sigma'}}
{\angles{r, \sigma} \leadsto \angles{r', \sigma'}}
{}
\]
For even further detail on the language specification, namely for the definition of contexts, please refer to~\citet{veracity}.

\else

\begin{figure}[t]\footnotesize
\[
		\begin{array}{rclrcl}
 			\angles{v, \sigma} & \leadsto & \angles{\sigma(v), \sigma} &
			\angles{\texttt{if(true)}\block{s_0}\texttt{else}\block{s_1}, \sigma} & \leadsto & \angles{s_0, \sigma} \\
			\angles{c_0 \text{ bop } c_1, \sigma} & \leadsto & \angles{eval(c_0 \text{ bop } c_1, \sigma), \sigma} &
			\angles{\texttt{if(false)}\block{s_0}\texttt{else}\block{s_1}, \sigma} & \leadsto & \angles{s_1, \sigma} \\
			\angles{v = c, \sigma} & \leadsto & \angles{\sskip, \sigma[v \mapsto c]} &
			\angles{\texttt{while}(e) \block{s}, \sigma} & \leadsto & \angles{\texttt{if}(e)\block{s; \texttt{while}(e)...}, \sigma} 
		\end{array}
\]
\vspace{-10pt}
\caption{\label{fig:reductions} Abridged reduction rules for the grammar (full rules \inApx{apx:redexes}).}
\end{figure}
\fi
We focus on \emph{conditional commutativity} and would like those conditions to be \emph{valid}. Suppose we are given program statements $s_0$ and $s_1$, and a predicate on states $\varphi^{s_0}_{s_1} : \Sigma \rightarrow \mathbb{B}$. 

\begin{definition}
	\label{Sufficient Commutativity Condition}
	$\varphi^{s_0}_{s_1}$ is a \emph{valid commutativity condition} for $s_0$ and $s_1$ when
for all $\sigma_0, \sigma_{m_0}, \sigma_{m_1}, \sigma_{f_0}, \sigma_{f_1}$,
if $\angles{s_0, \sigma_0} \leadsto^* \angles{\sskip, \sigma_{m_0}} \wedge \angles{s_1, \sigma_{m_0}} \leadsto^* \angles{\sskip, \sigma_{f_0}} \wedge \angles{s_1, \sigma_0} \leadsto^* \angles{\sskip, \sigma_{m_1}} \wedge \angles{s_0, \sigma_{m_1}} \leadsto^* \angles{\sskip, \sigma_{f_1}}$ then
$\varphi^{s_0}_{s_1}(\sigma_0) \implies \sigma_{f_0} = \sigma_{f_1}$
\end{definition}
\noindent
We will extend this to a similar notion fit for our non-local execution model in Sec.~\ref{Valid NCB Commutativity Condition}.

\medskip
{\bf Data Dependence and Program Dependence Graph (PDG)}.
Our algorithms build on the following established definitions.

\label{subsec:data}
\begin{definition}[Direct Data Dependence]

Let $s_1$ and $s_2$ be two statements in a program. We say that there exists a \emph{direct data dependence} from $s_1$ to $s_2$ if the following conditions hold:
\begin{enumerate}
    \item \textit{Define}: The statement $s_1$ defines the variable $x$ (e.g., through an assignment).
    \item \textit{Use}: The statement $s_2$ uses the variable $x$ (e.g., $y = x + 1$).
    \item \textit{Reachability}: The definition of $x$ in $s_1$ reaches $s_2$ along some execution path without any intermediate statement redefining $x$ between $s_1$ and $s_2$.
\end{enumerate}
\noindent
In this case, $s_2$ is said to be \emph{data-dependent} on $s_1$ due to the variable $x$.
\end{definition}

A Program Dependence Graph (PDG)~\cite{DBLP:journals/toplas/FerranteOW87} represents dependences between statements in a program, including both data and control dependences. 
The PDG is widely used for program analysis~\cite{DBLP:books/mk/AllenK2001}, including parallelization, optimization, and verification.

\begin{definition}
    \label{def:pdg}
    A PDG is a directed graph $G = (N, E)$ where:
    \begin{itemize}
        \item $N$ is a set of nodes, each $n \in N$ corresponding to an
        element of the program's AST.

        \item $E \subseteq N \times \mathcal{L} \times N$ is a set of
        labeled directed edges, where $\ell \in \mathcal{L}$ denotes a
        dependence type:
        \begin{enumerate}
            \item \emph{Data dependence}: $(n_i, \textsf{data}[vars^+], n_j) \in E$,
            written $n_i \xrightarrow{vars^+} n_j$, if each $x \in vars^+
            \subseteq \mathit{Varnames}$ is defined at $n_i$, used at $n_j$,
            and reaches $n_j$ without redefinition.

            \item \emph{Control dependence}: $(n_i, \textsf{control}, n_j) \in E$,
            written $n_i \dashrightarrow n_j$, if the execution of $n_j$ is
            conditionally dependent on the control predicate at $n_i$.
        \end{enumerate}
    \end{itemize}
\end{definition}

\emph{Loop-carried Dependence.}
	A loop-carried dependence (LCD) arises when the computation in one loop iteration relies on results produced by previous iterations. Given two memory access in a loop, a loop-carried dependence exists if they can access the same memory locations on different iterations and at least one access is a write operation.
An LCD enforces a sequential ordering between iterations $i$ and $j$, as iteration $j$ cannot execute until iteration $i$ completes the source access. For example, in \lstinline|for (i = 1; i < n; i++) {A[i] = A[i-1] + 1}|, each iteration depends on the value computed in the previous iteration, creating an LCD on \lstinline|A[i] = A[i-1] + 1|.
 
DOALL~\cite{DBLP:books/mk/AllenK2001} is a parallelization technique that can be used when there are no LCDs. Loop iterations are distributed across multiple threads since they can be executed independently.

\section{Global Commutativity Specifications}
\label{sec:ncb}

This section tackles \emph{how to specify the commutativity of nonadjacent NCB code fragments} so that, at runtime, those specifications can be evaluated to true/false to decide whether it is safe for the fragments to proceed in parallel.
Since correctness of these specifications is essential to sound execution,  Sec.~\ref{sec:verification} will then address how commutativity specifications can be automatically verified.

As we saw in Sec.~\ref{sec:overview}, there are two components: (1) the ability to \emph{name} regions of code with region-specific input variable names, and (2) the ability to write commutativity specifications as expressions over the shared state as well as the region-specific input variable names. 

A named commutative block (NCB) denoted $\nblockA{f}{$\bar{v}$}$ (where $\bar{v}$ is short for $v_1, ..., v_n$) is similar 
to the usual kind of program block, except that it is given a name (or ``label'') $\nblock{f}$ 
\begin{wrapfigure}[5]{r}{0.37\textwidth}
\begin{lstlisting}[style=vcy,language=VCY,basicstyle=\ttfamily\scriptsize]
	int main(int x, int y) {
		|$\nblockA{f}{x}$|:{ |a[x] = a[x] + 1;| }
		if (x<y) |$\nblockA{g}{x,y}$|:{ |a[y] = x;| }
	}
\end{lstlisting}
\end{wrapfigure}
and that label can refer to program variables such as \code{x} that are \emph{inputs} to the computation performed in the body of that block (read-only, as formalized below). To the right is a simple example.
(Here \code{a} is a global array.)
This program has two NCBs, $\nblockA{f}{x}$ has an input variable \code{x} and $\nblockA{g}{x,y}$ has two input variables \code{x} and \code{y}.
These NCB parameters should not be thought of as function arguments, but more akin to a meta syntax. They associate local variable names with positions in the parameterized NCB naming: above for $\nblock{f}$, local variable ``\code{x}'' is in the \emph{first} position.

Having established these variable positions, ``positional alias variables'' can be used in the commutativity specifications to avoid name clashes when two NCB fragments refer to their own input values of the same variable such as \code{x} in the example above (or in a loop index that changes between instances).
For example, one can write a commutativity condition 
$\varphi^{\nblockA{f}{x1}}_{\nblockA{g}{x2,y2}} := \codemath{(x1 != y2)}$. This specification (saying the fragments write distinct slots of \code{a}) uses \code{x1} as the positional alias for \code{x} in $\nblock{f}$, and \code{x2} as the positional alias for \code{x} in $\nblock{g}$. The positional aliases thus allow the commutativity specification to refer to the value of \code{x} in $\nblock{g}$ as distinct from the value of \code{x} in $\nblock{f}$. 

\emph{Self-commutativity.} In addition to enabling commutativity with \emph{other} fragments, providing a name enables commutativity between a fragment and \emph{itself}. For example, in the loop
\lstinline|for(int i=0;i<N;i++)$\nblockA{h}{i}$:{ arr[i] = i; }|,
NCB $\nblockA{h}{i}$ will allow for specifications of how 
the body of the loop may commute with itself for different iterations, \ie, for different NCB input values of \code{i}. This is achieved with an NCB commutativity condition such as:
$\varphi^{\nblockA{h}{i}}_{\nblockA{h}{j}} := \codemath{(i != j)}$
, indicating that the body of the loop commutes with itself on different iterations.

NCB commutativity specifications are evaluated at runtime. When a scheduler decides whether two such fragments may run in parallel (Sec.~\ref{sec:execution_model}), it evaluates their condition on the state at that moment, which has three parts: the variables shared by the pair, and each instance's snapshot of its own input parameters. The rest of this section makes this precise. Sec.~\ref{sec:ncb-blocks} adds NCBs to the language. Sec.~\ref{sec:ncb-partition} partitions the variables and the state of a pair of NCBs into a shared part and one local part per instance, and states the well-formedness conditions. Secs.~\ref{sec:ncb-spec-syntax} and~\ref{sec:ncb-spec-sem} give the syntax and semantics of conditions over this partition, and Sec.~\ref{sec:ncb-sound} defines when a condition is sound. Sec.~\ref{sec:verification} then shows how to verify soundness.

\subsection{Named commutative blocks}
\label{sec:ncb-blocks}

To support NCBs, we extend statements in the grammar of the language:
\[\boxed{\begin{array}{llll}
	stmt &::=& \cdots \mid  ncblabel \langle \textrm{varname}^* \rangle: \lbrace stmt^* \rbrace \\
\end{array}}\]
An $ncblabel$ is a simple label, and we assume it is distinct from other kinds of labels in the program.

We naturally also extend the reduction semantics (discussed in Sec.~\ref{sec:prelim}; full rules \POPL{\inApx{apx:redexes}}\ARXIV{in Fig.~\ref{fig:apx-redex}}) for this language extension, extending redexes to include 
${r ::= \cdots \mid ncblabel \langle \textrm{varname}^* \rangle: \lbrace stmt^* \rbrace}$.
The sequential semantics of this extended language is straightforward: the reduction simply ignores the label: 
${\angles{\mathit{ncblabel} \langle \textrm{varname}^* \rangle:\block{s}, \sigma}  \leadsto  \angles{s, \sigma}}$.
(In Sec.~\ref{sec:execution_model} we will discuss the concurrent semantics of NCBs in the execution model.)

\subsection{Variable and state partitioning}
\label{sec:ncb-partition}

Consider two NCBs $\nblockA{f}{$x_1,\ldots,x_n$}:\{s_\nblock{f}\}$ and $\nblockA{g}{$y_1,\ldots,y_m$}:\{s_\nblock{g}\}$ for which there is a commutativity specification entry (Sec.~\ref{sec:ncb-spec-syntax}).

\begin{definition}[Variable classes] 
\label{def:var-classes}
Variables fall into the following classes:
\begin{itemize}
\item Global variables ($\GVars{}$): the set of variables that are in the scope that encloses both  $s_\nblock{f}$ and $s_\nblock{g}$ (we sometimes casually refer to this as ``global variables'').
This set depends on the pair, and could be written $\GVars{}(\nblock{f},\nblock{g})$ but we simply write $\GVars{}$ when the pair is clear from context.
\item NCB block input and local variables ($\Varsf$): 
Each NCB fragment is invoked with its own \emph{input} parameters and its body $s_\nblock{f}$ may declare more variables (and $\Varsg$ for $\nblock{g}$ variables).
\end{itemize}
\ARXIV{\rocqref{Commutativity.ncb, Commutativity.is_global}}
\end{definition}

\ifARXIV
\begin{figure}
\input{varclasses}
\end{figure}
\else
\begin{wrapfigure}[12]{r}[1pt]{5.3cm}
\input{varclasses}
\end{wrapfigure}
\fi
\noindent
In the Fig.~\ref{fig:scopes} example,
$\Varsf=\{\texttt{i,u,x}\}$, $\Varsg=\{\texttt{i,y}\}$, and $\GVars=\{\texttt{a,b,n,i,u}\}$.
$\GVars{}$ is typically not disjoint from $\Varsf$ nor from $\Varsg$: an NCB parameter names a variable of the enclosing scope (e.g. \code{i} and \code{u}) so it belongs to $\GVars{}$, yet each execution instance also keeps its own copy of the parameter's value at entry, so it belongs to $\Varsf$ as well.

The two play different roles: the shared variable \code{i} lives in $\sigma_{gl}$ and keeps changing (\eg{} the loop increments it), whereas the instance's copy in $\sigma_\nblock{f}$ is fixed for the instance's lifetime, and reads in $s_\nblock{f}$ see the copy because they try $\sigma_\nblock{f}$ first. Only the variables declared inside an NCB body are disjoint from $\GVars{}$.
Since \code{i} is in both $\Varsf$ and $\Varsg$, and each instance executes with respect to its own copy, the name \code{i} alone would be ambiguous in a condition relating the two instances. Below, the positional aliases of commutativity specifications (Sec.~\ref{sec:ncb-spec-syntax}) will resolve this collision.

Note that, in addition to $\GVars{}$ and the $\nblock{f}$- and $\nblock{g}$-local variables, a program may have variables that are in scope at $s_\nblock{f}$ but not at $s_\nblock{g}$ (or vice versa). These form two further classes, $In_\nblock{f}$ and $In_\nblock{g}$, of inputs that the fragment may read but not write; we formalize them \inApx{apx:in-vars}. For simplicity, we assume here that $In_\nblock{f} = In_\nblock{g} = \emptyset$, counting such variables in $\GVars{}$ even if only one NCB accesses them.

\emph{Well-formedness.}
We require NCBs to be \emph{well-formed} in that every commutativity specification entry satisfies the following three conditions, all of which can be determined statically:
\begin{definition}\label{def:well-formed}
	For every commutativity specification entry over NCBs $\nblockA{f}{$\bar x$}:\{s_\nblock{f}\}$ and $\nblockA{g}{$\bar y$}:\{s_\nblock{g}\}$,
\begin{itemize*}
\item (W0) \emph{In scope}: variable names occurring as parameters in $\bar x$ are in scope at the program location of the NCB $\nblock{f}$ and likewise for $\bar y$ and $\nblock{g}$ (w.r.t. Fig.~\ref{fig:scopes}, \code{u} is in scope at $\nblock{f}$).
\item (W1) \emph{Parameters are read-only}: neither $s_\nblock{f}$ nor $s_\nblock{g}$ assigns a variable in $\bar x \cup \bar y$ (e.g. \code{i,u} are read-only in $\nblock{f}$).
\item (W2) \emph{No other non-shared writes}: every variable that $s_\nblock{f}$ assigns is declared in $s_\nblock{f}$ or belongs to $\GVars{}$, and likewise for $s_\nblock{g}$ (e.g. $\nblock{f}$ only assigns \code{x} and \code{a}).
\end{itemize*}
\ARXIV{\rocqref{Commutativity.W1, Commutativity.W2, Commutativity.wf_pair}}
\end{definition}
\noindent
Since $\GVars{}$ depends on the pair, an NCB that occurs in several entries must satisfy (W2) with respect to each of them.
Condition (W1) constrains only the NCB bodies, not the surrounding code. In \lstinline|for(int i=0;i<N;i++) $\nblockA{h}{i}$:\{...\}| (or in Ex.~\ref{ex:motivation} where \code{i=i+1;} occurs after $\nblock{g}$'s block) the increment is \emph{outside} every NCB body. Writes to a parameter \emph{between} instances are a reason for naming it: each instance must see its own entry value of \code{i}, not the current one.
\ARXIV{Under the assumption $In_\nblock{f} = In_\nblock{g} = \emptyset$ above, every variable that $s_\nblock{f}$ can assign is declared in $s_\nblock{f}$ or belongs to $\GVars{}$, so (W2) holds automatically; it becomes a real restriction only in the general setting of Apx.~\ref{apx:in-vars}.}

\begin{definition}[Partitioned state]
\label{def:state-triple}\label{def:partitioned-state}
A \emph{partitioned state} for an instance of $\nblock{f}$ and an instance of $\nblock{g}$ is a triple $(\sigma_{gl}, \sigma_\nblock{f}, \sigma_\nblock{g})$ of the shared state $\sigma_{gl}$ and the two instances' \emph{local states}.
The local state $\sigma_\nblock{f}$ is created when the instance is entered, as a copy $\sigma_\nblock{f} := \sigma{\upharpoonright}_{\{\bar x\}}$ of its parameters, where $\sigma$ is the state at that moment. While $s_\nblock{f}$ executes, $\sigma_\nblock{f}$ also holds the variables declared in $s_\nblock{f}$. Likewise for $\sigma_\nblock{g}$ and $\bar y$.
\ARXIV{\rocqref{Commutativity.frame, Commutativity.merge, Commutativity.run_ncb}}
\end{definition}

\noindent
The snapshot is what keeps instances apart. In Ex.~\ref{ex:motivation}, \code{i:=i+1} follows $\nblock{g}$'s block, so by the time two instances of $\nblock{g}$ are compared, the outer \code{i} has moved on, and each instance's own value of \code{i} is only in its local state.

An instance of $s_\nblock{f}$ runs on the sum $\sigma_l \oplus \sigma_{gl}$ of states (Sec.~\ref{sec:prelim}) of a fragment-local state $\sigma_l$ and the shared state $\sigma_{gl}$. Reads try $\sigma_l$ first, and an update goes to $\sigma_{gl}$ only if $\sigma_l$ does not define the variable and $\sigma_{gl}$ does. So a run ends in a sum $\sigma_l' \oplus \sigma_{gl}'$: variables declared in $s_\nblock{f}$ are in $\sigma_l'$, and every other write is to $\sigma_{gl}'$. While $s_\nblock{f}$ executes, $dom(\sigma_l)$ is $\{\bar x\}$ together with the variables declared so far in $s_\nblock{f}$. As discussed, it may overlap $\GVars{}$ when a parameter names a variable of $\GVars{}$; the shadowed shared variable then keeps its value.

\subsection{Commutativity specifications}
\label{sec:ncb-spec-syntax}

Commutativity specifications can include many pairs of NCBs and within a pair's condition $\varphi_{\nblockA{f}{$a_1,\ldots,a_n$}}^{\nblockA{g}{$b_1,\ldots,b_m$}}$, variables may be global variables or positional alias variables such as $a_k$, interpreted according to the appropriate portion of the partitioned state.

\begin{definition}[Commutativity specification syntax]
	\label{def:commutativity-spec-syntax}
\[\boxed{\begin{array}{llllll}
	gc\text{-}specs & ::= & \texttt{commutativity} \; \lbrace\; \lbrace \; \lbrace ncblabel \langle \textrm{varname}^* \rangle \rbrace + : \varphi_{\bullet}^{\bullet} \rbrace^* \; \rbrace \; \\
	\varphi_{\nblockA{f}{$a_1,\ldots,a_n$}}^{\nblockA{g}{$b_1,\ldots,b_m$}} &::=& \texttt{true} \mid \neg \varphi \mid \varphi \vee \varphi \mid ncb\text{-}e \circlelt ncb\text{-}e \\
	ncb\text{-}e &::=& c \mid ncb\text{-}e \;\; bop \;\; ncb\text{-}e 
	\mid ncb\text{-}e[ncb\text{-}e]
	\mid ncb\text{-}e.id
	\mid \ldots
	\mid v  \\
    prog &::=& gc\text{-}specs \;\; gdecls
\end{array}}\]
\ARXIV{\rocqref{Commutativity.cond}}
\end{definition}

Above $gc\text{-}specs$ specifies a collection of global commutativity specification entries, each over multiple labeled NCBs. For each specification, set of NCBs to be considered is given ($\bullet$ are instantiated to NCB labels) and then an expression for their pairwise commutativity.
The commutativity condition $\varphi_\nblock{f}^\nblock{g}$ is a boolean expression which can include boolean relations on (restricted, pure) value expressions $ncb\text{-}e$, or else standard boolean connectors.
Those $ncb\text{-}e$ expressions are comprised of standard expressions from the language such as constants, standard unary/binary operations, array index expressions, structure field dereference, read-only operations on built-in abstract data types such as dictionaries, etc. 
An $ncb\text{-}e$ expression can also be a variable $v$, whose value is interpreted according to the appropriate portion of the partitioned state ($\GVars$ or $\Varsf$ or $\Varsg$, as formalized below).
We use $bop$ to range over binary operations on NCB expressions.
\POPL{The full grammar is given \inApx{apx:comm-spec-grammar}.}
\ifARXIV
The full grammar is given in Fig.~\ref{fig:apx-comm-grammar}. A specification block $gc\text{-}specs$ consists of the keyword \texttt{commutativity} followed by zero or more entries; each entry lists one or more labeled NCBs with their parameters, and a condition that is interpreted pairwise over the corresponding NCB instances.
\begin{figure*}[p!thb]
	\begin{center}
	$\begin{array}{llllll}
		\mathit{id} &::=& \texttt{[a-z\_][a-zA-Z\_0-9]*} \\
		ncb\text{-}e &::=& c \mid v \mid \mathit{uop} \: ncb\text{-}e \mid ncb\text{-}e \;\mathit{bop}\; ncb\text{-}e \mid (ncb\text{-}e) \\
		& & \mid ncb\text{-}e_0 ? ncb\text{-}e_1 : ncb\text{-}e_2 \mid ncb\text{-}e[ncb\text{-}e] \mid ncb\text{-}e.id \\
		\varphi_{\nblockA{f}{x$_\nblock{f}$}}^{\nblockA{g}{x$_\nblock{g}$}} &::=& ncb\text{-}e \circlelt ncb\text{-}e \mid \texttt{true} \mid \neg \varphi \mid \varphi \vee \varphi \mid \varphi \wedge \varphi \\
		gc\text{-}specs & ::= & \texttt{commutativity} \; \lbrace\; \lbrace \; \lbrace ncblabel \langle \textrm{varname}^* \rangle \rbrace + : \varphi_{\bullet}^{\bullet} \rbrace^* \; \rbrace \; \\
		prog &::=& gc\text{-}specs \;\; gdecls
	\end{array}$
	\end{center}
	\caption{Complete grammar of the commutativity specification language.}
	\label{fig:apx-comm-grammar}
\end{figure*}
\fi
For simplicity we represent values as simple integers.
\ARXIV{Finally, a program $prog$ is extended to include global commutativity specification preludes.}

A specification entry $\nblockA{f}{$a_1,\ldots,a_n$}, \nblockA{g}{$b_1,\ldots,b_m$} : \varphi$ for the NCBs fixed in Sec.~\ref{sec:ncb-partition} names their parameters by \emph{positional aliases}, which will be used for the semantics of a condition $\varphi$.

\begin{definition}[Variable association]
\label{def:var-assoc}
For a specification entry $\nblockA{f}{$a_1,\ldots,a_n$}, \nblockA{g}{$b_1,\ldots,b_m$} : \varphi$
corresponding to NCB blocks $\nblockA{f}{$x_1,\ldots,x_n$}:s_\nblock{f}$, and 
block $\nblockA{g}{$y_1,\ldots,y_m$}:s_\nblock{g}$, 
the \emph{alias maps} $A_\nblock{f} : \{\bar{a}\} \to \{\bar{x}\}$ and $A_\nblock{g} : \{\bar{b}\} \to \{\bar{y}\}$ map each positional alias to the NCB parameter in its corresponding position, $A_\nblock{f}(a_k) = x_k$ and $A_\nblock{g}(b_k) = y_k$. We require $dom(A_\nblock{f}) \cap dom(A_\nblock{g}) = \emptyset$.
\ARXIV{\rocqref{Commutativity.cond_ok, Commutativity.Pi}}
\end{definition}

\noindent
\name{} uses this association throughout: verification (Sec.~\ref{sec:verification}) translates aliases to fresh variables and $\GVars{}$ to the shared {\sc Servois2} state, and {\sc ForkTask} (Sec.~\ref{sec:execution_model}) captures an NCB's parameters into the new job's local state at creation.

\subsection{Semantics of commutativity conditions}
\label{sec:ncb-spec-sem}

\begin{definition}
A commutativity condition $\varphi_\nblock{f}^\nblock{g}$ is interpreted over a partitioned state $(\sigma_{gl}, \sigma_\nblock{f}, \sigma_\nblock{g})$ by two mutually defined functions:
a boolean interpretation $\sem{\cdot}_\mathbb{B}$ of conditions $\varphi$, and an
integer interpretation $\sem{\cdot}_\mathbb{N}$ of NCB expressions $ncb\text{-}e$.
Selected cases:
\[
\begin{array}{lll}
\sem{\texttt{true}}_\mathbb{B}(\sigma_{gl}, \sigma_\nblock{f}, \sigma_\nblock{g}) &=& \texttt{true}\\
\sem{\varphi_1 \vee \varphi_2}_\mathbb{B}(\sigma_{gl}, \sigma_\nblock{f}, \sigma_\nblock{g}) &=& \sem{\varphi_1}_\mathbb{B}(\sigma_{gl}, \sigma_\nblock{f}, \sigma_\nblock{g}) \vee \sem{\varphi_2}_\mathbb{B}(\sigma_{gl}, \sigma_\nblock{f}, \sigma_\nblock{g})\\
\sem{ncb\text{-}e_1 \circlelt ncb\text{-}e_2}_\mathbb{B}(\sigma_{gl}, \sigma_\nblock{f}, \sigma_\nblock{g}) &=& \sem{ncb\text{-}e_1}_\mathbb{N}(\sigma_{gl}, \sigma_\nblock{f}, \sigma_\nblock{g}) \circlelt \sem{ncb\text{-}e_2}_\mathbb{N}(\sigma_{gl}, \sigma_\nblock{f}, \sigma_\nblock{g})\\[4pt]
\sem{v}_\mathbb{N}(\sigma_{gl},\sigma_\nblock{f},\sigma_\nblock{g}) &=& 
  \left\{ 
	\begin{array}{ll}
		\sigma_{gl}(v) & \text{if } v \in \GVars\\
		\sigma_{\nblock{f}}(A_\nblock{f}(v))  & \text{if } v \in dom(A_\nblock{f})\\
		\sigma_{\nblock{g}}(A_\nblock{g}(v))  & \text{if } v \in dom(A_\nblock{g})
	\end{array}
  \right. \\
\sem{\textsf{v}_{gl}}_\mathbb{N}(\sigma_{gl},\sigma_\nblock{f},\sigma_\nblock{g}) &=& \sigma_{gl}(\textsf{v}_{gl})\\
\sem{ncb\text{-}e_1 \;\;bop\;\; ncb\text{-}e_2}_\mathbb{N}(\sigma_{gl},\sigma_\nblock{f},\sigma_\nblock{g}) &=& \sem{ncb\text{-}e_1}_\mathbb{N}(\sigma_{gl},\sigma_\nblock{f},\sigma_\nblock{g}) \;\;bop\;\; \sem{ncb\text{-}e_2}_\mathbb{N}(\sigma_{gl},\sigma_\nblock{f},\sigma_\nblock{g})
\end{array}
\]
\POPL{Unabridged definition \inApx{apx:commspecinterp}.}
\ARXIV{The remaining cases, writing $\bar\sigma$ for $(\sigma_{gl}, \sigma_\nblock{f}, \sigma_\nblock{g})$ and treating an array value as a map from indices to values are as follows.
\[
\begin{array}{lll}
\sem{\texttt{false}}_\mathbb{B}(\bar\sigma) &=& \texttt{false}\\
\sem{\neg \varphi}_\mathbb{B}(\bar\sigma) &=& \neg \sem{\varphi}_\mathbb{B}(\bar\sigma)\\
\sem{\varphi_1 \wedge \varphi_2}_\mathbb{B}(\bar\sigma) &=& \sem{\varphi_1}_\mathbb{B}(\bar\sigma) \wedge \sem{\varphi_2}_\mathbb{B}(\bar\sigma)\\
\sem{c}_\mathbb{N}(\bar\sigma) &=& c\\
\sem{\mathit{uop}\; ncb\text{-}e}_\mathbb{N}(\bar\sigma) &=& \mathit{uop}\; \sem{ncb\text{-}e}_\mathbb{N}(\bar\sigma)\\
\sem{ncb\text{-}e_0 \,?\, ncb\text{-}e_1 : ncb\text{-}e_2}_\mathbb{N}(\bar\sigma) &=& \sem{ncb\text{-}e_1}_\mathbb{N}(\bar\sigma) \text{ if } \sem{ncb\text{-}e_0}_\mathbb{N}(\bar\sigma) \neq 0, \text{ else } \sem{ncb\text{-}e_2}_\mathbb{N}(\bar\sigma)\\
\sem{ncb\text{-}e_0[ncb\text{-}e_1]}_\mathbb{N}(\bar\sigma) &=& \sem{ncb\text{-}e_0}_\mathbb{N}(\bar\sigma)\big(\sem{ncb\text{-}e_1}_\mathbb{N}(\bar\sigma)\big)
\end{array}
\]
A field access $ncb\text{-}e.id$ selects the field $id$ of the record value of $ncb\text{-}e$.}
The boolean interpretation $\sem{\cdot}_\mathbb{B}$ \ARXIV{(also written as simply $\sem{\cdot}$)} includes the standard connectives, as well as boolean relations $\circlelt$ over integer expressions which are interpreted by $\sem{\cdot}_\mathbb{N}$.
The integer semantics $\sem{\;}_\mathbb{N}$ is also as expected. 
For variables, it interprets positional alias variables, via the appropriate association $A_\nblock{f}$ or $A_\nblock{g}$ to its corresponding local variable.

\ARXIV{\rocqref{Commutativity.cond_eval, Commutativity.cond_holds}}
\end{definition}
\noindent
Every free variable of a well-formed $\varphi$ is a positional alias or a variable of $\GVars{}$ (Def.~\ref{def:var-assoc}). So $\varphi$ never reads body locals, and its value does not depend on how far the two instances have run. {\sc DepElim} (Sec.~\ref{sec:execution_model}) evaluates $\varphi$ on $(\gstate{}, \sigma_l, \sigma_l')$, the shared state and the local states of the two jobs.

\subsection{Sound commutativity conditions}
\label{sec:ncb-sound}

We require commutativity conditions be a sound approximation of the commutativity of NCBs.
Intuitively, commutativity means that if the two NCB bodies are executed in any context where the condition $\varphi$ holds (thus placing constraints on the global variables and fragment-local inputs), their effects on the global variables will be the same in either order.
Note that this is a property of the two fragment bodies, rather than a requirement that the program ever executes them as if they appeared back-to-back in the source.
We formalize this as follows:

\begin{definition}%
	\label{Valid NCB Commutativity Condition}
	$\varphi^{\nblock{f}}_{\nblock{g}}$ is a \emph{sound commutativity condition} for well-formed (Def.~\ref{def:well-formed}) NCBs $\nblockA{f}{$\bar{x}$}\{ s_0 \}$ and $\nblockA{g}{$\bar{y}$} \{ s_1 \}$ when
	\[
	\begin{array}{llll}
			\multicolumn{3}{l}{\forall \sigma_{gl}^0, \sigma_{gl}^1, \sigma_{gl}^{1'},
			\sigma_{gl}^2, \sigma_{gl}^{2'}, \sigma_{\nblock{f}}, \sigma_{\nblock{f}}',
			\sigma_{\nblock{f}}'', \sigma_{\nblock{g}}, \sigma_{\nblock{g}}', \sigma_{\nblock{g}}'':} \\
			\;\;\;\angles{s_0, \sigma_{\nblock{f}} \oplus \sigma_{gl}^0} \leadsto^* \angles{\sskip, \sigma_{\nblock{f}}' \oplus \sigma_{gl}^1} &\wedge&
			\angles{s_1, \sigma_{\nblock{g}} \oplus \sigma_{gl}^1} \leadsto^* \angles{\sskip, \sigma_{\nblock{g}}' \oplus \sigma_{gl}^2} &\wedge \\
			\;\;\;\angles{s_1, \sigma_{\nblock{g}} \oplus \sigma_{gl}^0} \leadsto^* \angles{\sskip,  \sigma_{\nblock{g}}'' \oplus \sigma_{gl}^{1'}} & \wedge&
			\angles{s_0, \sigma_{\nblock{f}} \oplus \sigma_{gl}^{1'}} \leadsto^* \angles{\sskip, \sigma_{\nblock{f}}'' \oplus \sigma_{gl}^{2'}} &\wedge \\
			\multicolumn{3}{l}{\;\;\;\sem{\varphi^{\nblock{f}}_{\nblock{g}}}(\sigma_{gl}^{0},\sigma_{\nblock{f}},\sigma_{\nblock{g}}) = \texttt{true} \implies \sigma_{gl}^2 = \sigma_{gl}^{2'}}
		\end{array}
    \]
\ARXIV{\rocqref{Commutativity.sound_comm_cond}}
\end{definition}
\noindent
This definition has a similar shape to Def.~\ref{Sufficient Commutativity Condition}, except that it allows for NCB-specific input states $\sigma_{\nblock{f}}$ and $\sigma_{\nblock{g}}$, and the commutativity condition is evaluated under the partitioned state $(\sigma_{gl}^{0}, \sigma_{\nblock{f}}, \sigma_{\nblock{g}})$.
Local states $\sigma_{\nblock{f}}$ and $\sigma_{\nblock{g}}$ provide NCB input, but their final values obtained in either order
(i.e. states $\sigma_{\nblock{f}}',\sigma_{\nblock{g}}',\sigma_{\nblock{f}}'',\sigma_{\nblock{g}}''$) are irrelevant.
Discarding them loses nothing, because of (W1) and (W2):
\begin{lemma}[Local states can be discarded]\label{lem:discard}
Let $\nblockA{f}{$\bar{x}$}\{s_0\}$ be well-formed, let $\sigma_{gl}$ be any shared state (not necessarily restricted to $\GVars{}$), and let $\angles{s_0, \sigma_{\nblock{f}} \oplus \sigma_{gl}} \leadsto^* \angles{\sskip, \sigma_{\nblock{f}}' \oplus \sigma_{gl}'}$. Then
(i) $\sigma_{\nblock{f}}'$ agrees with $\sigma_{\nblock{f}}$ on $\{\bar x\}$, and
(ii) $\sigma_{gl}'$ agrees with $\sigma_{gl}$ outside $\GVars{}$.
\ARXIV{\rocqref{Commutativity.ncb_readonly, Commutativity.ncb_writes_only_global}}
\end{lemma}
\ARXIV{{\begin{proof}
A run never changes a variable that its statement does not assign. By (W1), $s_0$ assigns no parameter, which gives (i). By (W2), it assigns only variables declared in $s_0$ and variables of $\GVars{}$. The former are in $dom(\sigma_{\nblock{f}})$, so they never reach $\sigma_{gl}'$, which gives (ii). For the consequence, the final shared states agree on $\GVars{}$ by Def.~\ref{Valid NCB Commutativity Condition}, and elsewhere both equal the initial shared state, by (ii) applied to each of the four runs.
\end{proof}
\noindent
Because $In_\nblock{f} = In_\nblock{g} = \emptyset$, a variable that a fragment writes but does not declare is in $\GVars{}$, so Def.~\ref{Valid NCB Commutativity Condition} compares its final value in both orders. Apx.~\ref{apx:in-vars} shows that (W2) is necessary once $In$ variables are allowed.}}

\noindent
Consequently, if $\varphi^{\nblock{f}}_{\nblock{g}}$ is sound for well-formed $\nblock{f}$ and $\nblock{g}$, the two orders in Def.~\ref{Valid NCB Commutativity Condition} end in equal shared states, not merely in states that agree on $\GVars{}$.
\POPL{The proof is \inApx{apx:proof-discard}.}

\emph{Stability.} Def.~\ref{Valid NCB Commutativity Condition} justifies swapping one pair. At run time, a job may overtake several jobs (Sec.~\ref{sec:execution_model}). Moving it back behind them in program order then swaps it with each of them in turn, each time in the state after the earlier ones have run, so each condition must still hold there. We write $\varphi^{\nblock{g}}_{\nblock{f}}$ for the condition under which $\nblock{g}$ may run before $\nblock{f}$, where $\nblock{f}$ precedes $\nblock{g}$ in program order.
\begin{definition}[Stable conditions]\label{def:stable}
Let NCBs $\nblock{h}$, $\nblock{f}$, $\nblock{g}$ (in program order) have bodies $s_\nblock{h}$, $s_\nblock{f}$, $s_\nblock{g}$ and entries for all three pairs. Their conditions are \emph{stable} if, for all $\sigma_{gl}^0, \sigma_{gl}^1, \sigma_\nblock{h}, \sigma_\nblock{h}', \sigma_\nblock{f}, \sigma_\nblock{g}$,
\[\begin{array}{l}
\angles{s_\nblock{h}, \sigma_\nblock{h} \oplus \sigma_{gl}^0} \leadsto^* \angles{\sskip, \sigma_\nblock{h}' \oplus \sigma_{gl}^1}
\wedge \sem{\varphi^{\nblock{f}}_{\nblock{h}}}(\sigma_{gl}^0, \sigma_\nblock{f}, \sigma_\nblock{h})
\wedge \sem{\varphi^{\nblock{g}}_{\nblock{h}}}(\sigma_{gl}^0, \sigma_\nblock{g}, \sigma_\nblock{h})
\wedge \sem{\varphi^{\nblock{g}}_{\nblock{f}}}(\sigma_{gl}^0, \sigma_\nblock{g}, \sigma_\nblock{f})\\
\implies \sem{\varphi^{\nblock{g}}_{\nblock{f}}}(\sigma_{gl}^1, \sigma_\nblock{g}, \sigma_\nblock{f})\end{array}
\]
A specification is stable if the conditions of every three NCBs (not necessarily distinct) with entries for all three pairs are stable.
\ARXIV{\rocqref{Stability.stable3}}
\end{definition}
That is, running one of three jobs whose conditions all hold keeps the condition between the other two. Like Def.~\ref{Valid NCB Commutativity Condition}, this is a property of the code and the specification, and can be checked in the same way. Without it, overtaking is unsound even for sound conditions: with $\nblockA{f}{k}:\{$\code{arr[0] = arr[0] - k;}$\}$, $\nblock{g}:\{$\code{if (arr[0] > 0) x = 1;}$\}$ and the sound entry $\nblock{g}, \nblockA{f}{k} : 0 < \texttt{k} \wedge \texttt{k} < \texttt{arr[0]}$, from \code{arr[0] = 5} a job of $\nblock{g}$ may overtake two jobs of $\nblockA{f}{3}$ (the condition holds for each), ending with \code{x = 1}, whereas program order ends with \code{x = 0}. The specification of Ex.~\ref{ex:motivation} is stable: $\nblock{f}$ only adds \code{x*x} to \code{arr[i]}, so \code{arr[i] > 0} stays true, and $\nblockA{g}{c}$ changes only \code{arr[c]}, which another instance $\nblockA{g}{a}$ with $a \neq c$ does not read. Both facts are mechanized (details \inApx{apx:rocq-examples}).

\section{Verifying commutativity specifications}
\label{sec:verification}

If $\varphi_\nblock{f}^\nblock{g}$ is not a valid commutativity condition, then non-commutative NCBs may be executed concurrently, leading to executions that are not equivalent to some sequential execution (unserializable). Moreover, writing valid commutativity conditions can often be difficult for users. Thus this section describes a method for verifying \ekremoved{and sometimes synthesizing}commutativity conditions. \ekremoved{Although synthesis perhaps seems the more appealing option, for several reasons synthesis is difficult and incomplete so it is sometimes helpful to manually provide conditions that are then verified.}

Prior work~\cite{veracity} showed that program code can be translated to logical pre/post conditions, for which the tool {\sc Servois2}~\cite{DBLP:conf/atva/ChenFNPBK23} can synthesize or verify commutativity conditions. For example, two adjacent code fragments 
\code{\{ x:=x+1; \}\{ if(x>0) y:=y-1; \}} can be translated to post conditions 
of the form \code{x'=(x+1)} and \code{y'=(x>0 ? y : (y-1))}, with SSA-like \code{let} clauses being used for more complicated code. {\sc Servois2} can then verify or synthesize the commutativity condition \code{x>0}.

Unfortunately this prior translation does not immediately support NCBs because it assumes the code fragments share a common scope, are distinct from each other (i.e. not self-commuting fragments such as in loop bodies), and do not have a way to refer to variables through their NCB naming.
We thus  describe how to translate named commutative blocks $\nblockA{f}{x}\texttt{\{ $stmt_\nblock{f}$ \}}$ and $\nblockA{g}{y}\texttt{\{ $stmt_\nblock{g}$ \}}$ into logical method pre/post conditions,
along with a translation of the commutativity condition
$\nblockA{f}{i},\nblockA{g}{j} : \varphi_{\nblock{f}}^{\nblock{g}}$, into an ADT input for {\sc Servois2}, such that a {\sc Servois2}-verified translated commutativity condition implies that $\varphi_{\nblock{f}}^{\nblock{g}}$ is a valid commutativity condition for $\nblock{f}$ and $\nblock{g}$.

Intuitively, the translation ensures that the occurrences of variables in the code and specifications are correctly partitioned into states $(\sigma_{gl}, \sigma_\nblock{f}, \sigma_\nblock{g})$, by isolating the local environments of $\nblockA{f}{i}$ and $\nblockA{g}{i}$ and producing SMT expressions for transition relations that may update the global variables, but cannot depend on the opposite fragment's local input variables. This state partitioning is also saved so that the runtime can evaluate the commutativity condition correctly.

\input{translation}

\section{Program Analysis}
\label{sec:analysis}

\newcommand\pdgcAnalysis{{\sc PDG$_C$Analysis}}
\newcommand\mkTaskDepGraph{{\sc buildParGraph}}
\newcommand\mkJobGenProg{{\sc buildJobGenProg}}
\newcommand\jgTaskGraph{\mathcal{P}_{\textsf{jg}}}

In this section we discuss dependence analysis and parallelization procedures for programs with named commutative blocks. 
The main steps are shown in Algorithm~\ref{alg:analysis}\POPL{ to the}
\ifARXIV
\begin{figure}
\input{mainalg}
\end{figure}
\else
\begin{wrapfigure}[11]{r}[1pt]{7.2cm}
\begin{minipage}{7.0cm}
\vspace{-10pt}
\input{mainalg}
\end{minipage}
\end{wrapfigure}
\fi
\POPL{right}; this section discusses each step in turn.
The output is a parallelized \emph{job-generating program} $\jgTaskGraph=(initTask, T)$,
where $initTask$ constructs the initial configuration and $T$ is a set of tasks that can asynchronously invoke each other.
$\jgTaskGraph{}$ is input to the execution model (Sec.~\ref{sec:execution_model}) that schedules jobs (\ie{} instances of tasks with runtime values) that may generate other jobs, while abiding dependence and commutativity constraints.
At the end of this section we will show that the resulting $\jgTaskGraph{}$ satisfies a \emph{faithfulness} condition that relates the task graph back to the original program. Ultimately, faithfulness will be combined with pairwise serializability (due to commutativity soundness and synchronization) to prove the equivalence between concurrent and sequential executions. 

An example of the analysis steps applied to Ex.~\ref{ex:motivation} is given in Fig.~\ref{fig:transformation-process}.
We now discuss the three analysis steps, referring to corresponding to the three subsequent subfigures of Fig.~\ref{fig:transformation-process}.

\smallskip
\noindent
{\bf (1) Commutativity-Dependence Analysis (\pdgcAnalysis)}.
As with many dependence-based parallelization techniques, we continue to leverage the program dependence graph (PDG) to represent dependences between program statements.
However, to support parallel execution of nonadjacent commutative blocks, the input program has \ncb{}s $\nblockA{f}{i}$, $\nblockA{g}{j}$, etc.
We would like these \ncb{}s to be considered for parallelization as single units so that their commutativity can permit them to execute concurrently.
Therefore, we extend the program dependence graph (cf. Def.~\ref{def:pdg}):
\begin{definition}
A \emph{commutativity PDG} (PDG$_C$) is like a PDG graph $G=(V,E)$, but:
\begin{itemize}
\item A node can be a named commutative block $\nblock{f}$ (\ie{} a compound statement) in addition to being an individual statement $s$.
\item In addition to data and control, edges can alternatively be 
$(\nblockA{f}{i},[\varphi_{\nblock{f}}^{\nblock{g}}],\nblockA{g}{j})$, which includes the commutativity condition $\varphi_{\nblock{f}}^{\nblock{g}}$, also denoted $\nblockA{f}{i} \xrightarrow{\varphi_{\nblock{f}}^{\nblock{g}}} \nblockA{g}{j}$.
\end{itemize}
\end{definition}
\noindent
Permitting PDG nodes that are NCBs is a departure from traditional PDG-based parallelization~\cite{ps-dswp,DBLP:books/mk/AllenK2001}, where the algorithm would recursively traverse into each block for deeper parallelization. By contrast, NCBs explicitly \emph{prevent} deeper traversal for two reasons:
(1) by using NCBs, programmers can control the granularity of parallelization, halting it from going deeper into specific regions of the code, thereby allowing more intentional management of parallel tasks;
(2) commutativity-based conflict is more general than read/write dependences which sometimes can appear to conflict despite operations being commutative, \eg{}~two NCBs that increment the value of an integer variable. 

The algorithm for constructing the PDG$_C$ is similar to standard algorithms~\cite{DBLP:journals/toplas/FerranteOW87,ps-dswp,DBLP:books/mk/AllenK2001} except that, when reaching an AST node for an NCB, we create a single node for it and ensure that the nodes related to the NCB are never merged with other nodes. We also add commutativity edges between NCB nodes based on the global commutativity specification.

\newcommand\ncbf{\nblock{f}}
\newcommand\ncbg{\nblock{g}}
\newcommand\varsubfrag[2]{\code{#1}$_{\nblock{#2}}$}
\newcommand\varsubfragprime[2]{\code{#1}$_{\nblock{#2}'}$}
\newcommand\isubf{\varsubfrag{i}{f}}
\newcommand\isubg{\varsubfrag{i}{g}}
\newcommand\isubfprime{\varsubfragprime{i}{f}}
\newcommand\isubgprime{\varsubfragprime{i}{g}}

An example PDG$_C$ is given in Fig.~\ref{fig:pdg}. Nodes are freshly labeled $n_i$ and correspond to either a single program expression/statement or a named block: $\nblock{f}$ or $\nblock{g}$.
The dashed grey edges are traditional PDG control-dependence edges, and the single-point block edges are traditional PDG data-dependence edges labeled with the dependent variables. The three bolded green boxes/edges show the new commutativity-labeled edges. For example, the edge
$
\nblock{f}
\xrightarrow{
  \text{\isubf\code{!=}\isubg $\mid\mid$ (\isubf\code{==}\isubg \&\& \code{arr}[\isubf]\code{>0})}
}
\nblock{g}
$
records the commutative-dependence between 
$\nblock{f}$ and $\nblock{g}$.

\begin{figure}[t]
    \centering
    \begin{subfigure}[b]{.95\textwidth}
        \centering
		\includegraphics[width=1.0\textwidth]{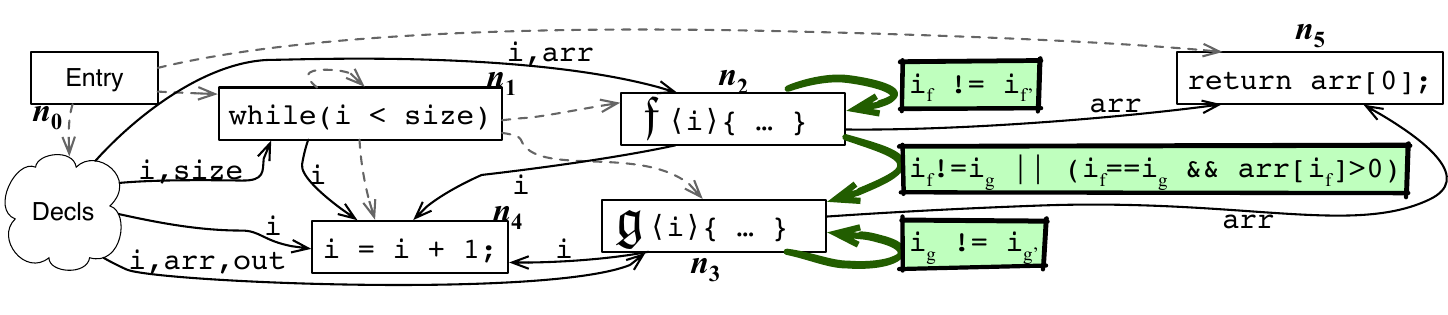}
        \caption{Commutative Program Dependence Graph (PDG$_C$). Edges are for data dependences with variable names (solid black), control dependences (dashed grey), and commutativity conditions (bold green).}
        \label{fig:pdg}
    \end{subfigure}

    \begin{subfigure}{0.42\textwidth}
		\centering
        \begin{tikzpicture}[
            node distance=.5cm and .5cm,
            every node/.style={font=\small},
            tasknode/.style={draw, minimum size=0.4cm},
            subtask/.style={-Stealth, thin, draw=black},
            scale=0.8, transform shape
        ]
        \node[tasknode] (init) {\begin{minipage}[t]{0.7in} {\bf Entry} \ignore{\newline Globals: \code{i},\code{arr}, \code{size}, \code{out}} \newline cpdg:$\{\text{(decls)}\}$ \end{minipage}};
		\node[tasknode] (s0) [right=1.5cm of init] {\begin{minipage}[t]{0.5in} {\bf Seq} \newline cpdg:$\{n_5\}$ \end{minipage}};
		    \node[right=0.05cm of s0, anchor=west] {{$v_4$}};
        \node[tasknode] (s1) [below=0.7cm of init] {\begin{minipage}[t]{0.7in} {\bf Seq} \newline cpdg:$\{n_1,n_4\}$ \end{minipage}};
            \node[left=0.05cm of s1, anchor=east] {{$v_1$}};
        \node[tasknode] (s2) [below=0.7cm of s1] {\begin{minipage}[t]{0.7in} {\bf Async} \newline cpdg:$\{n_2\}$ \end{minipage}};
            \node[left=0.05cm of s2, anchor=east] {{$v_2$}};
		\node[tasknode] (s3) [right=1.5cm of s1] {\begin{minipage}[t]{0.5in} {\bf Async} \newline cpdg:$\{n_3\}$\end{minipage}};
            \node[right=0.05cm of s3, anchor=west] {{$v_3$}};
        
		\draw[subtask, dashed] (init) to[bend right=15] (s0);
		\draw[subtask, dashed] (init) to[bend right=24] (s1);
		\draw[subtask, dashed] (s1) to[bend right=20] (s2);
		\draw[subtask, dashed] (s1) to[bend right=15] (s3);

		\draw[subtask] (init) -- node[draw=none, above] {\texttt{arr}} (s0);
        \draw[subtask] (init) -- node[draw=none, right] {\texttt{i,size,arr}}(s1);
        \draw[subtask] (s1) -- node[draw=none, right] {\texttt{i}} (s2);
		\draw[subtask] (s1) -- node[draw=none, above] {\texttt{i}} (s3);
		\draw[subtask] (init) to[bend right=82] node[draw=none, above, pos=0.3pt] {\texttt{arr}} (s2);
		\draw[subtask] (init) -- node[draw=none, above, pos=0.9pt] {\texttt{arr,out}} (s3);
		
		\draw[<->] (s2.east) to[bend right=30] node[draw=none, pos=0.8pt, fill=green!20] {\isubf{}\texttt{!=}\isubg{} $\mid\mid$ (\isubf\texttt{==}\isubg{} \&\& \texttt{arr}[\isubf]\texttt{>0})} (s3);

		\path[<->] (s2.east) edge[loop, in=-30, out=-60, distance=8mm] node[draw=none, right, fill=green!20] {\isubf\texttt{!=}\isubfprime} (s2.east);
		\path[<->] (s3.north) edge[loop, in=25, out=60, distance=8mm] node[draw=none, right, fill=green!20] {\isubg{!=}\isubgprime} (s3.north);

        \end{tikzpicture}
		\caption{Parallelization Graph (generalizes merged PDG of \citet{ps-dswp}). Each vertex contains multiple PDG$_C$ nodes, and is marked as either Sequential or Asynchronous.}
		\label{fig:taskgraph}
	\end{subfigure}
	\hfill
    \begin{subfigure}{0.56\textwidth}
		\centering
		\begin{tikzpicture}[
            node distance=.3cm and .3cm,
            every node/.style={draw, minimum size=1.4cm, font=\small},
            subtask/.style={-Stealth, thin, draw=black},
            scale=0.8, transform shape
        ]
		
		\node (init) {\begin{minipage}[t]{1.5in} {\bf Init Task}: \newline \lstinline|decls: [...];| \newline \lstinline|{$\newJob$(T1,[i;size;arr]);| \newline \lstinline|$\;\;\newJob$(T4,[arr]);}| \end{minipage}};
		\node (s1) [below=of init] {\begin{minipage}[t]{1.5in} {\bf Task1}:\newline  \lstinline|while(i < size){| \newline \lstinline|$\;\;\newJob$(T2,[i]);| \newline \lstinline|$\;\;\newJob$(T3,[i]); i=i+1;}| \newline Deps: \{T2:Vars(\code{i}), T3:Vars(\code{i})\} \end{minipage}};
		\node (s0) [below=of s1] {\begin{minipage}[t]{1.5in} {\bf Task4}:\newline \lstinline|{return arr[0];}| \newline Deps:\{Init:Vars(\code{arr})\} \end{minipage}};
		\node (s2) [anchor=north west] at ([xshift=0.25cm]init.north east) {\begin{minipage}[t]{2.0in} {\bf Task2}: \newline \lstinline|$\nblockA{f}{i}\block{...}$| \newline Deps: \newline \{T1:Vars(\code{i}), \newline T2:Com(\isubf\code{!=}\isubfprime),\newline  T3:Com(\isubf\code{!=}\isubg $\mid\mid$ (\isubf\code{==}\isubg{} $\&\&$ \code{arr}[\isubf]\code{>0})))\} \end{minipage}};
		\node (s3) [below=of s2] {\begin{minipage}[t]{2.0in} {\bf Task3}: \newline \lstinline|$\nblockA{g}{i}\block{...}$| \newline Deps:\{T1:Vars(\code{i}), \newline T2:Com(\isubf\code{!=}\isubg{} $\mid\mid$ (\isubf\codemath{==}\isubg \&\& \codemath{arr}[\isubf]\codemath{>0})), \newline T3:Com(\isubg\code{!=}\isubgprime)\}\end{minipage}};

        \end{tikzpicture}
		\caption{Resulting Job-Generating Program $\jgTaskGraph$. AST subtrees are reconstructed with asynchronous calls to other tasks. Data and/or commutative dependences are listed to inform the execution model as to what waiting is needed.}
		\label{fig:jobgenprog}
	\end{subfigure}
	
    \caption{Example of the analysis from the original \name{} program, to a collection of job-generating tasks.}
    \label{fig:transformation-process}
\end{figure}

\begin{figure}[t]
    \begin{algorithm}[H]
    \caption{Parallelization Graph Construction}
    \label{alg:Task-Gen}
    \begin{minipage}[t]{0.44\textwidth}
        \begin{algorithmic}
        \Procedure{\mkTaskDepGraph}{cpdg}
            \State $sccG \gets SCC(cpdg)$
            \State $cpdg1 \gets MarkAllVsAsync(sccG)$
            \State $cpdg2 \gets AdjustVLabels(cpdg1)$
            \State $cpdg3 \gets \Call{MergeAsync}{cpdg2}$
            \State $cpdg4 \gets \Call{MergeSeq}{cpdg3}$
            \State \Return $cpdg4$
        \EndProcedure
        \Procedure{CanMergeAsync}{$v_1, v_2$}
            \State \Return ($v_{1}.label = v_{2}.label = \text{async-vertex}$ \textbf{and} $v_{1}, v_2 \; \text{not NCBs}$ \textbf{and} $\neg\textit{ExistsIntermediateBlock}(v_1, v_2)$ \textbf{and} $\neg\textit{HasLoopCarriedArcs}(v_1, v_2)$)
        \EndProcedure
        \end{algorithmic}
    \end{minipage}
    \hfill
	\vline
    \begin{minipage}[t]{0.55\textwidth}
        \begin{algorithmic}
        \Procedure{MergeAsync}{$G$}
            \If{$\nexists(v_1,v_2)$ s.t.~$\Call{CanMergeAsync}{v_1, v_2}$} G \Else
            \; pick $(v_1,v_2)$ s.t.~$\Call{CanMergeAsync}{v_1, v_2}$
            \State \Call{MergeAsync}{\text{Merge}$(G, v_1, v_2)$}
            \EndIf    
        \EndProcedure
        \Procedure{MergeSeq}{$G$}
            \If{$\nexists(v_1,v_2)$ s.t.~$\text{AreSeqAdjacent}(v_1, v_2)$} G \Else
            \; pick $(v_1,v_2)$ s.t.~$\text{AreSeqAdjacent}(v_1, v_2)$
            \State \Call{MergeSeq}{\text{Merge}$(G, v_1, v_2)$}
            \EndIf    
        \EndProcedure
        \end{algorithmic}
    \end{minipage}
    \end{algorithm}
\end{figure}

\smallskip
\noindent
{\bf (2) Parallelization Graph Construction (\mkTaskDepGraph)}.
The next step is to determine, based on the PDG$_C$,  how to group code into fragments and to decide whether each fragment can be executed asynchronously or must be sequential.
PS-DSWP~\cite{ps-dswp} performs a similar procedure, but creates only a single ``doall'' fragment for loop parallelization and then partitions across a parametric number of threads. By contrast, we allow an arbitrarily nested degree of async-vs-sequential behavior, and assume a thread pool for execution.
To distinguish from a PDG$_C$ \emph{node}, we here refer to parallelization graphs as having \emph{vertices}.

Algorithm~\ref{alg:Task-Gen} describes how a parallelization graph is constructed from the PDG$_C$. First, a strongly-connected component algorithm is used, except that vertices corresponding to NCBs \emph{are not} merged with vertices corresponding to other statements. 
Nodes that are merged in general may introduce dependence cycles in the graph; we discuss how to overcome such cycles in Sec.~\ref{sec:execution_model}.
The next step involves labeling vertices to determine whether code segments can be executed asynchronously or must be sequential. We begin by labeling all vertices in the strongly-connected graph as $Async$. We continue by analyzing each node's internal edge connections. If we detect a loop-carried dependence, we relabel that node as $Seq$ (short for ``sequential'').

After labeling, we optimize by merging vertices with identical labels where possible. 
The algorithm consolidates vertices, combining $Seq$ vertices with other $Seq$ vertices, and $Async$ vertices with other $Async$ vertices, while keeping NCBs as separate vertices.
Two $Async$ vertices, $v_1$ and $v_2$, are eligible for merging if they satisfy a few criteria ({\sc CanMergeAsync}):
(i) We do not merge NCBs.
(ii) In the SCC graph, we do not merge if there exists an intermediate block $v_3$ that creates a path from $v_1$ to $v_2$ %
Such an intermediate block would induce cyclic dependences between blocks after merging, potentially creating thread-level dependence cycles.
Although cyclic dependences are handled (Sec.~\ref{sec:execution_model}), avoiding them can yield more parallelism.
(iii) None of the dependence edges between PDG$_C$ nodes in $v_1$ and $v_2$ should be loop-carried.
Loop-carried dependences inherently restrict parallel execution by creating sequential constraints between iterations, which would violate the properties required for safe merging of $Async$ vertices. 
(iv) Any dependences related to control flow must be respected within the merged structure.
The algorithm ensures that control dependences between $v_1$ and $v_2$ are maintained, avoiding conflicts that could result from merging nodes that rely on specific control flows.

The algorithm merges greedily until no further valid merges can be performed. This approach effectively groups parallelization graph vertices to maximize parallel execution potential while preserving essential program dependences.

The parallelization graph generated for the PDG$_C$ in Fig.~\ref{fig:pdg} is shown in Fig.~\ref{fig:taskgraph}. This graph introduces new merged vertices that capture the asynchronous or sequential properties of the program regions (\eg{}~vertex $v_1$ is a $Seq$ vertex resulting from merging PDG$_C$ nodes $n_1$ and $n_4$), with edges reflecting the dependences and commutativity relationships, and all global variable declarations integrated into a single $Entry$ node.

\smallskip
\noindent
{\bf (3) Translation to a Job-Generating Program (\mkJobGenProg)}.
Next the parallelization graph is used to decompose a program into a collection of synchronous or asynchronous \emph{tasks}, each of which may have a dependence or commutative relationship with other tasks. An instance of a task (which has runtime values for variables) is called a \emph{job}. For example, in the body of the loop
\texttt{for(i=0;i<n;i++)$\nblockA{f}{i}$:\{...\}}, 
there is only one \emph{task} $\nblock{f}$, but an execution of this loop could create \texttt{n} \emph{jobs}, each with a different input value for \texttt{i}.
This allows for each job to perform a different computation (\eg{} different iterations of a loop), while having a small number of tasks with predictable dependences.

\begin{definition}[Task]
	A task $T$ is a triple $(tid, s, D)$ where $tid \in \mathbb{N}$ is a unique task identifier, $s$ is a statement, and $D$ is a set of dependences. For each dependence $(tid', V, \varphi \textsf{ option}) \in D$, $tid'$ identifies the dependent task, $V$ is a set of variable data dependences, and $\varphi$ specifies when task $tid$ may commute with $tid'$ despite the $V$ dependences. 
	\label{def:task}
\ARXIV{\rocqref{Jobs.task}}
\end{definition}
\begin{definition}[Job]
	A job $(jid, tid, \sigma, s)$ has an id $jid$, task id $tid$, local state $\sigma$, and statement $s$.
	\label{def:jobs}
\ARXIV{\rocqref{Jobs.job}}
\end{definition}

\noindent	
\mkJobGenProg{} transforms a parallelization graph into a job-generating program (collection of tasks).
\mkJobGenProg{} operates on each vertex of the parallelization graph and 
reconstructs the program AST from the vertex's statement list. However, whenever task $i$ has some subtask $j$, execution of a job of task $i$ must be able to spawn a job of task $j$. For example, Fig.~\ref{fig:jobgenprog} shows the job-generating program (which is a set of tasks $T$ for Ex.~\ref{ex:motivation}).
Notice that {\bf Task1} may invoke {\bf Task2} with a $\newJob$ command:
since executing a job of one task may spawn jobs of other tasks, we introduce a statement to indicate that to the runtime:
\[\boxed{\begin{array}{llllll}
	stmt &::= \cdots \mid \newJob \; (tid, (t,argument)^*) %
\end{array}}\]
The $\newJob$ statement allows us to replace an AST subtree with a call to an asynchronously invoked child task $tid$ (consisting of said subtree), passing the necessary dependences. 

In the \mkJobGenProg{} algorithm (details \POPL{\inApx{apx:Job-Gen}}\ARXIV{below}), we construct tasks by traversing the
\begin{wrapfigure}[7]{r}[5pt]{4.0cm}
    \vspace{-10pt}\centering
    \begin{tikzpicture}[
		node distance=0.3cm,
		every node/.style={circle, draw, minimum size=0.05cm, inner sep=0.05cm},
		box/.style={rectangle, draw, rounded corners, inner sep=0.5cm}
	]
	
	\node (n2) {$n_2$};
	\node (n3) [below=of n2] {$n_3$};
	\node (n4) [below left=of n3] {$n_4$};
	\node (n5) [below left=of n4] {$n_5$};
	\node (n6) [below right=of n4] {$n_6$};
	\node (n7) [below right=of n3] {$n_7$};
	\node (n8) [below right=of n7] {...};
	
	\node[box, fit=(n7) (n8), label={[xshift=+0.08cm, yshift=-0.62cm]above:asyncJob}, xshift=0.05cm, yshift=-0.1cm, xscale=0.6, yscale=0.75] {};
	\node[box, fit=(n3) (n4) (n5) (n6) (n7), xshift=0.5cm, yshift=-0.1cm, xscale=1.05, yscale=0.75, label={[xshift=-0.5cm, yshift=-0.25cm]above right:Task$_i$}] {};
	
	\draw[->] (n2) -- (n3);
	\draw[->] (n3) -- (n4);
	\draw[->] (n4) -- (n5);
	\draw[->] (n4) -- (n6);
	\draw[->] (n3) -- (n7);
	\draw[->] (n7) -- (n8);
	
	\end{tikzpicture}
\end{wrapfigure}
original abstract syntax tree and using the parallelization graph to decide where to carve out subtrees into distinct tasks. When we reach a subtree that should be separate, we replace that subtree with a 
$\newJob$ statement, specifying the ID of the task corresponding to that subtree and the necessary data dependences.
To the right, we illustrate how this process works. In this example, a vertex in the
parallelization graph includes nodes $(n_3, n_4, n_5, n_6)$ from the original AST. Node $n_7$ was also part of the AST within this subtree but not part of the list for this parallelization graph vertex. Therefore, when applying the reconstruction algorithm to create a task from this parallelization graph vertex, node $n_7$ is replaced by a $\newJob$ statement. 

\ifARXIV
The \textit{Translation to a Job-Generating Program} algorithm constructs a job-generation program from a parallelization graph of input program, enabling effective parallelization by creating distinct tasks and dependences based on the program's abstract syntax tree (AST) and a parallelization graph (result of Alg.~\ref{alg:Task-Gen}). This algorithm consists of two main stages: reconstructing the AST and generating job tasks from the program's parallelization graph.

\paragraph{Reconstructing the AST: \textproc{ReconstructAST}.}
The \textproc{ReconstructAST} procedure recursively traverses the AST to identify and position tasks' bodies based on the parallelization graph.
Starting from AST entry node and the target vertex from parallelization graph, the function explores each node in the AST to determine if it is part of the current vertex, or it should be processed as an independent job.\\
{\bf Finding and Checking Subtrees in $V$}: The algorithm first attempts to locate the subtree within the AST that belongs to $V$ ($parGraph$'s vertex which is a set of AST's element) and could be separated as an independent task. Once the subtree is identified, it verifies that all nodes in this subtree are present in $V$. The process is as follows:
\begin{itemize}
	\item If the \texttt{foundSubtree} flag is \texttt{false}, the algorithm continues traversing down to sub-nodes. However, when it encounters named commutative blocks (NCBs) in the AST, the traversal stops at these nodes. This is because the NCBs are forced to remain as single nodes in the parallel dependence graph (PDG$_C$) to preserve their commutativity properties.
	\item If \texttt{foundSubtree} is \texttt{true} and some nodes in the subtree are not part of $V$, the algorithm replaces the missing part of the subtree with a $\newJob$ statement. This $\newJob$ statement represents an asynchronous job that encapsulates the excluded parts, enabling parallel execution while handling dependences. Dependences are calculated using \textproc{GetDataDependences} and are linked to this job, ensuring that data flows are managed correctly for the asynchronously executing portions of the AST.
\end{itemize}

\paragraph{Generating the job program: {\mkJobGenProg}.}
The {\mkJobGenProg} function is the core procedure that creates the job-generation program, $\jgTaskGraph$ which includes the \texttt{initTask} (the initial configuration) and a set of tasks, by leveraging the AST reconstruction. Here's its structure:

\begin{itemize}
    \item \textbf{Initialization Task}: It begins by creating an \texttt{initTask}, representing the entry point of the parallelized program. This task is generated based on the entry node of the parallelization graph and contains all the global declarations and a set of jobs that has to be created from that point.

    \item \textbf{Task Construction}: A set $T$ is initialized to hold all generated tasks. For each vertex in parallelization graph:
        \begin{itemize}
            \item A new task is created with a unique \texttt{id}.
            \item The task's body is constructed by calling \textproc{ReconstructAST} with the AST root and the specific node in the parallelization graph, indicating where this node should branch off as a new task in the AST.
            \item Dependences are calculated for the task based on \textproc{GetDependences}, capturing any necessary data flows between tasks.
        \end{itemize}
\end{itemize}
\fi

\noindent
{\bf (4) Lock Synthesis for pair-wise atomicity ({\sc LockSynth}).}
Concurrent jobs may mutate global state, so even when jobs commute sequentially, synchronization is necessary to ensure atomicity despite their interleaving. Note that this does not require full mutual exclusion: pair-wise atomicity can also be addressed with two-phase locking, optimistic concurrency control, etc.
Our implementation automatically synthesizes sufficient synchronization, through a static analysis similar to~\citet{DBLP:conf/popl/EmmiFJM07}. For lack of space, see Sec.~\ref{sec:implementation}\POPL{ and more details \inApx{apx:locksynth}}. More complicated synchronization synthesis~\cite{DBLP:conf/popl/EmmiFJM07,DBLP:conf/pldi/FlanaganQ03,DBLP:conf/popl/VechevYY10,DBLP:conf/ppopp/Golan-GuetaRSY15,DBLP:conf/pldi/CheremCG08,DBLP:conf/popl/McCloskeyZGB06,hicks2006lock,DBLP:journals/jpdc/DinizR98,DBLP:journals/concurrency/DinizR99,diniz1997dynamic} is orthogonal to this work and \Tool{} users can also provide their own synchronization.
{Formally, we require that a job and a job it overtook execute atomically with respect to each other, \ie{} their concurrent steps are equivalent to running one entirely before the other; that the other jobs running concurrently are independent is derived (assumption (S) in Sec.~\ref{sec:correctness}).} This requirement can be discharged by whichever synchronization synthesis algorithm is instantiated. %

\emph{Faithfulness.}
The algorithms above make choices over the dependency
graph when deciding what to parallelize. These choices  affect performance but
not soundness: every program they produce is \emph{faithful} to the original,
meaning that the job-generating program, when inlined back into a single AST,
has reduction semantics equivalent to those of the
original program. 
We sketch the faithfulness of the job-generating programs\ARXIV{ produced by our algorithms} \POPL{in~\cite{technicalreport}}\ARXIV{in Sec.~\ref{sec:correctness-proof}}\ARXIV{ (Lemma~\ref{lem:faithful-outline})}.

\newcommand\Pdswp{\mathcal{P}_\textsf{dswp}}

\section{Execution Model}
\label{sec:execution_model}

We now discuss how the transformed job-generating programs $\jgTaskGraph{}$ of Sec.~\ref{sec:analysis} are concurrently executed while respecting dependences and utilizing commutativity. 
We also discuss a proof sketch of correctness: that concurrent execution ends in an equivalent global state as sequential execution.

The runtime execution model consists of transitions on configurations.
A \emph{configuration} $(\jobs{}^R, \jobs{}^W, \gstate{})$ includes running jobs $\jobs{}^R$, waiting jobs $\jobs{}^W$, and global state $\gstate{}$. Transitions (discussed below) may also refer to the set of tasks $\jgTaskGraph{}.T$ from the job-generating program $\jgTaskGraph{}$ generated in Sec.~\ref{sec:analysis}. 
Elements $(jid,tid,\sigma_l,s)\in\jobs{}^R$ are jobs, as in Def.~\ref{def:jobs} with $\sigma_l$ being the job-local state and $s$ being the remaining statement to execute.
Elements of $\jobs{}^W \subseteq \jobs{} \times \mathcal{P}(JobId) \times \mathcal{P}(TaskId)$ (where $\jobs{}$ is the set of jobs) are jobs each with a set of job ids and a set of task ids to wait on. 
The following illustrates the lifetime of an individual job.
\begin{center}
\begin{tikzpicture}[font=\scriptsize,
    st/.style={draw,rounded corners,thick,minimum height=4.5mm,inner sep=2.5pt},
    wait/.style={st,fill=blue!10},
    arr/.style={-{Latex[length=1.6mm]},thick},
    every loop/.style={min distance=4mm,looseness=6}]
  \node[wait] (jw) at (0,0)   {Waiting on jobs};
  \node[wait] (tw) at (4.0,0) {Waiting on tasks};
  \node[st,fill=green!12,draw=green!60!black] (run) at (7.8,0) {Running};
  \node[st,fill=gray!15,draw=gray!60]        (done) at (10.4,0) {Done};
  \fill ([yshift=5mm]tw.north) circle (1.2pt);
  \draw[arr] ([yshift=5mm]tw.north) -- node[right,pos=0.4] {{\sc ForkTask}} (tw.north);
  \draw[arr] (tw.north west) to[bend right=20] node[above] {{\sc WaitNewTask}} (jw.north east);
  \draw[arr] (jw.east) to[bend left=20] node[below] {no jobs left} (tw.west);
  \draw[arr] (jw) edge[loop above] node[left,pos=0.2] {{\sc DepElim}} (jw);
  \draw[arr] (tw) to[bend left=12] node[above,pos=0.45] {{\sc FinishWait}} (run);
  \draw[arr] (run) edge[loop above] node[left,pos=0.2] {{\sc Local}} (run);
  \draw[arr] (run) to[bend left=12] node[above] {{\sc Terminate}} (done);
\end{tikzpicture}
\end{center}
\noindent
We now describe the execution model. (The full formal model is given \POPL{\inApx{apx:model}}\ARXIV{{in Sec.~\ref{sec:formal-model}}}.) {From $\jgTaskGraph$, we construct the initial configuration $(\{j_0\}, \emptyset, \gstate{}_0)$, whose single running job $j_0$ runs the initialization task $\jgTaskGraph{}.initTask$: it sets the global variables, giving $\gstate{}_0$, and spawns the initial jobs ({\sc ForkTask}), each waiting on the tasks it depends on.}
Transitions between configurations are given by the following rules:

{\bf Rule} {\sc ForkTask}: \emph{immediately} (\ie{} synchronously) obtain currently running dependences that are not spawned from an additional dependence, initialize local state with passed parameters, and add the new job to the configuration.
Since parameters are read-only (W1), this copy is never stale, and $\sigma_l{\upharpoonright}_{\bar x}$ stays fixed for the job's whole lifetime. So {\sc DepElim} may evaluate $\varphi$ on it even after the job has started. Variables of $\GVars{}$ are not copied: they reside in $\gstate{}$.

{\bf Rule} {\sc WaitNewTask}: queries for all existing jobs of the specified task, and mark them for waiting.

{\bf Rule} {\sc DepElim}: once any data or commutative dependency (job) has terminated, or can be bypassed due to commutativity, remove it from the waiting list.
A job may be bypassed (\emph{overtaken}) whether it is still waiting or already running. Whether the waiting job may run before all the jobs it overtook is decided by the re-check at its first step (discussed below) together with stability (Def.~\ref{def:stable}).
The condition read here only decides when the job stops waiting. Correctness rests on the re-check before its first local step ({\sc local}, below). In a serializable execution, that step comes after every step of an overtaken job that had already started, which then precedes it in program order anyway. So overtaking a running job needs no further premise.

{\bf Rule} {\sc FinishWait}: start running a job that is waiting on nothing.

{\bf Rule} {\sc local}:  take a job-local step, possibly manipulating job-local state and/or the global state.
Conditions may change between overtaking and running, so before its first local step a job \emph{re-checks}, in the current state, that its condition still holds for every job it overtook that has not terminated. It also re-checks that the jobs of its \emph{closure} 
(the jobs it overtook, together with every job they still wait for or have themselves overtaken and so on)
are (i) pairwise unordered and that (ii) none of these still waits for a task whose jobs may not exist yet. Otherwise it does not step yet; the overtaken jobs precede it in program order and do not wait on it, so they eventually terminate. The closure is needed because a job that the waiting job does not depend on, but that an overtaken job still waits for, could otherwise run in between and change the overtaken jobs' own conditions. When each of the job's conditions for the jobs it overtook reads only the two jobs' parameters, the closure is not checked: no job can change such a condition, so the overtaken jobs may run in any order among themselves, and the waiting job's conditions for them stay true. The re-check is part of the job's first step, so an implementation must perform it atomically with that step, under the same locks as the job's body. If a condition reads only parameters, the re-check of that condition cannot fail, since by (W1) both jobs' parameters stay fixed.

{\bf Rule} {\sc terminate}: a job's statement is fully reduced; remove it from the configuration.

\smallskip
\emph{Jobs and waiting.}
During execution, a program configuration can have any number of running jobs which can nondeterministically be scheduled. 
A key feature of the runtime semantics is to execute jobs concurrently while ensuring that jobs correctly wait.
There are a few requirements:
\begin{enumerate}
\item If a job $i$ depends on a \emph{running} or \emph{waiting} job $j$, then $i$ must wait.
\item A job may depend on a job that has yet to be created so we lift dependences to \emph{order dependences} that wait on jobs that may create additional jobs that will be dependences. \label{job-laterdep}
\item We wish to exploit our \emph{commutative dependences}. \label{job-commute}
\end{enumerate}

\noindent
The structure of waiting on tasks (bringing new sets of jobs in via {\sc WaitNewTask}) accomplishes (\ref{job-laterdep}).
By having a disjunct in the premise for {\sc DepElim}, we may consider the case where there is a commutative dependence with a satisfied commutativity condition.
Thus by considering a satisfied commutativity condition in place of a terminated job, we may accomplish  (\ref{job-commute}).
\POPL{A concrete runtime example for Ex.~\ref{ex:motivation} is given \inApx{apx:ex21-runtime}.}

\ARXIV{\input{ex21-runtime}}

\emph{Waiting in topological order.}
Dependent tasks must be waited upon in topological order relative to the spawning graph, as a job of one task may spawn additional instances of others that require waiting.
However, this alone is insufficient, as a \emph{descendant} of a task may be the one to spawn the dependent job.
As such we must lift the edges in the graph\footnote{Here we consider commutative dependences to be labeled data dependences. Indeed, data dependences may be considered commutative dependences labeled with the condition \false{}}.
Consider two graphs: the data dependence graph, and the spawning graph.
	Suppose $t_1, t_2, t_3$ are distinct tasks.
	We say task $t_2$ is \emph{order-dependent} on $t_1$ if either:
		(1) $t_2$ has a data dependence on $t_1$ (note this is a necessary for $t_1$ creating $t_2$), or
		(2) there exists $t_3$ such that $t_2$ is data dependent on $t_3$, and $t_1$ is an ancestor of $t_3$ and not an ancestor of $t_2$ in the spawning graph. 
When (1) is used to create an order dependence, and the original dependence had a commutativity condition, then we retain that condition on the lifted order dependence.
An example requiring this restriction is given \POPL{\inApx{apx:model}}\ARXIV{{in Fig.~\ref{fig:psdswp}}}.
\ifARXIV
\paragraph{Example: waiting in topological order.}

\begin{figure}[t]
	\begin{minipage}{\columnwidth}
		\begin{minipage}{\columnwidth}
				\begin{lstlisting}[style=vcy,language=VCY,basicstyle=\ttfamily\scriptsize]
					commutativity { |$\nblockA{f}{rowi},\nblockA{f}{rowj}$|: rowi != rowj; }
					int main(int[] visited, int[] pqFlag, int[] pValue, int pSize) {
						while(p < pSize){
							id = pIds[p];
							if(!visited[id]) {
								visited[id] = true;
								row = p; int q = 0; 
								|$\nblockA{f}{row}$|:{while (q<99 && !pqFlag[row][q]){
										q = q+1; 
										if(q<100) sum = sum+pValue[row];}}
							}
							p = p + 1; }
						return sum;
					}
				\end{lstlisting}
			\end{minipage}
			\vspace{0.5em}
			\begin{minipage}{\columnwidth}
				\begin{tikzpicture}[
					node distance=.3cm and .3cm,
					every node/.style={draw, minimum size=0.8cm, minimum width=4.5cm, font=\small},
					subtask/.style={-Stealth, thin, draw=black},
					scale=0.8, transform shape
					]
					\begin{scope}[shift={(0.25,0)}]
						\node[anchor=north west] (init) at (0, 0) {\begin{minipage}[t]{1.4in} {\bf Init Task}: \newline
								\lstinline|decls: [...];| \newline
								\lstinline|{$\newJob$(1,[]);| \newline
									\lstinline|$\;\;\newJob$(3,[]);}| \end{minipage}};
						\node (s0) [below=of init] {\begin{minipage}[t]{1.4in} {\bf Task2}:\newline
								\lstinline|{ while (q < 99 && !pqFlag[row][q % 100]) {| \newline
										\lstinline| q = q + 1; }| \newline
									\lstinline| if (q < 100) { sum = (sum + pValue[row]); } }| \newline
								Deps: \{T1: \code{q, row, init, pqValue, pqFlag}\} \end{minipage}};
						\node[anchor=north west] (s1) at (4.85, 0) {\begin{minipage}[t]{1.5in} {\bf Task1}:\newline
								\lstinline|{ while (p < pSize) { |\newline 
										\lstinline|  id = pIds[p];| \newline
										\lstinline| if (!visited[id % 100]) {| \newline
											\lstinline|  visited[id % 100] = true;| \newline
											\lstinline|  row = p; q = 0;| \newline
											\lstinline|  $\newJob$(2, [deps...]);| \newline
											\lstinline|} p = p + 1; } }| \newline
								Deps: \{ \}  \end{minipage}};
						\node (s2) [below=of s1] {\begin{minipage}[t]{1.4in} {\bf Task3}:\newline
								\lstinline|{return sum;}|\newline
								Deps: \{T1: $\emptyset$, T2: \code{sum}\} \end{minipage}};
					\end{scope}
				\end{tikzpicture}
				\vspace{0.25in}
			\end{minipage}
		\end{minipage}
		\caption{\label{fig:psdswp} Nested loop array example (adapted from~\cite{ps-dswp}), used to illustrate how jobs/tasks must wait.}
	\end{figure}
	
See Fig.~\ref{fig:psdswp}.
The dependence from {\bf Task3} on {\bf Task2} is lifted to a dependence from {\bf Task3} on {\bf Task1}.
Furthermore, as we wait in topological order, the job for {\bf Task3} must first wait for {\bf Task1} to fully execute.
When the job for {\bf Task1} terminates, all jobs for {\bf Task2} will have been created; this ensures that {\bf Task3} properly waits on all jobs of {\bf Task2}, instead of only waiting on a subset.
\fi

\ifARXIV
\subsection{Formal model}
\label{sec:formal-model}
This section gives the formal details of the model described above.

\begin{figure*}[tp]
\begin{minipage}{\linewidth}
\begin{gather*}
	\infer[\textrm{\sc local}]{
		\angles{ jid, tid,\sigma_L, s  } \uplus \jobs{}^R, \jobs{}^W, \gstate{}
		\;\;\hookrightarrow\;\;
		\angles{ jid, tid,\sigma_L', s' } \uplus \jobs{}^R, \jobs{}^W, \gstate{}'
	}{
		\angles{s, \sigma_L\oplus\gstate{}} \leadsto_{seq} \angles{s', \sigma_L'\oplus\gstate{}'}
		{{} \wedge still\_ok(jid, \gstate{})} {{} \wedge quiet(jid)}{, \text{ and then } w_o(jid) := \emptyset}
	}
\\
\\
	\infer[\textrm{\sc terminate}]{
		\angles{ jid, tid,\sigma_L, \texttt{skip} } \uplus \jobs{}^R, \jobs{}^W, \gstate{}
		\;\;\hookrightarrow\;\;
		\jobs{}^R, \jobs{}^W, \gstate{}
	} { }
\\
\\
\infer[\textrm{\sc WaitNewTask}]{
	\begin{array}{@{}c@{}}
	\jobs{}^R, (j, \emptyset, \{tid\} \uplus w_t) \uplus \jobs{}^W, \gstate{}
	\;\;\hookrightarrow\;\; \jobs{}^R, (j, w_j, w_t) \uplus \jobs{}^W, \gstate{}
	\end{array}
} { 
	\begin{array}{@{}c@{}}
  	  {\text{no job of } jobs(\jobs{}^R, \jobs{}^W) \text{ other than } j \text{ may spawn a job of } tid \text{, directly or through the jobs it spawns}} \\
	  w_j = \{jid \mid (jid, tid, \_, \_) \in jobs(\jobs{}^R, \jobs{}^W) \land jid \ne j.jid \}
	\end{array}
}
\\
\\
\infer[\textrm{\sc DepElim}]{
	\jobs{}^R, (\angles{jid, tid, \sigma_l, s}, jid' \uplus w_j, w_t) \uplus \jobs{}^W, \gstate{}
	\;\;\hookrightarrow\;\;
	\jobs{}^R, (\angles{jid, tid, \sigma_l, s}, w_j, w_t) \uplus \jobs{}^W, \gstate{}
} {
	\bigvee \left\{ \begin{array}{l}
	(jid',\_,\_,\_) \notin jobs(\jobs{}^R, \jobs{}^W)\\
	(jid',tid', \sigma_l', \nblock{g}) \in jobs(\jobs{}^R, \jobs{}^W) \wedge s=\nblock{f} \wedge \sem{\varphi^{\nblock{f}}_{\nblock{g}}} (\gstate{}, \sigma_l, \sigma_l') = \mathtt{true} \\
	{\quad {} \text{ and then } w_o(jid) := w_o(jid) \cup \{jid'\}}
	\end{array} \right\}
}
\\
\\
\infer[\textrm{\sc FinishWait}]{
\jobs{}^R, (j, \emptyset, \emptyset) \uplus \jobs{}^W\added{,} \gstate{}
\;\;\hookrightarrow\;\;
j \uplus \jobs{}^R, \jobs{}^W, \gstate{}
}{}
\\
\\
\infer[\textrm{\sc ForkTask}]{
	\angles{ jid, tid,\sigma_L, \newJob{}(tid_j, \sigma_j); s } \uplus \jobs{}^R, \jobs{}^W, \gstate{}
	\;\;\hookrightarrow\;\;
	\angles{ jid, tid,\sigma_L, s } \uplus \jobs{}^R, \alpha \uplus \jobs{}^W , \gstate{}
} {\begin{array}{l}
	(tid_j, s_j) \in \jgTaskGraph \land
	\sigma_j \subseteq \sigma_L \land
	preds = pred_{\jgTaskGraph.T}(tid_j) \\
	\alpha = (\angles{ fresh(), tid_j,\sigma_j, s_j }, init(preds, jobs(\jobs{}^R, \jobs{}^W)), {later(tid_j)})
\end{array}
}
\\
\begin{array}{rrl}
\text{Where:} &&
\\
& pred_{\jgTaskGraph.T}(tid) = & \text{task ids of the predecessors of the task with id } tid
\\
& &  \text{ in } \jgTaskGraph.T \text{(lifted to order dependence),}
\\
& jobs(\jobs{}^R, \jobs{}^W) = & \jobs{}^R \cup \{ j \mid (j, \_, \_) \in \jobs{}^W \} \text{ is the set of all running and}
\\
& & \text{waiting jobs in a configuration,}
\\
& init(tids, jobs) = & {\text{ the jobs, other than the spawning job, whose task is in } tids,}
\\
& {later(tid)} = & {\text{ the predecessors of } tid \text{ whose jobs may still be spawned, chosen by}}
\\
& & {\text{ the analysis; the new job waits for them with {\sc WaitNewTask}.}}
\\
& {unordered(j_1, j_2)} = & {\text{ the dependence analysis relates neither job to the other,}}
\\
& & {\text{ or both orientations of their condition hold in } \gstate{},}
\\
& {still\_ok(jid, \gstate{})} = & {\text{ the condition of } jid \text{ holds in } \gstate{} \text{ for each job of } w_o(jid) \text{ that still}}
\\
& & {\text{ exists; and, unless each of these conditions reads only parameters,}}
\\
& & {\text{ the jobs of its closure (the jobs of } w_o(jid) \text{, and recursively the jobs}}
\\
& & {\text{ they wait for or have overtaken) are pairwise unordered, and none}}
\\
& & {\text{ of them still waits for a task.}}
\\
& {quiet(jid)} = & {\text{ no other running job's remaining code writes a shared variable that}}
\\
& & {\text{ a condition read by } still\_ok(jid, \gstate{}) \text{ reads.}}
\end{array}
\end{gather*}
\end{minipage}
\caption{Rules for \name{}.}
\label{fig:rules}
\end{figure*}

\begin{definition}[Configuration]
A configuration ($\jobs{}^R, \jobs{}^W, \gstate{}$) is a set of running jobs $\jobs{}^R$, a set of waiting jobs with what to wait on $\jobs{}^W$, and global state $\gstate{}$.
Elements of $\jobs{}^W \subseteq \jobs{} \times \mathcal{P}(JobId) \times \mathcal{P}(TaskId)$ (where $\jobs{}$ is the set of jobs) are jobs each with a set of job ids and a set of task ids to wait on. 
\ARXIV{\rocqref{Jobs.config, Jobs.config_step}}
\end{definition}

The configuration \emph{transition} relation is denoted $\hookrightarrow$ between program configurations, i.e., 
$
(\jobs{}^R, \jobs{}^W, \gstate{})
\hookrightarrow
(\jobs{}'^R, \jobs{}'^W, \gstate{}')
$.
The relation is defined as the smallest relation that satisfies the inductive rules (given in Fig.~\ref{fig:rules}).
Transitions may refer to the job-generating program's task graph $\jgTaskGraph$.

{\bf Rules.}
\begin{itemize}
	\item {\sc local} -- allows for reductions of program statements within a single thread. Defers to the sequential semantics $\leadsto{}$.
 {A job that overtook others first re-checks, at its first local step, that the overtaking is still justified in the current state ($still\_ok$); the step then empties $w_o(jid)$, so later steps need no check.}
	\item {\sc terminate} -- removes jobs whose statement is \texttt{skip} from the program configuration.
	\item {\sc WaitNewTask} -- sets the job under consideration to wait on all of the jobs that match the given task id $tid$ from currently existing jobs. {It does so only once no other job may spawn a job of $tid$, so that every job of $tid$ that precedes it exists. The implementation waits for the spawners in the same way (\texttt{wait\_complete}).}
	\item {\sc DepElim} (``dependence eliminate'') -- removes a dependent job from the set of jobs to wait on once it's valid to do so. The premise allows for two cases: either the job $jid'$ is not in the configuration (because it has terminated), or it is the job of a task whose commutativity condition $\varphi_{\nblock{f}}^{\nblock{g}}$ is satisfied, where $(\gstate{}, \sigma_l, \sigma_l')$ is the partitioned state of Def.~\ref{def:state-triple}. Note data and commutative dependences can be removed via termination, but only commutative dependences can be prematurely eliminated by satisfying the commutativity condition.
 {The map $w_o$ records, for each job, the jobs it has overtaken; it starts empty for every job. {A job may be overtaken while it is waiting or running.} {(The premise $(jid', \_, \_, \_) \in jobs(\emptyset, \jobs{}^W)$, that the job has not started, is dropped: the proofs use only the re-check in {\sc local}; see rocq/Jobs.v.)}}
	\item {\sc FinishWait} -- A job has no more dependences to wait upon so it can be moved to the running set.
	\item {\sc ForkTask} -- a new job is initialized from the statement in the task it's created from, with a fresh job id. Dependences that are guaranteed to exist (which correspond to those who do not have an ancestor in the spawning graph) must be retrieved synchronously; these dependences are the set of jobs the job is initialized with to wait on. The new job also tracks all the other tasks it is dependent upon, to be waited upon later via {\sc WaitNewTask}. {The snapshot covers the jobs of every predecessor task, including other jobs of the spawning job's own task; the tasks of $later(tid_j)$ are waited for again with {\sc WaitNewTask}. In the implementation, $later(tid_j)$ is the set of predecessors that the analysis does not relate to $tid_j$ through spawning, that is, those that are not spawning dependences on an ancestor of $tid_j$.}
\end{itemize}
\fi

\section{Correctness}
\label{sec:correctness}

Correctness relates every complete concurrent execution of $\jgTaskGraph{}$ to the original sequential program $s$. We have mechanized it in Rocq.

\newcommand\correctnessstatement[1]{Let $\jgTaskGraph$ be the job-generating program for $s$, and let $C_0 = (\{j_0\}, \emptyset, \sigma_{gl}^0)$, where the initial job $j_0$ runs $initTask$. Under #1, for every complete execution $C_0 \hookrightarrow^* (\emptyset, \emptyset, \sigma_{gl}^f)$ there is a state $\sigma$ such that $\angles{s, \sigma_{gl}^0} \leadsto^* \angles{\sskip, \sigma}$ and $\sigma$ agrees with $\sigma_{gl}^f$ outside the variables local to NCB bodies.}
\begin{theorem}[Correctness]\label{thm:correctness}
\correctnessstatement{the assumptions \POPL{summarized below}\ARXIV{(V), (D), (O), (S), (T), (A1) and (A2) of Sec.~\ref{sec:correctness-assumptions}}}
\ARXIV{\rocqref{Correctness.main_theorem_sync}}
\end{theorem}

\POPL{{The theorem rests on the following assumptions. The commutativity conditions are sound, which {\sc Servois2} checks, and \emph{stable}: running one of three NCBs whose conditions all hold keeps the condition between the other two. The dependence analysis is sound: jobs that it does not relate do not conflict. Spawning is structured so that jobs are created in program order, the pre-order of the fork tree; \name{} checks this and otherwise falls back to a sequential task graph. A job and a job it overtook execute atomically with respect to each other, \ie{} the execution is equivalent to one in which their steps do not interleave, as two-phase locking or optimistic concurrency control ensures; this is the obligation of the synchronization layer and is orthogonal to the scheduling studied here, and \name{} can enforce it directly. Each job, and the sequential run of $\jgTaskGraph{}$, terminates. Finally, two syntactic conditions relate the task graph to $s$: after its first $\newJob$, a spawning task does not interfere with the jobs it can spawn, which \name{} checks, and code outside NCB bodies does not access NCB-local variables. The proof shows that the task graph is faithful to $s$, linearizes an execution, sorts its jobs into program order using the commutativity conditions and their stability, and relates the resulting sequential run to $s$. The assumptions are stated formally, with a proof outline, in the extended version of this paper~\cite{technicalreport}.}}

\ifARXIV
\input{correctness}
\fi

\section{Implementation}
\label{sec:implementation}

We implemented named commutative blocks, verification (Sec.~\ref{sec:verification}), the analysis (Sec.~\ref{sec:analysis}) and runtime model (Sec.~\ref{sec:execution_model}) in a new tool called \tool{} (\name). \name is written in OCaml and builds on Veracity~\cite{veracity} and Servois2~\cite{DBLP:conf/atva/ChenFNPBK23}. See Sec.~\ref{sec:data-availability} for links to source code and benchmarks.

\name{} generalizes Veracity's implementation to arbitrary, nonadjacent regions identified by named blocks and their associated commutativity conditions. Second, \name{} can verify conditions across these nonadjacent named blocks, including self-commutativity.
The program analysis, parallelization and translation to a job-generating program are also new to \name{}.

The runtime of \name{} is also new, and uses
a shared queue of jobs for running $\mathfrak{J}^R$ and waiting $\mathfrak{J}^W$ jobs.
The queue is protected by a mutex, and we have not found performance loss from over-accessing the jobs queue.
The runtime performs all necessary waiting for dependences and commutativity condition checks.
{Recall that for correctness, a job and a job it overtook must execute atomically with respect to each other ((S) in Sec.~\ref{sec:correctness}).} \name{} addresses this with a simple program analysis  for lock synthesis (similar to~\citet{DBLP:conf/popl/EmmiFJM07}) applied post-parallelization: given the job-generating program, the analysis identifies shared memory regions accessed by non-ancestor jobs and inserts mutex locks {around conflicting accesses} (details \POPL{\inApx{apx:locksynth}}\ARXIV{{below}}). 
Often locks are not necessary at all if a fragment only has a single global memory access.
This is a lightweight, conservative approach; more sophisticated techniques~\cite{DBLP:conf/popl/EmmiFJM07,DBLP:conf/pldi/FlanaganQ03,DBLP:conf/popl/VechevYY10,DBLP:conf/ppopp/Golan-GuetaRSY15,DBLP:conf/pldi/CheremCG08,DBLP:conf/popl/McCloskeyZGB06,hicks2006lock,DBLP:journals/jpdc/DinizR98,DBLP:journals/concurrency/DinizR99,diniz1997dynamic} could reduce lock granularity further.
\ARXIV{{
We implemented a static lock synthesis analysis that automatically inserts lock/unlock
calls around conflicting
shared-variable accesses in programs that have been parallelized by our algorithms in Sec.~\ref{sec:analysis}.
Such lock synthesis analyses can be seen for example in~\citet{DBLP:conf/popl/EmmiFJM07} (as well as numerous other related works~\cite{DBLP:conf/pldi/CheremCG08,DBLP:conf/popl/McCloskeyZGB06,hicks2006lock,DBLP:journals/jpdc/DinizR98,DBLP:journals/concurrency/DinizR99,DBLP:conf/popl/VechevYY10,diniz1997dynamic}).
The lock synthesis analysis operates in two phases: a \emph{pre-parallelization} phase analyses the
source AST to identify
variables written by concurrent NCB instances,
then groups consecutive conflicting statements—including read-then-write
pairs—under a single critical section per variable; a \emph{post-parallelization} 
phase extends this to the generated task graph by computing concurrent
task pairs (those with no directed path in the dependency DAG) and
merging the resulting inter-task conflict set with the NCB set before
transforming each task body.
A single shared ``lock-state'' is used across all program transformations so that
each shared variable receives one stable integer lock ID across every
task. Meanwhile, the scheduler's waiting
subsumes locks for tasks that are already serialized by explicit data
dependencies.
{An array access is locked per element only when its index is a parameter that every specification entry of the block keeps distinct, as \code{i != j} does; if some entry allows equal indices, as the condition between \code{f1} and \code{f2} in the running example does, the whole array is locked. The locks provide (S) of Sec.~\ref{sec:correctness} when the accesses of two blocks that conflict form one such group, or touch different array elements. \texttt{--atomic-overtaking} enforces (S) in general: a job holds the locks of the jobs it overtook while it runs.} {The tasks emitted for the running example, with their synthesized locks, are shown in Apx.~\ref{apx:emitted-tasks}.}}}

{\emph{Overtaking.} {\sc DepElim} allows a job to stop waiting for a job whose condition holds, whether that job is waiting or running; correctness rests on the re-check of {\sc local}. When to stop waiting is therefore a performance choice, and the runtime makes it per condition. A condition that reads only parameters and variables that no task writes cannot change, so the job may overtake any job for which it holds. A condition that reads written state may overtake a waiting job, or a running job until that job first writes a variable that some condition reads: the analysis marks each such write, the job sets a flag at the mark under a per-job lock, and the condition is evaluated under the same lock. Before its first step, a job that overtook others re-checks $still\_ok$ {and $quiet$} under the synthesized locks of the variables the conditions read, and if the re-check fails it waits for every job it overtook, as if it had never overtaken them. The closure is tested pair by pair, grouped by task: a condition that reads no variable is evaluated once per pair of tasks. When every condition the job checks reads only parameters, as in \texttt{banking}, the closure is skipped. {For $quiet$, a running job's whole task body stands in for its remaining code, and a job that the re-checking job overtook, or that overtook it, is not checked: by (S) it counts as done before the re-checking job starts.} Without this, two overtaken deposits to the same account fail the closure test and the job waits for both. One simplification remains: the re-check releases its locks before the job's first statement takes its own, so the two are not one atomic step as {\sc local} requires. Closing this window needs the re-check and the body to share their locks. All experiments below use the default.}
{Against stopping to wait without any re-check (\texttt{--no-recheck}), the re-check costs under 1\% on \texttt{motivation}, \texttt{simple-io}, \texttt{multi-blocks} and \texttt{vote}, and 10\% on \texttt{banking} (1.98\,s against 1.80\,s at $n = 5\times10^7$). On \texttt{erc20} it costs $5.8\times$ (0.64\,s against 0.11\,s): the transfer condition reads balances, 38 of its 57 re-checks fail the closure test, and those jobs wait.}

\ARXIV{Some benchmarks involve I/O, so \name{} includes basic operations such as file copying, reading, and writing. To support verification, we extend the SMT program model to include a filesystem. Filesystems are treated as a mapping from file names to a mapping from line numbers to strings, as well as a set of opened files, and a map from opened files to current line numbers. This simple representation is sufficient to reason about simple operations and test performance.}

\newcommand{\cmark}{\textcolor{darkgreen}{\ding{51}}}%
\newcommand{\xmark}{\textcolor{darkred}{\ding{55}}}%
\newcommand\nowhy[1]{\xmark & (#1)}

\begin{figure}[t]
	\centering
	\small
	\captionof{table}{\label{tab:summary} Comparison between \name{} and prior works on parallelization of benchmarks. 
	The comparison indicates various techniques' ability to parallelize benchmarks, with reason for inability: $(1)$: parallelization target is body of a loop, $(2)$: parallelization requires commutativity, $(3)$: commutativity is conditional on dynamic state, $(4)$: blocks to parallelize are non-local, $(5)$: blocks are nested. $(*)$: Example uses interprocedural calls, which are not supported, but the calls can be inlined.}
	\begin{tabular}{lc|rl|rl|rl|rl}
		\toprule
		{\bf Benchmark} & {{\bf {\#}Stmts}} & \multicolumn{2}{c|}{\bf PS-DSWP} & \multicolumn{2}{c|}{\bf CommSet} & \multicolumn{2}{c|}{\bf Veracity} & \multicolumn{2}{c}{\bf \name{}}
		\\
		\midrule
		\texttt{psdswp} & 31 & \cmark & & \cmark & & \nowhy{1} & \cmark \\
		\midrule
		\texttt{commset-md5} & 23 & \nowhy{2} & \cmark & & \nowhy{1} & \cmark \\
		\texttt{commset-kmeans} & 38 & \nowhy{2} & \cmark & & \nowhy{1} & \cmark \\
		\texttt{commset-potrace} & 31 & \nowhy{2} & \cmark & & \nowhy{1} & \cmark \\
		\midrule
		\texttt{vcy-nested} & 16 & \nowhy{2} & \nowhy{3} & \cmark & & \nowhy{5} \\
		\texttt{vcy-crowd} & 93 & \nowhy{2} & \nowhy{3} & \cmark & & \cmark & (*) \\
		\texttt{(24 more vcy-*...)} & ... & \nowhy{2} & \nowhy{3} & \cmark & & \cmark \\
		\midrule
		\texttt{motivation} & 27 & \nowhy{2} & \nowhy{3} & \nowhy{1} & \cmark \\
		\texttt{simple-io} & 25 & \nowhy{2} & \nowhy{3} & \cmark & & \cmark \\
		\texttt{multiblocks} & 40 & \nowhy{2} & \nowhy{3} & \nowhy{4} & \cmark \\
		\texttt{vote} & 64 & \nowhy{2} & \nowhy{3} & \cmark & & \cmark \\
		\texttt{erc20} & 53 & \nowhy{2} & \nowhy{3} & \nowhy{1} & \cmark \\
		\texttt{banking} & 90 & \nowhy{2} & \nowhy{3} & \nowhy{1} & \cmark \\
		\toprule
	\end{tabular}
	\vspace{0.5\baselineskip}
	\begin{subfigure}[b]{0.32\textwidth}
		\centering
		\includegraphics[width=\textwidth,trim={0 23bp 0 0},clip]{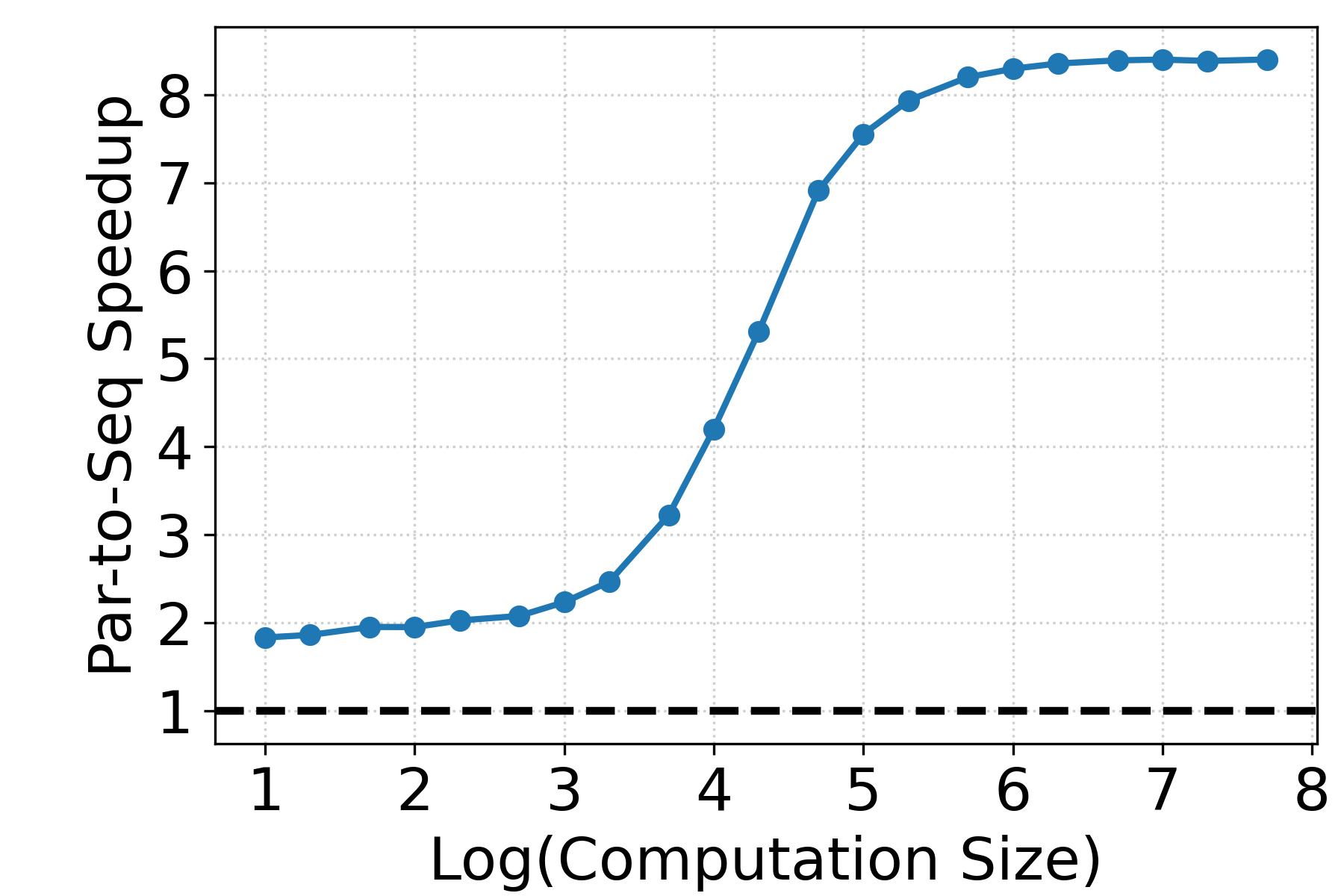}
		\caption{\texttt{psdswp}}
		\label{fig:summary-psdswp}
	\end{subfigure}
	\hfill
	\begin{subfigure}[b]{0.32\textwidth}
		\centering
		\includegraphics[width=\textwidth,trim={0 23bp 0 0},clip]{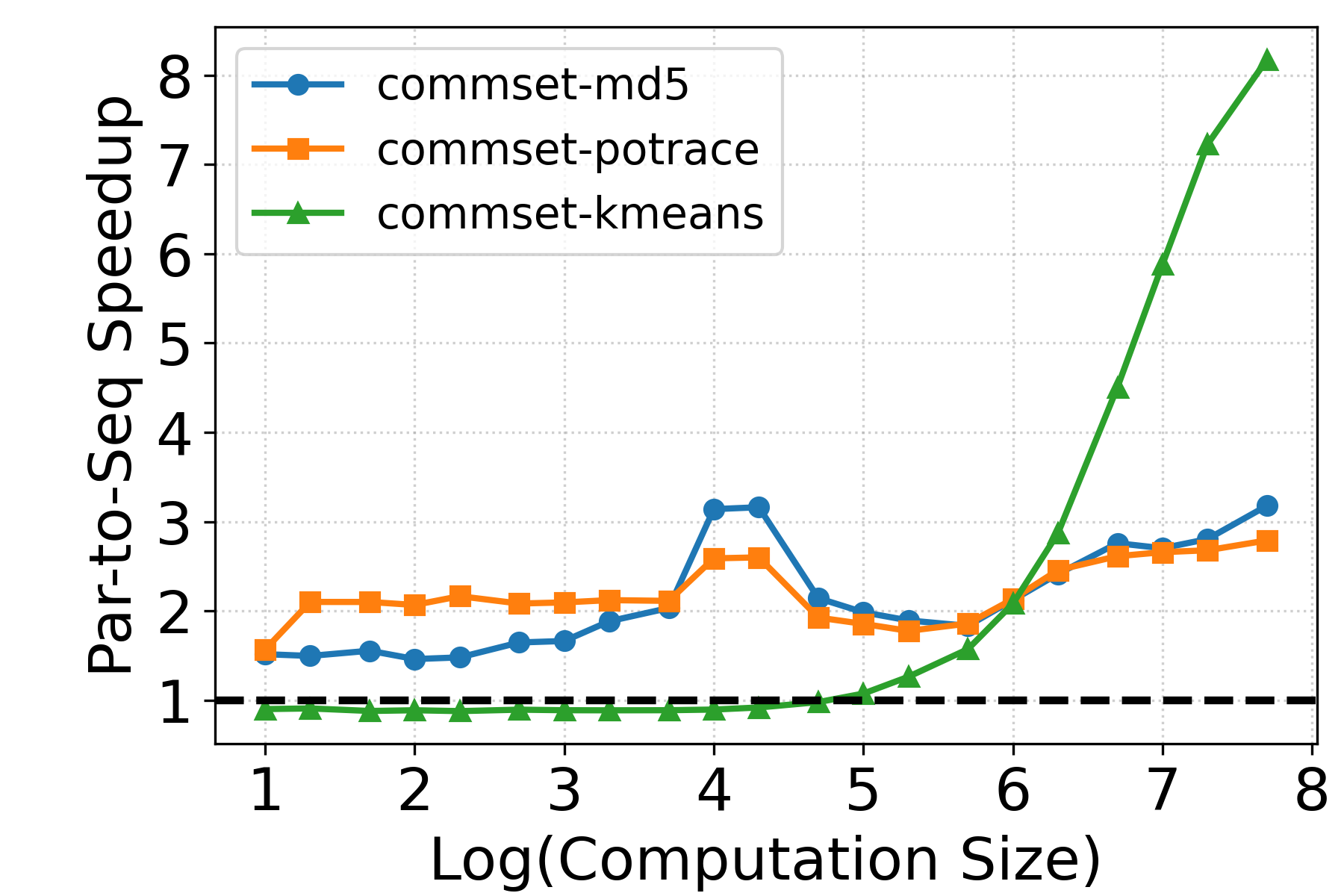}
		\caption{\texttt{commset-*}}
		\label{fig:summary-commset}
	\end{subfigure}
	\hfill
	\begin{subfigure}[b]{0.32\textwidth}
		\centering
		\includegraphics[width=\textwidth,trim={0 23bp 0 0},clip]{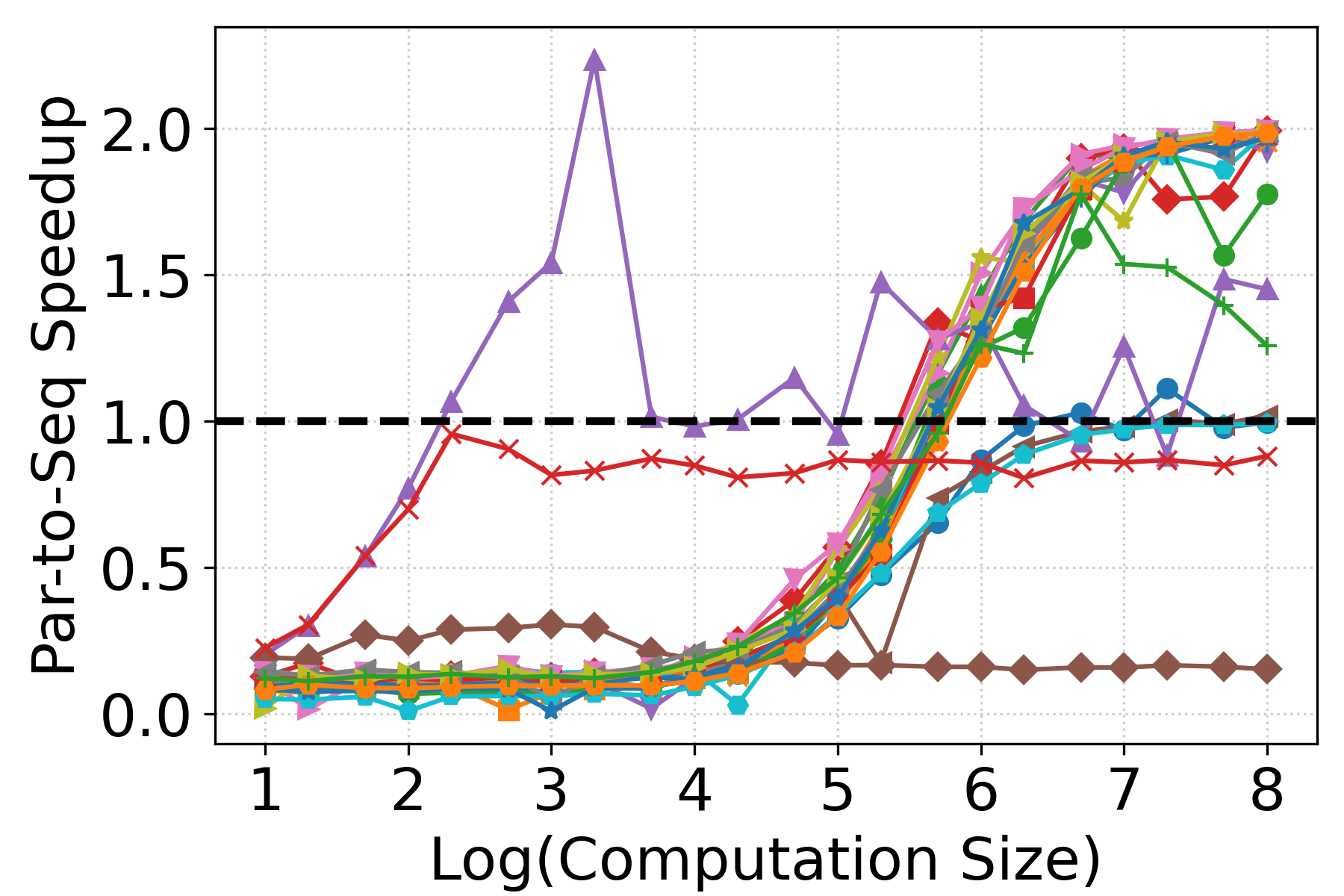}
		\caption{\texttt{vcy-*}}
		\label{fig:summary-vcy}
	\end{subfigure}
	\captionof{figure}{\label{fig:summary-speedups} Parallel-to-sequential speedup of \name{} on the benchmarks from prior work in Table~\ref{tab:summary}, versus the log of the computation size, with all \name{} features enabled, including, where applicable, runtime commutativity checking and concurrent execution of commutative NCB jobs. The dashed line marks speedup $=1$.}
	\vspace*{-1.5\baselineskip}
\end{figure}

\section{Evaluation}
\label{sec:eval}

We evaluate whether our approach
{\bf (i)} can also be applied to examples from the most closely related works,
{\bf (ii)} enables parallelization that could not previously be achieved, and 
{\bf (iii)} enables commutativity specifications of NCBs to be automatically verified.

We build on Veracity, so we can perform a direct comparison against Veracity per below. PS-DSWP and CommSet used an older version of LLVM and are no longer maintained or publicly available. Also, these tools were implemented at the IR/backend level (rather than at the interpreter level), further impeding an apples-to-apples comparison. As discussed below, \name{} generalizes these tools, so we can perform a more meaningful comparison by disabling the relevant features of \name{}, thus focusing the evaluation on whether the \emph{novel} aspects of our work offer performance improvement by keeping other variables (interpreter-vs-compiler, AST-vs-IR, etc.) the same. 
All experiments were run on an x86\_64 AMD EPYC 7452 32-Core with 125GB of RAM.

{\bf (i) Supporting examples from prior works.} We compare against Veracity~\cite{veracity}, CommSet~\cite{commset}, and PS-DSWP~\cite{ps-dswp}.
A summary is given in Table~\ref{tab:summary}, illustrating which benchmarks could be parallelized by which methods. This table provides a ``diff'' with existing work by quantifying what prior parallelization can be expressed with \ncb{}s, as well as the limits of those prior works.
In Table~\ref{tab:summary}, \texttt{psdswp} is the motivating example in~\citet{ps-dswp}, which requires global parallelization. The second set of benchmarks from \citet{commset} (\texttt{commset-*}) require commutativity reasoning \emph{and} global parallelization, ruling out PS-DSWP and Veracity. The third set from Veracity  (\texttt{vcy-*}) require \emph{dynamic} commutativity checking. The final set are new benchmarks that require global parallelization and dynamic commutativity checking. As seen in the table, \name{} is the only technique that can support (nearly) all of these. 

While NCBs are indeed expressive enough to capture these prior works, we also wanted to confirm that our parallelization could indeed scale on those benchmarks.
PS-DSWP reported an average speedup of 114\% with 6 threads.
CommSet reported best speedups on 8 threads of $7.6\times$ for \texttt{commset-md5}, $5.5\times$ for \texttt{commset-potrace}, and $5.2\times$ for \texttt{commset-kmeans}.
Veracity speedups approached $2\times$ for two-fragment commute blocks as the computation size grows.

PS-DSWP and CommSet are implemented at the LLVM layer (and using a very old version of LLVM), while Veracity and \name{} are implemented as interpreters over ASTs. 
Despite these differences, we still evaluated whether \name{} can achieve similar scaling behavior.
To this end, across the benchmarks listed in Table~\ref{tab:summary}, we executed \name{} with varied the computation size (CPU load). 

Fig.~\ref{fig:summary-speedups} summarizes the results (more detail \inApx{apx:comparison}), with all \name{} features enabled, including, where applicable, runtime commutativity checking and concurrent execution of commutative NCB jobs. \name{} achieves speedup on these benchmarks once the computation size is large enough to amortize the parallelization overhead. This includes \texttt{psdswp}, which contains no NCBs: there, the speedup comes from dependence-based parallelization alone, confirming that \name{}, like PS-DSWP, achieves dependence-based scaling. Furthermore, our experiments confirm scaling similar to CommSet on \texttt{commset-potrace} and \texttt{commset-kmeans} and similar to Veracity on \texttt{vcy-*}, and a more moderate $\sim$3$\times$ speedup on \texttt{commset-md5}.
This shows that \name{} obtains the prior works' forms of parallelization, as well as new forms (e.g. the last six rows of Table~\ref{tab:summary}) discussed next.

{\bf (ii) New forms of parallelization.}
\name{} is uniquely able to parallelize non-adjacent code regions that require dynamic commutativity checks.
To our knowledge, there are no existing benchmarks specifically designed for whole-program dynamic commutativity.
We thus constructed a set of 6 benchmarks\ekremoved{: \texttt{motivation}, \texttt{simple-io}, \texttt{multi-blocks}, \texttt{banking}, \texttt{vote}~\cite{voting}, and \texttt{erc20}~\cite{erc20}}.
These benchmarks, listed in Table~\ref{table:benchmarks}, cover a variety of program shapes needing dependence-based  parallelization, and code regions that commute depending on runtime conditions.
\emph{Existing techniques cannot parallelize these code regions.}
Some benchmarks are computationally bound, and some cpu bound.
Briefly, \texttt{simple-io} runs both a copy of a file, and a partial copy (via read and write) of another file. \texttt{multi-blocks} has several \ncb{}s that each perform hashtable and array operations.
The \texttt{erc20} adapts the ERC20 token standard~\cite{erc20} which simulates token operations with approval phase and transfer transactions between agents. It implements commutative token operations optimized for parallel execution, allowing concurrent token approvals and transfers while maintaining transaction integrity across multiple agent accounts.
The \texttt{banking} simulates a financial system with parallel account transactions including deposits, withdrawals, and transfers between accounts. It leverages explicit commutativity conditions to enable concurrent processing of operations when they access different accounts, while maintaining transaction integrity and providing final balance statistics.

Table~\ref{table:benchmarks} also lists the number of NCBs, gives an example of the commutativity condition and indicates whether the condition could be verified automatically (discussed below).

\newcommand*\benchtablerowstretch{1.4} %
\newcommand*\benchtabletabcolsep{2pt} %

\begin{table}[t]
	\renewcommand{\arraystretch}{\benchtablerowstretch}\setlength{\tabcolsep}{\benchtabletabcolsep}
		\centering
		\smaller
	\caption{\label{table:benchmarks}
    New benchmarks to which previous tools cannot be applied
	(see Table~\ref{tab:summary}; detailed comparisons to prior work are \inApx{apx:comparison}).
	 Included are the number of unique NCBs (\#NCBs), commutativity conditions (\#$\varphi$'s), an example condition, and an indication as to whether the condition was verified by \name{}.}
		\begin{tabular}{lp{1.77in}ccp{1.55in}c}
		\toprule
		{\bf Name} & {\bf Features} & {\bf {\#}NCBs} & {\bf {\#}$\varphi$'s} & {\bf Example commu.~cond.} $\varphi$ & {\bf {Verified}} \\
		\midrule
		\texttt{motivation} & Ex.~\ref{ex:motivation} loop with 2~NCBs. & 2 & 3 & (\isubf$\neq$\isubg$\lor$(\isubf=\isubg$\land$\code{arr}[\isubf]\code{>0}))${}^{\nblock{g}}_{\nblock{f}}$ & \checkmark\\
		\texttt{simple-io} & Copies a file while reading/writing others. & 2 & 1 & ${(dstA \neq srcB \land dstA \neq}$ ${dstB \land srcA \neq dstB)^{f_1}_{f_2}}$ & \checkmark \\
		\texttt{multi-blocks} & Hash table or vector ops. & 5 & 11 & ${(files[i_{4'}] = files[i_{4}])}^{f_4}_{f_4}$ & \checkmark \\
		\texttt{vote} & Voting smart contract~\cite{voting}. & 4 & 6 & ${(voterID_{1} \neq voterID_{2})}^{vote_1}_{vote_2}$ & \checkmark \\
		\texttt{erc20} & ERC20~\cite{erc20} token approvals and transfers & 2 & 3 & {$(i_1 \neq i_2 \vee account[i_1] \geq amount_1 + amount_2)^{transfer}_{transfer}$} & \checkmark \\
		\texttt{banking} & Txns w/ disjoint-acct~parallelism. & 3 & 5 & $(acc\_id_{1} \neq acc\_id_{2})^{deposit}_{deposit}$ & \checkmark \\
		\bottomrule
		\end{tabular}
	\end{table}

\begin{figure*}[tb]
    \centering
    \captionsetup[subfigure]{skip=1pt}
	\begin{subfigure}[b]{0.32\textwidth}
		\centering
		\includegraphics[width=\textwidth,trim={0 29bp 0 0},clip]{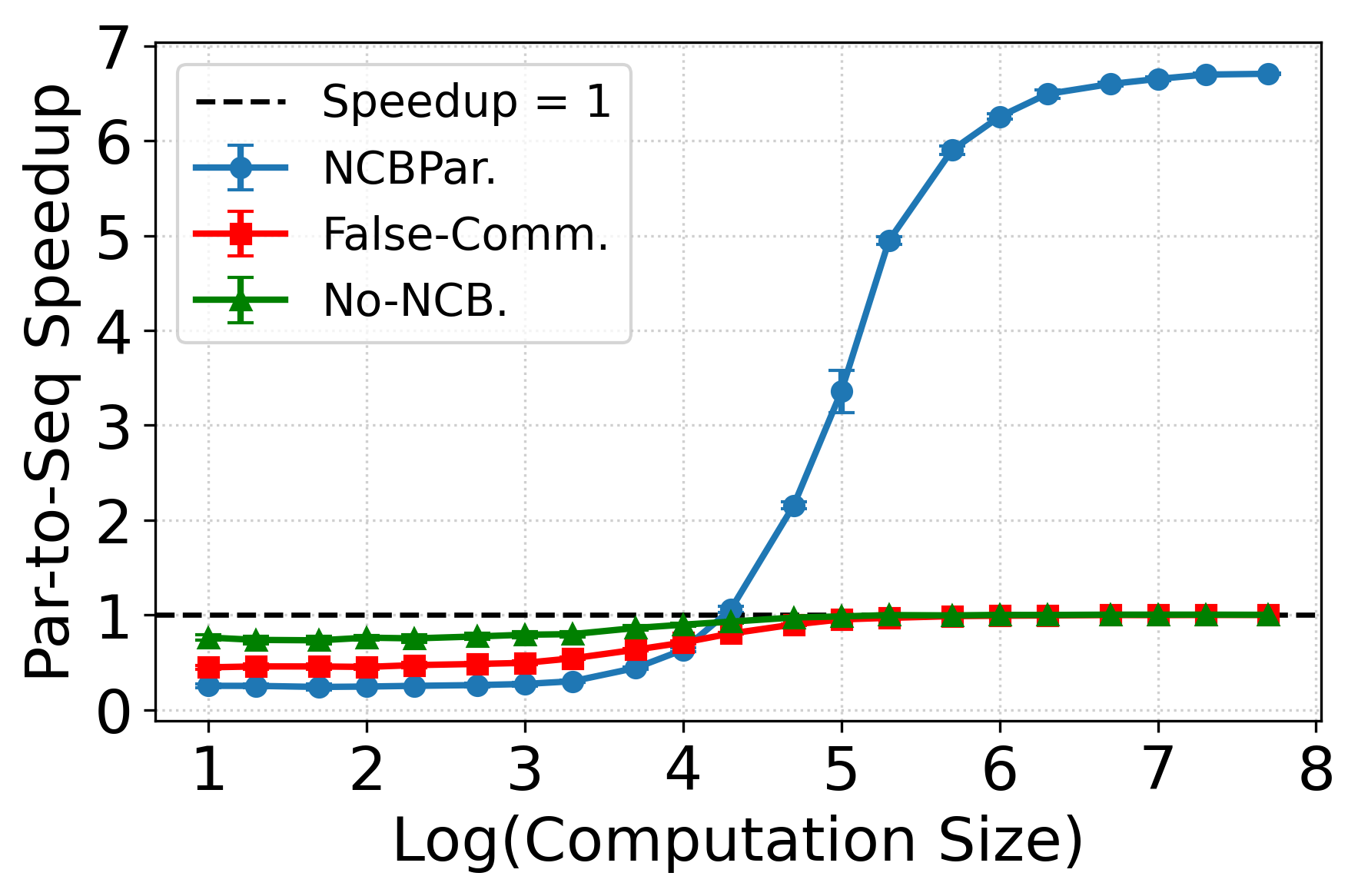}
		\caption{\texttt{motivation}}
	\end{subfigure}
    \begin{subfigure}[b]{0.32\textwidth}
        \centering
        \includegraphics[width=\textwidth,trim={0 29bp 0 0},clip]{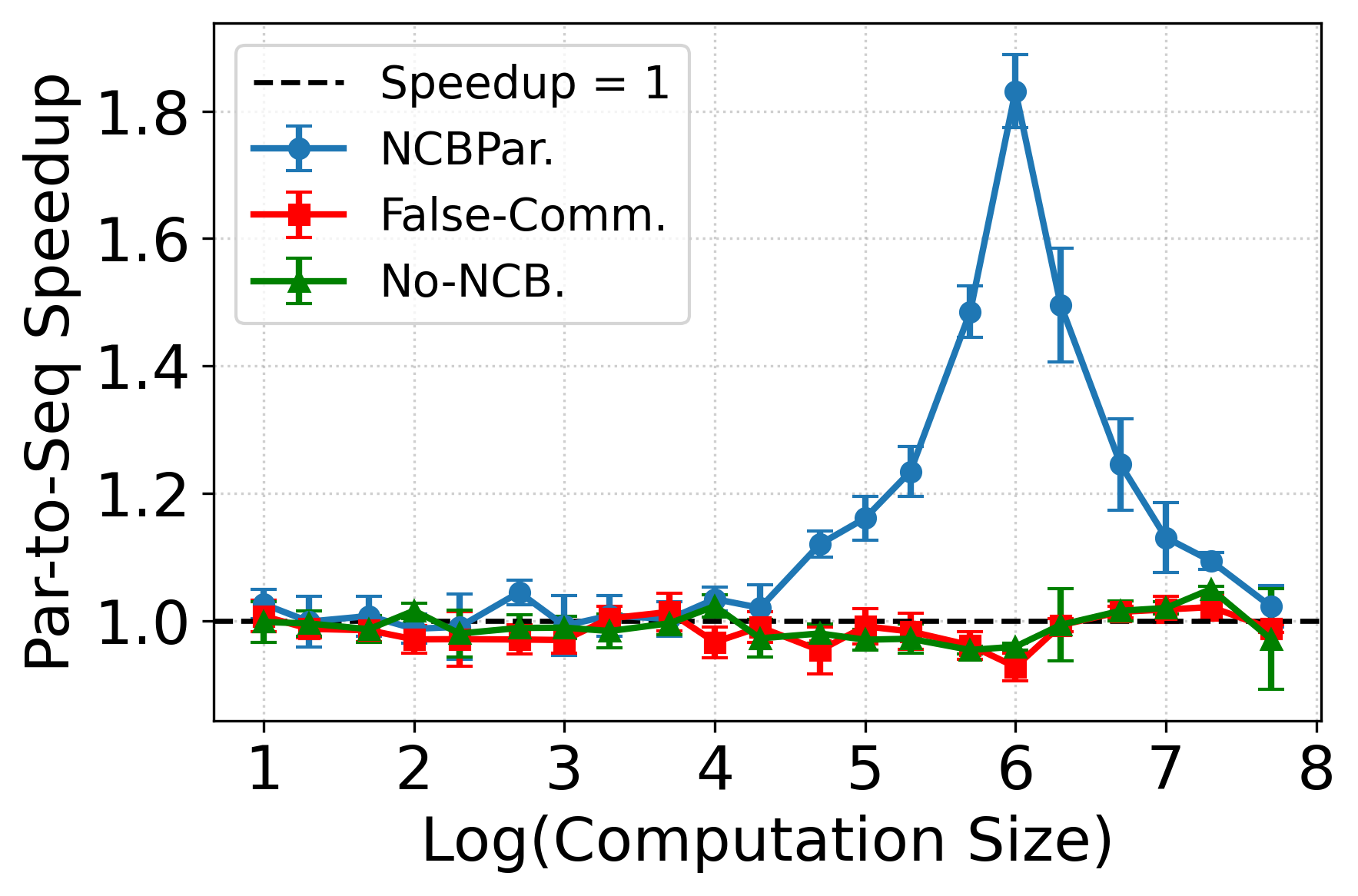}
        \caption{\texttt{simple-io}}
    \end{subfigure}
	\begin{subfigure}[b]{0.320\textwidth}
        \centering
        \includegraphics[width=\textwidth,trim={0 29bp 0 0},clip]{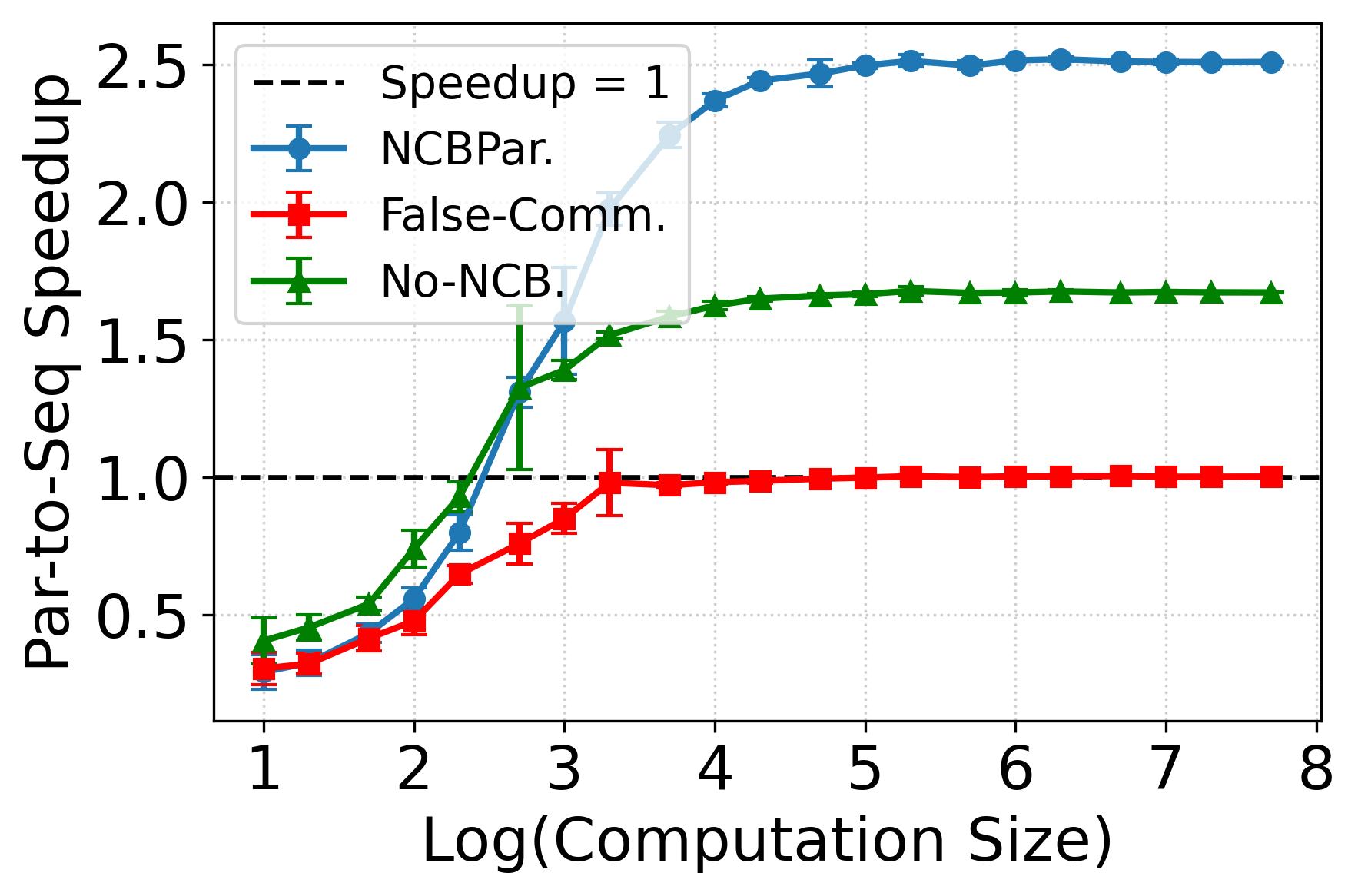}
        \caption{\texttt{multi-blocks}}
    \end{subfigure}
	
    \begin{subfigure}[b]{0.32\textwidth}
        \centering
        \includegraphics[width=\textwidth,trim={0 29bp 0 0},clip]{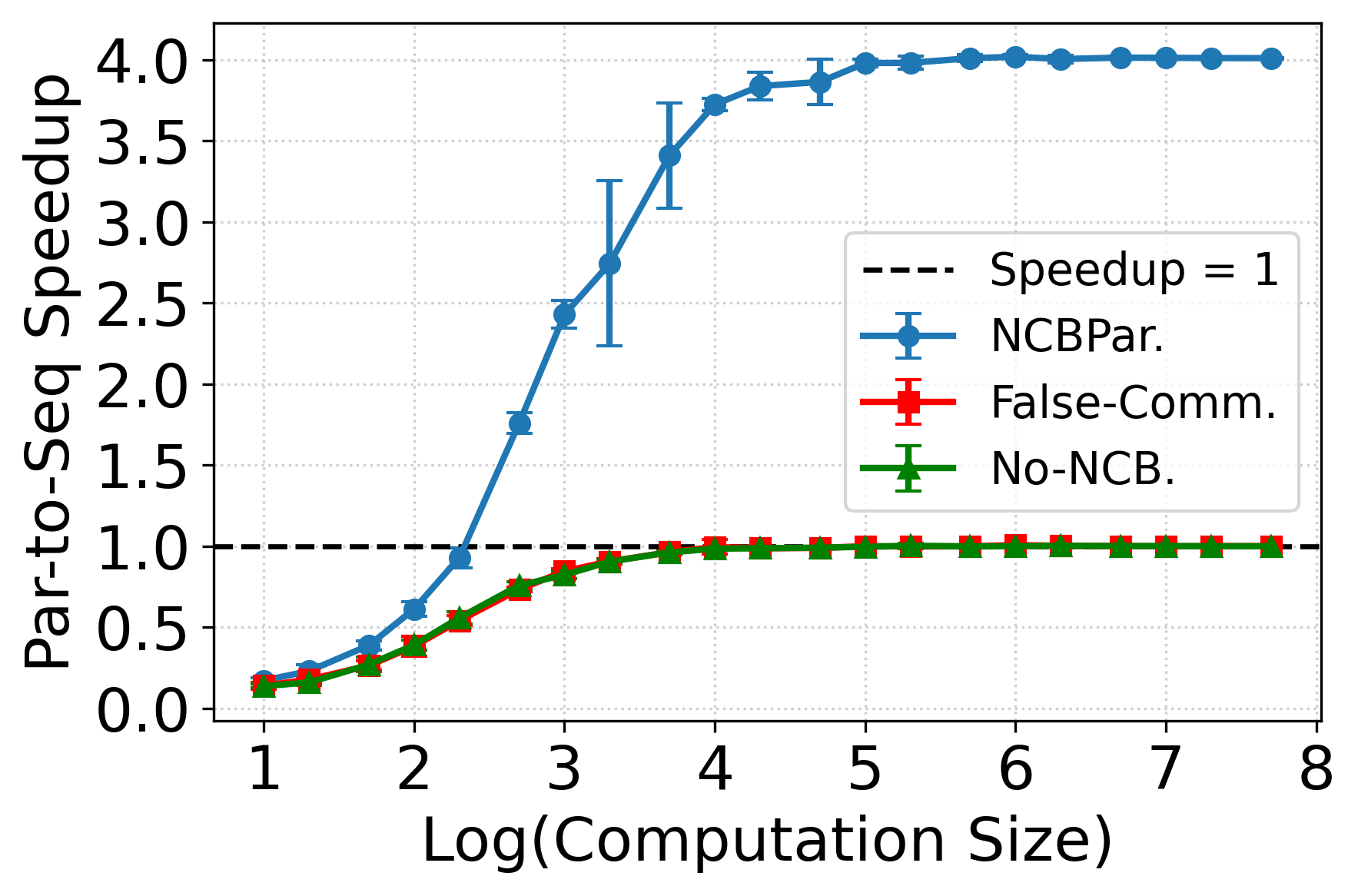}
        \caption{\texttt{vote}}
    \end{subfigure}
	\begin{subfigure}[b]{0.32\textwidth}
        \centering
        \includegraphics[width=\textwidth,trim={0 29bp 0 0},clip]{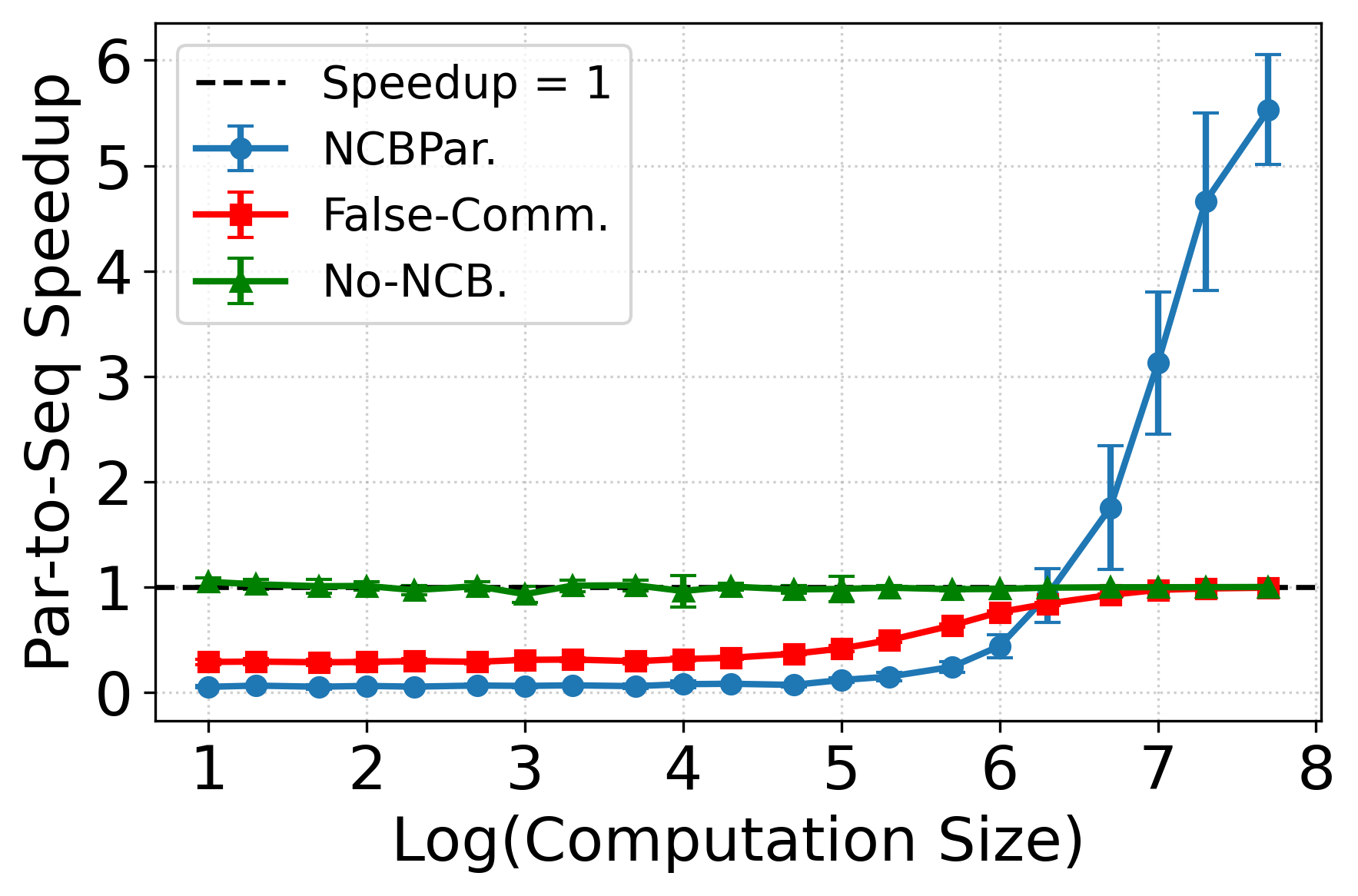}
        \caption{\texttt{erc20}}
    \end{subfigure}
	\begin{subfigure}[b]{0.32\textwidth}
        \centering
        \includegraphics[width=\textwidth,trim={0 29bp 0 0},clip]{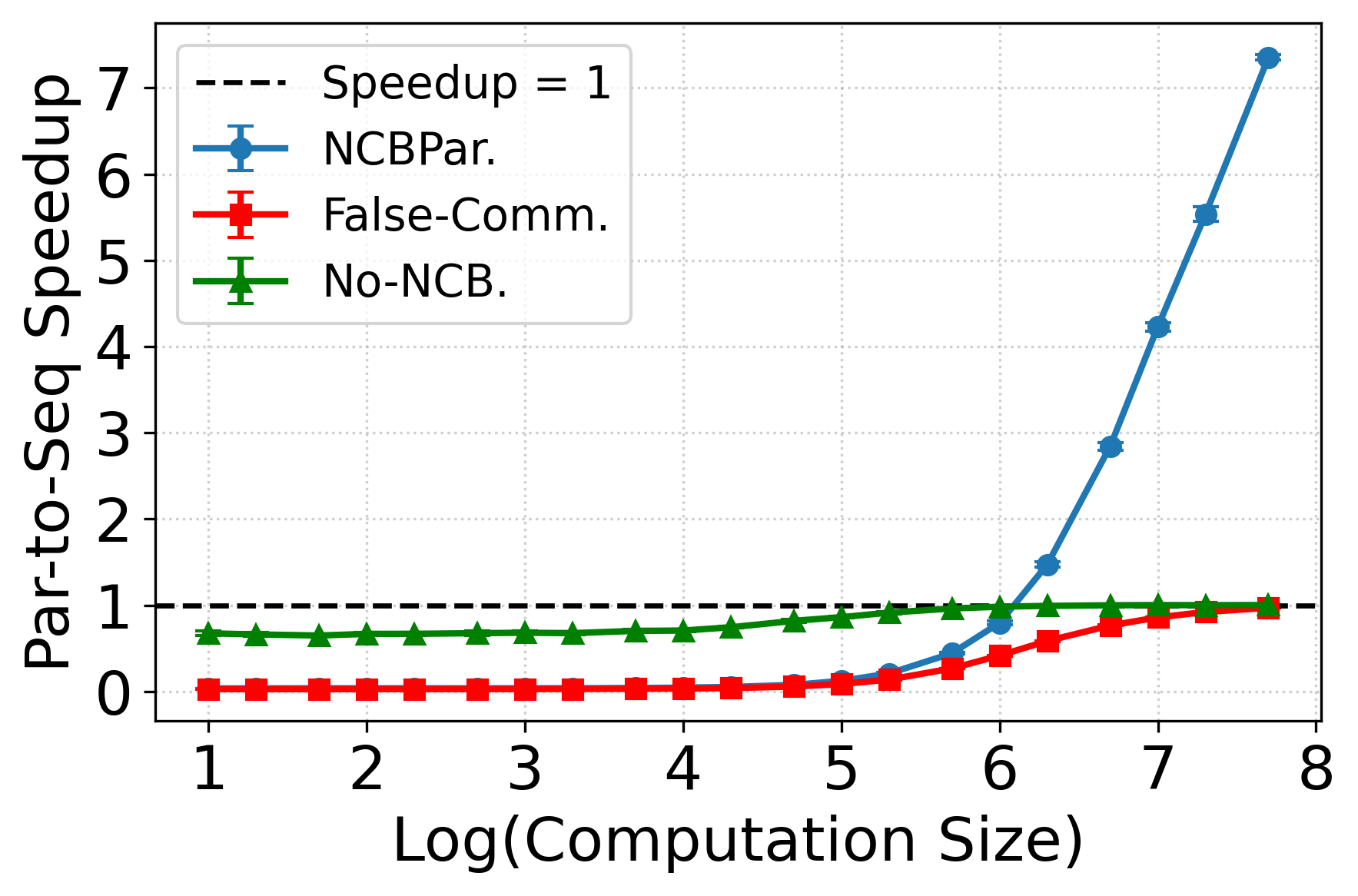}
        \caption{\texttt{banking}}
    \end{subfigure}

    \caption{Speedup comparison under the three configurations described across all benchmarks. 
    Each subplot shows the parallel-to-sequential speedup ratio against the log of computation size.
	Points show the geometric mean speedup over 6 trials; error bars show the standard deviation.}
	Prior tools cannot obtain speedup on these examples (see Table~\ref{tab:summary}; detailed comparisons to prior work are \inApx{apx:comparison}).
    \label{fig:speedup-comparison}
\end{figure*}

Each benchmark includes an artificial busy wait whose duration is an input parameter $n$ to the benchmark, which we refer to as the ``computation size''. Our experiments vary this input parameter to measure how speedup varies with computation size, plotting $\mathrm{log}(n)$ on the x-axes of Fig.~\ref{fig:speedup-comparison}.

All experiments use a fixed pool of 8 worker threads (OCaml~5 domainslib). Since the main thread also executes jobs while awaiting results, at most 9 jobs run concurrently. We fix the pool size rather than use one thread per core because (i) several benchmarks expose little parallelism beyond this (\texttt{vote}, \texttt{multi-blocks}, and \texttt{simple-io} never have more than 4 compute-heavy jobs ready at once) and (ii) it keeps results comparable to commodity 8--10 core machines.
The pool size is a runtime parameter. Rerunning all benchmarks with 16 worker threads confirms (1): \texttt{vote}, \texttt{multi-blocks}, and \texttt{simple-io} do not improve, while the loop-based \texttt{motivation} and \texttt{banking}, which spawn one job per iteration, continue to scale (up to ${\sim}10\times$ and $13\times$, respectively). {\texttt{erc20} also spawns one job per iteration but stays at $1.5\times$, because its overtaking jobs fail the re-check (Sec.~\ref{sec:implementation}).}

We ran all benchmarks under several configurations to evaluate the speedup attributable to NCB's. The graphs in Fig.~\ref{fig:speedup-comparison} include all three of these configurations, which remove one ingredient at a time: Cfg2 drops commutativity but keeps the NCBs, and Cfg3 also drops the NCBs themselves:

\begin{description}
	\item[Cfg1:] \textcolor{blue}{{\bf Blue $\bullet$ circle}}. All \name{} features enabled, including runtime commutativity checking and concurrent execution of commutative NCB jobs.
	This is the speedup a programmer gets from writing NCBs and commutativity specifications.
	\item[Cfg2:] \textcolor{red}{{\bf Red $\blacksquare$ square}}. 
	All commutativity conditions are set to \texttt{false}, so potentially commutative NCB pairs fall back on their read/write dependences, and each NCB remains a single PDG node. This does not look inside NCBs as dependence-based methods (\eg{}~\cite{ps-dswp}) would. Since the program and its PDG are otherwise identical to Cfg1, the gap between Cfg1 and Cfg2 is the speedup due to commutativity alone.
	\item[Cfg3:] \textcolor{teal}{{\bf Green $\blacktriangle$ triangle}}. NCBs are treated as regular blocks (no commutativity), so parallelization is only achieved through PDG node merging \`{a} la PS-DSWP~\cite{ps-dswp}. This is a baseline for dependence-based parallelization without \name{}: the PDG is built over individual statements, so the parallelizer may split or regroup the code of an NCB. Comparing Cfg3 with Cfg2 shows whether coarse NCB nodes cost any parallelism, and comparing it with Cfg1 shows the net benefit of NCBs over a PS-DSWP-style approach.
\end{description}

Each configuration and computation size was run 6 times, and we report the geometric mean speedup with standard deviation.
Figure~\ref{fig:speedup-comparison} presents results of executing each configuration, with computation size $n$ varying from $10^1$ to $5 \times 10^7$ (plotted as $\textrm{log}(n)$).
As expected, parallelization benefit outweighs its cost as the computation size increases. In most examples we start to see benefits at a computation size that corresponds to a runtime of about 0.01s.
Gains are more modest in the I/O-bound \texttt{simple-io}, with some of the largest examples of \texttt{simple-io} reaching a bottleneck.
In almost all benchmarks, parallelization of commutative NCBs ({\bf Cfg1}) outperforms non-commutative NCBs ({\bf Cfg2}) and even NCB-free variants ({\bf Cfg3})\footnote{{\bf Cfg3} does not capture when blocks may statically commute, as in CommSet~\cite{commset}, so it may not offer the best comparison for specifically the \texttt{commset} benchmark.}.
\texttt{vote}, \texttt{simple-io}, \texttt{erc20}, \texttt{banking}, and \texttt{motivation} only show speedup with dynamic
commutativity checks ({\bf Cfg1}), demonstrating that conditional commutativity enables parallelization in more situations.

\begin{table}[t]
	\renewcommand{\arraystretch}{0.7}
		\setlength{\tabcolsep}{{\benchtabletabcolsep}}
		\centering
		\small
		\setlength\extrarowheight{2mm}
	\caption{\label{table:execution}
	Runtime execution statistics and performance for the new benchmarks. See inline prose for column descriptions.
    ``Speedup'' compares theoretical and observed speedups over sequential execution.
	}
		\begin{tabular}{l|cc|ccc|rr|rr|rr}
		\toprule
		 & \multicolumn{2}{c|}{{\bf PDG$_C$ stats}} & & & & \multicolumn{2}{c|}{\bf Avg Job Exec(s)} & \multicolumn{2}{c|}{\bf Total Exec(s)} & \multicolumn{2}{c}{\bf Speedup} \\
		 {\bf Name} & {\bf Nodes} & {\bf Depth} & {\bf Tasks} & {\bf Locks} & {\bf Jobs} & {\bf Pure} & {\bf +Deps} & {\bf Seq} & {\bf Par} & {\bf Thry.} & {\bf Obsrv.} \\
		\midrule
		\texttt{motivation} & 20 & 3 & 5 & 1 & 23 & 1.32 & 2.39 & 30.68 & 4.57 & 8$\times$ & 6.71$\times$ \\
		\texttt{simple-io} & 15 & 2 & 3 & 0 & 3 & 0.03 & 0.03 & 0.08 & 0.04 & 2$\times$ & 1.83$\times$ \\
		\texttt{multi-blocks} & 25 & 2 & 9 & 3 & 9 & 0.17 & 0.27 & 1.53 & 0.61 & 3$\times$ & 2.51$\times$ \\
		\texttt{vote} & 7 & 2 & 6 & 1 & 6 & 10.19 & 10.19 & 61.29 & 15.29 & 4$\times$ & 4.01$\times$ \\
		\texttt{erc20} & 44 & 4 & 6 & 5 & 83 & 0.01 & 0.03 & 0.92 & 0.17 & 9$\times$ & 5.53$\times$ \\
		\texttt{banking} & 57 & 5 & 10 & 0 & 3005 & 0.01 & 1.30 & 15.36 & 2.09 & 9$\times$ & 7.35$\times$ \\
		\bottomrule
		\end{tabular}
\end{table}

Table~\ref{table:execution} reports runtime execution statistics for these benchmarks, for a fixed computation size $n$.
We report PDG characteristics, including the number of nodes ({\bf Nodes}) prior to transformation and control-dependence depth ({\bf Depth}), defined as the length of the longest control-dependence chain in the PDG starting from the entry node.
We also report the number of locks automatically synthesized ({\bf Locks}).
A value of 0 indicates that the corresponding task performs only a single access to shared state (e.g., one read or write), which is already atomic and therefore does not require a synthesized lock.
{\bf Tasks} and {\bf Jobs} denote the number of static tasks generated from the parallelization graph and the total dynamic job instances executed at runtime, respectively. {\bf Avg Job Exec} shows the average execution time of a job in seconds, both for pure computation and including dependency-induced waits. {\bf Total Exec} lists sequential (Seq) and parallel (Par) runtimes, while {\bf Speedup} columns compare theoretical ({\bf Thry.}) and observed ({\bf Obsrv.}) speedups. 
We manually wrote down the theoretical speedup, which indicates the maximum possible speedup, given the number of threads available, the number of jobs spawned, and any constraints that prevent continual full utilization of threads.
These results highlight the effectiveness of \name{} in enabling dependence-based parallelization: observed speedups are generally close to theoretical predictions{, except for \texttt{erc20}, whose transfers mostly fail the re-check and wait (Sec.~\ref{sec:implementation})}, and even benchmarks with complex commutativity dependencies achieve substantial parallel performance. This demonstrates that dynamic commutativity checks can uncover concurrency opportunities that conservative static analyses would otherwise miss.

{\bf (iii) Verifying commute conditions.}
To address evaluation goal (ii), we applied our work from Sec.~\ref{sec:verification} to verify our provided commutativity conditions in our conditionally commutative benchmarks.
As can be seen in Table~\ref{table:benchmarks}, for all 6 benchmarks we were able to successfully verify all commutativity conditions (29 in total), with each verification taking fractions of a second (the times were not longer than the ones in the Servois2 benchmarks~\cite{servois}).
\ekremoved{Commutativity of loops is left as  future work.}
We can verify all CommSet examples, as well as the examples that Veracity was able to verify, since our implementation extends Veracity.
Additionally, the commutativity conditions of all of the examples used in Sec.~\ref{sec:overview} were verified with our tool, with similarly quick verification times.

\emph{Impact of the size of an NCB.}
In our current work, the programmer decides how many statements are enclosed within an NCB. (Note future work in Sec.~\ref{sec:conclusion}.) We observed this while using NCB ourselves. There are some performance trade-offs to consider when making such a choice\POPL{ including (i) the complexity of the condition, (ii) synchronization overhead for larger NCBs, (iii) smaller NCBs have fewer dependencies, and (iv) larger NCBs have less forking overhead}.
\ARXIV{These include:
	 (i) \emph{Complexity of commutativity conditions:} Larger NCBs tend to have more complex commutativity conditions, as they may manipulate more variables in more complex ways within a single NCB. This complexity can make it harder to identify truly commutative instructions, potentially limiting the degree of parallelization.
	 (ii) \emph{Synchronization overhead:} Larger NCBs require fewer synchronization points, reducing overhead and potentially improving performance. In contrast, smaller NCBs introduce more frequent synchronization, which can negate the benefits of higher commutativity by adding more coordination overhead.
	 (iii) \emph{Runtime commutativity checking:} In smaller NCBs, commutativity conditions have fewer dependencies and can be evaluated more quickly.
	 (iv) \emph{Code coverage, instruction grouping, and memory access:} Larger NCBs cover more instructions within a single block, reducing the total number of blocks required and simplifying program structure. When a program has limited memory access, forming larger blocks is particularly advantageous, as fewer memory dependences reduce data conflicts and simplify commutativity conditions. Smaller NCBs, however, allow finer control over parallelization by isolating dependences, though they require more blocks to achieve full program coverage.
}

{\emph{Writing NCBs for concurrency.}
Beyond their size, how NCBs and their conditions are written decides how often a job may run ahead of an earlier one. Measuring our benchmarks under several scheduling rules suggests the following guidelines.
(i) \emph{Let conditions read only parameters where possible.} Each per-instance value that a condition needs, such as an index or an identifier, should be named as a parameter, and parameters are read-only (W1, Def.~\ref{def:well-formed}). No running job can then change the condition, so the runtime may act on it as soon as both jobs exist, even if the earlier job is already running, and when the job checks the condition again before its first step, it still holds. The conditions of \texttt{vote} (${voterID_{1} \neq voterID_{2}}$), \texttt{commset} and \texttt{simple-io} are of this form, and they are the benchmarks whose speedup depends most on overtaking running jobs.
(ii) \emph{Keep written state out of the parameters.} A variable that a block assigns, including an array or table that it stores into, belongs to the shared state and is not a parameter. A condition that needs its value reads it as shared state.
(iii) \emph{When a condition must read shared state, write that state late.} A running job can then still be overtaken until it first writes a variable that the condition reads, so long computation should come before such writes. A block that starts with \code{c = c + 1;} makes every later job whose condition reads \code{c} wait for it.
(iv) \emph{Keep the shared reads of a condition narrow.} Our implementation tracks an array as one variable, so a write to \code{arr[j]} counts as a write to everything a condition over \code{arr[i]} reads. Conditions over separate variables interfere less.
(v) \emph{Prefer conditions that the blocks can only reinforce.} In Ex.~\ref{ex:motivation}, $\nblock{f}$ only adds \code{x*x} to \code{arr[i]}, so once \code{arr[i] > 0} holds, running $\nblock{f}$ keeps it true. Such a condition is more likely to still hold when other jobs run in between and the job checks it again before its first step.}

\section{Related Work}
\label{sec:relwork}

As discussed in Sec.~\ref{sec:intro} there are almost no existing automatic parallelization works that exploit verified commutativity conditions at runtime.

Some prior works exploit (unverified) commutativity conditions for parallel performance at runtime.
The Jade~\cite{jade} system provided a declarative language for specifying how tasks access shared objects. Jade expressed commutativity through commuting access specifications to an object \texttt{o}, written \texttt{cm(o)}, indicating that the task's accesses to \texttt{o} belong to a class of accesses declared to commute with one another.
Furthermore, runtime conditions could be written to determine whether a commuting access specification applied (roughly following the form \texttt{withonly if($conditions$) cm(o)}).
Finally, Jade's runtime scheduler used per-object queues to enforce access dependencies, allowing commuting accesses to execute in different serial orders while exploiting parallelism among independent tasks.
Galois~\cite{galois} exploited commutativity of method calls on graph-like ADTs to achieve data-parallelism in irregular programs via speculation and rollback.

Other works provided annotations for operations that \emph{always} commute, such as 
commutative function annotations for DSWP~\cite{bridges-micro07}, and ``commutes with'' class method annotations in Deterministic Parallel Java~\cite{DBLP:conf/oopsla/BocchinoADAHKOSSV09}. Unlike \name{}, \ARXIV{these works do not support commutativity \emph{conditions}:} operations must always commute so they miss cases where operations only sometimes commute. These works also do not verify commutativity and instead rely on the programmer.
\citet{commset} describe CommSet, which uses commutativity pragma annotations in a dependence-based  parallelization (such as PS-DSWP~\cite{ps-dswp} or DOALL~\cite{DBLP:books/mk/AllenK2001}). These annotations can have predicates that are resolved statically. 
\ARXIV{\name{} goes a step further by exploiting commutativity for concurrent execution of program fragments that are only conditionally commutative, and thus those conditions must be checked at runtime. As compared to CommSet, \name{} also integrates commutativity more directly into the language and can verify commutativity conditions. Finally, NCBs can be created with arbitrary depth and are transformed into an arbitrary mixture of asynchronous-vs-sequential subcomputations, whereas CommSet is focused on parallelizing the most critical loop (via DOALL or PS-DSWP).}
\POPL{\name{} goes a step further by verifying commutativity conditions and evaluating them at runtime.}
See also Sec.~\ref{sec:eval} in which we experimentally compared these works against \name{} (see a summary in Table~\ref{tab:summary} and details \inApx{apx:comparison}) showing that \name{} largely subsumes those works and enables new forms of parallelism. 

Veracity~\cite{veracity} exploits commutativity of more limited kinds of programs, i.e., those in which commuting fragments are directly adjacent. 
In contrast, \name{} allows for unrestricted placement of commutative blocks and exploits the commutativity of those nonadjacent blocks in a dependence-based parallelization. \name{} can thus also exploit self-commutativity (e.g. different iterations of the same loop body).  
\name{} subsumes Veracity, since NCBs can be adjacent and use no parameters\ARXIV{, \ie{}, $\nblockA{f}{\;}:\{s_1\}\;\nblockA{g}{\;}:\{s_2\}$ with a commutativity condition provided at the top of the program (see also Sec.~\ref{sec:eval})}.

\emph{Commutativity verification.}
\citet{kim-rinard-pldi11} manually verify commutativity conditions of ADT specifications, a line of work that led to further automated synthesis/verification works including {\sc Servois}~\cite{DBLP:conf/tacas/BansalKT18} and {\sc Servois2}~\cite{DBLP:conf/atva/ChenFNPBK23}. Those works do not parallelize or even execute programs. 
Veracity~\cite{veracity} extended commutativity verification to program fragments, via a translation to {\sc Servois2}~\cite{DBLP:conf/atva/ChenFNPBK23}, and \name{} extends verification to pairs of arbitrary code regions.

Other works verify commutativity of specific objects or in specific domains.
\citet{DBLP:conf/tacas/ChenHSW15} focus on Map-Reduce programs and discuss undecidability for checking the commutativity of integer reducers.
In the setting of CRDTs, \citet{DBLP:conf/forte/ZellerBP14} verify convergence using semilattice properties including commutativity of merge operations, and \citet{DBLP:journals/pacmpl/GomesKMB17} prove commutativity of certain classes of CRDT objects via their abstract convergence theorem.
\citet{DBLP:conf/cav/NagarJ19} 
introduce the concept of commutativity modulo consistency policy for CRDTs allowing convergence even when certain operations do not commute
and \citet{DBLP:journals/pacmpl/LiuPRK0V20} verify commutativity of RDTs in LiquidHaskell.
Finally, \citet{DBLP:conf/pldi/Pirlea0S21} described an abstract interpretation that tracks the linearity of assignments to determine when operations always commute.

OpenMP~\cite{openmp} is a well known API 
that provides the programmer with primitives to manage parallelism such as marking asynchronous regions, managing local/shared data access, synchronization, etc. 
\ARXIV{By contrast, \name{} (and CommSet, Veracity, PS-DSWP) intentionally shield the programmer from such primitives and instead automatically parallelize.
Additionally, \name{} verifies commutativity conditions and \name{}'s built-in runtime support for conditional commutativity allows some loop iterations to be parallelized with some others, rather than an all-or-nothing approach.}
OpenMP~5.0 \texttt{mutexinoutset}~\cite{openmp5} let tasks on the same data run in either order (though not at the same time), but again unconditionally and as declared by the programmer.
Others also encapsulated code fragments into tasks~\cite{starPU,dask}. 
StarPU's \texttt{STARPU\_COMMUTE} access mode lets writes to the same data run in any order, but unconditionally.

Dependence-based parallelization has been active for decades. 
\citet{DBLP:books/mk/AllenK2001} provide a comprehensive view.
\citet{DBLP:journals/toplas/FerranteOW87} introduced the program dependence graph and describe how it can be used for optimization. 
\citet{DBLP:journals/toplas/PinterP94} describe ``idioms''  for parallelization based on a PDG.
\citet{DBLP:conf/issta/HarroldMR93} describe how a PDG can be built efficiently.
PS-DSWP~\cite{ps-dswp} and DOALL~\cite{DBLP:books/mk/AllenK2001} (discussed throughout this paper), and many other works including \citet{DBLP:conf/micro/OttoniRSA05,DBLP:conf/cgo/HuangRJZHA10,DBLP:journals/taco/RanganVOA08} are more recent examples of PDG-based parallelization.

\emph{Reasoning methods} pertaining to commutativity is also an active field, with applications to proofs~\cite{DBLP:conf/cav/LetteF23,DBLP:journals/pacmpl/FarzanKP23,DBLP:journals/pacmpl/FarzanKP24,DBLP:journals/pacmpl/EneaK24}, 
race detection~\cite{DBLP:conf/pldi/DimitrovRVK14},
predictive analysis~\cite{DBLP:journals/pacmpl/FarzanM24}, 
distributed computing~\cite{DBLP:conf/eurosys/NajafzadehGYFS16,DBLP:journals/pacmpl/HoushmandL19},
etc.
It is also used in reduction-based proof techniques~\cite{DBLP:journals/cacm/Lipton75,DBLP:conf/popl/ElmasQT09} and induction~\cite{DBLP:conf/pldi/KraglEHMQ20}.

\section{Conclusion}
\label{sec:conclusion}
We have presented a method of automatic parallelization that allows programmers to write commutativity conditions over arbitrary regions of code and have those conditions be verified and exploited at runtime.
\name{} introduces new parallelization opportunities over those afforded by prior works
and allows for a geomean speedup of \finalSpeedup{}$\times$  attributable to runtime commutativity.

\emph{Future work.} 
Although commutativity conditions can usually be verified and sometimes synthesized, there is room to improve commutativity condition synthesis in the future. 
In the long run, improved synthesis can be used in concert with profiling information, to make decisions about where to place NCBs could be done (semi{-})automatically in a compiler.

\section{Disclosure of Use of AI in Research}
\label{sec:ai-disclosure}
The core ideas of this work, the language design, the parallelization analysis, the execution model, the vast majority of the implementation of \tool{}, and the initial writing of the paper were carried out by without the help of AI, before AI coding assistants were in common use.
More recently, we used the AI coding assistant Claude Code (Anthropic) 
to help implement lock synthesis (Sec.~\ref{sec:implementation}\POPL{; details \inApx{apx:locksynth}}), following the design described in this paper.

\section{Data Availability Statement}
\label{sec:data-availability}

Our implementation of \tool{} (Sec.~\ref{sec:implementation}) and all benchmark programs are available at \url{https://github.com/veracity-lang/veracity2g}.
The repository contains the source of the OCaml ($\geq$5) interpreter, including the commutativity verification (via Servois2), the parallelization analysis, lock synthesis and the multicore runtime, together with all benchmark programs, the Python scripts used to run the experiments of Sec.~\ref{sec:eval}, and directions for reproducing the results.
\POPL{Furthermore, additional detail is in the extended version of this paper~\cite{technicalreport}.}

\begin{acks}
We thank the anonymous reviewers for their feedback on earlier versions of this paper. 
This material is based upon work supported by 
\grantsponsor{GSNSF}{NSF}{https://www.nsf.gov} award \grantnum{GSNSF}{CCF-2008633} and 
\grantsponsor{GSNSF}{NSF}{https://www.nsf.gov} award \grantnum{GSNSF}{CCF-2315363}.
\end{acks}


\ifARXIV
\appendix

\section{General variable classes}
\label{apx:in-vars}

In Sec.~\ref{sec:ncb} we assumed $In_\nblock{f} = In_\nblock{g} = \emptyset$, counting every variable declared outside both NCB bodies in $\GVars{}$. Here we lift this assumption. Fix a specification entry and NCBs as in Sec.~\ref{sec:ncb}.

\begin{definition}[General variable classes]
\label{def:var-classes-gen}
$\GVars{}$ is the set of variables of the scopes enclosing the outermost loop that contains both $s_\nblock{f}$ and $s_\nblock{g}$ (the common enclosing scope when there is no such loop), together with globals and memory locations.
$In_\nblock{f}$ is the set of variables declared outside $s_\nblock{f}$, in scope at $\nblock{f}$, but not in $\GVars{}$.
The $\nblock{f}$-local variables are those declared in $s_\nblock{f}$.
$In_\nblock{g}$ and the $\nblock{g}$-local variables are defined symmetrically.
\ARXIV{\rocqref{Commutativity.ncb}}
\end{definition}

\noindent
Instances from different iterations of a loop that contains both NCBs do not share its body's variables, so these are in $In_\nblock{f}$ and $In_\nblock{g}$ rather than $\GVars{}$. A loop-header variable such as \code{i} in \code{for(int i=...)} belongs to the enclosing scope, hence to $\GVars{}$. Variables of $In_\nblock{f}$ are declared, e.g., in the loop body or in a block enclosing $\nblock{f}$ but not $\nblock{g}$; by (W2), $s_\nblock{f}$ may only read them. Parameters are not a further class: by (W0), each $x_k$ is in $\GVars{}$ or $In_\nblock{f}$ (Fig.~\ref{fig:scopes-gen}).

\begin{figure}[h]
\begin{minipage}{0.4\linewidth}
\begin{lstlisting}[style=vcy,language=VCY,basicstyle=\ttfamily\scriptsize]
int[] a, b;
for (int i = 0; i < n; i++) {
  int u = h(i);
  |$\nblockA{f}{i}$|:{ int x = u*u; a[i] = x; }
  int t = u + 1;
  |$\nblockA{g}{i}$|:{ int y = t*2; b[i] = y; }
}
\end{lstlisting}
\end{minipage}\hfill
\begin{minipage}{0.6\linewidth}
\centering
\begin{tikzpicture}%
  \draw[rounded corners] (0,0) rectangle (6,3.1);
  \node[anchor=north west] at (0,3.1) {$\GVars{}$: \texttt{a}, \texttt{b}, \texttt{n}, \texttt{i}};
  \draw[rounded corners, dashed] (0.25,0.2) rectangle (5.75,2.55);
  \node[anchor=north west] at (0.25,2.55) {loop body: \texttt{u} $\in In_\nblock{f} \cap In_\nblock{g}$};
  \draw[rounded corners, fill=gray!12] (0.45,0.4) rectangle (2.55,1.75);
  \node[align=center] at (1.5,1.07) {locals of $s_\nblock{f}$\\[1pt]\texttt{x}};
  \draw[rounded corners, dotted] (2.75,0.3) rectangle (5.6,2.15);
  \node[anchor=north west] at (2.75,2.15) {after $\nblock{f}$: \texttt{t} $\in In_\nblock{g}$};
  \draw[rounded corners, fill=gray!12] (2.95,0.45) rectangle (5.4,1.6);
  \node[align=center] at (4.175,1.02) {locals of $s_\nblock{g}$\\[1pt]\texttt{y}};
\end{tikzpicture}
\end{minipage}
\caption{Variable classes for the pair $\nblockA{f}{i}, \nblockA{g}{j} : \texttt{i != j}$.
Here $In_\nblock{f} = \{\texttt{u}\}$ and $In_\nblock{g} = \{\texttt{u}, \texttt{t}\}$, since \texttt{t} is not in scope at $\nblock{f}$.
The shared state $\sigma_{gl}$ holds $\GVars{}$. Each instance's local state holds
entry snapshots of its parameters and $In$ variables (read-only by (W1), (W2)) and its
body locals. The parameter \texttt{i} names a $\GVars{}$ variable; its snapshot is what
the condition refers to as \texttt{i} or \texttt{j}.}
\label{fig:scopes-gen}
\end{figure}

\noindent
Variable association (Def.~\ref{def:var-assoc}) is unchanged, so a condition still cannot mention variables of $In_\nblock{f}$ and $In_\nblock{g}$ other than parameters. Local states (Def.~\ref{def:partitioned-state}) now also copy the $In$ variables: $\sigma_\nblock{f}$ is created as $\sigma{\upharpoonright}_{\{\bar x\} \cup In_\nblock{f}}$, and likewise for $\nblock{g}$.
Variables of $In_\nblock{f}$ are entry snapshots just like parameters, but only parameters can appear in $\varphi$. While $s_\nblock{f}$ runs, $dom(\sigma_l)$ in Sec.~\ref{sec:ncb-partition} is $\{\bar x\} \cup In_\nblock{f}$ together with the variables declared in $s_\nblock{f}$. Def.~\ref{Valid NCB Commutativity Condition} is unchanged. Lemma~\ref{lem:discard} holds with (i) strengthened to: $\sigma_{\nblock{f}}'$ agrees with $\sigma_{\nblock{f}}$ on $\{\bar x\} \cup In_\nblock{f}$. In the proof, the variables that $s_0$ may assign by (W2) are not in $In_\nblock{f}$, by definition and by the naming-apart convention. In the execution model, {\sc ForkTask} copies $\sigma_L{\upharpoonright}_{\{\bar x\} \cup In_\nblock{f}}$ into the new job's local state. By (W1) and (W2) these variables are read-only in $s_\nblock{f}$, so the copy is the value the body reads.

\noindent
(W2) is necessary. Without it, suppose $\nblock{g}$ writes a variable \code{k} that is declared between the common scope and $\nblock{g}$, e.g., \code{\{ int k; $\nblock{g}$:\{ k = arr[0]; \} use(k); \}}, and $\nblock{f}$ writes \code{arr[0]}. Then the final \code{k} depends on the order, yet \code{k} is outside $\sigma_{gl}$, so the definition would accept \code{true}.
Treating such writes as per-fragment outputs, compared in both orders like ADT return values, is left to future work.

\section{Parallelization opportunities in comparison to prior work}
\label{apx:comparison}

We now discuss the detailed experiments that address evaluation goal (i).

{\bf PS-DSWP (dependence-based  parallelization; no commutativity).}
Parallel-Stage Decoupled Software Pipelining (PS-DSWP)~\cite{ps-dswp} identifies strongly connected components in a 
program's data dependence graph and executes them in
parallel across multiple threads.
PS-DSWP focuses primarily on identifying a single heavy top-level loop to parallelize, creating a pipeline of stages that can execute concurrently.
Conceptually, \name{} differs from PS-DSWP in its approach to handling NCBs. While PS-DSWP is mostly limited to parallelizing a single main loop, \name{} can identify parallelizable blocks throughout the program structure.
To compare with PS-DSWP, we recreated their primary motivating example (using a 2-dimensional array instead of a linked list). This specific benchmark demonstrates dependence-based  parallelization without commutativity.
Figure ~\ref{fig:ps-dswp-2darray} plots parallel-to-sequential speedup ratio against the logarithm of computational size. The results show that \name{} scales like PS-DSWP in the absence of NCBs, confirming that \name{}, like PS-DSWP, achieves dependence-based scaling. {This benchmark lies outside the scope of the mechanized theorem, which has no commutativity conditions to exploit here. Its pipeline also relies on a runtime mechanism the mechanization does not model: the row search's jobs run in parallel, and each waits only at the point where it spawns a job of the next stage until the earlier row searches have finished, so that the next stage's jobs are created in program order. The mechanization instead covers spawners whose jobs run one at a time, which here would serialize the row search.}

\begin{figure}[H]
	\centering
	\includegraphics[width=0.5\linewidth]{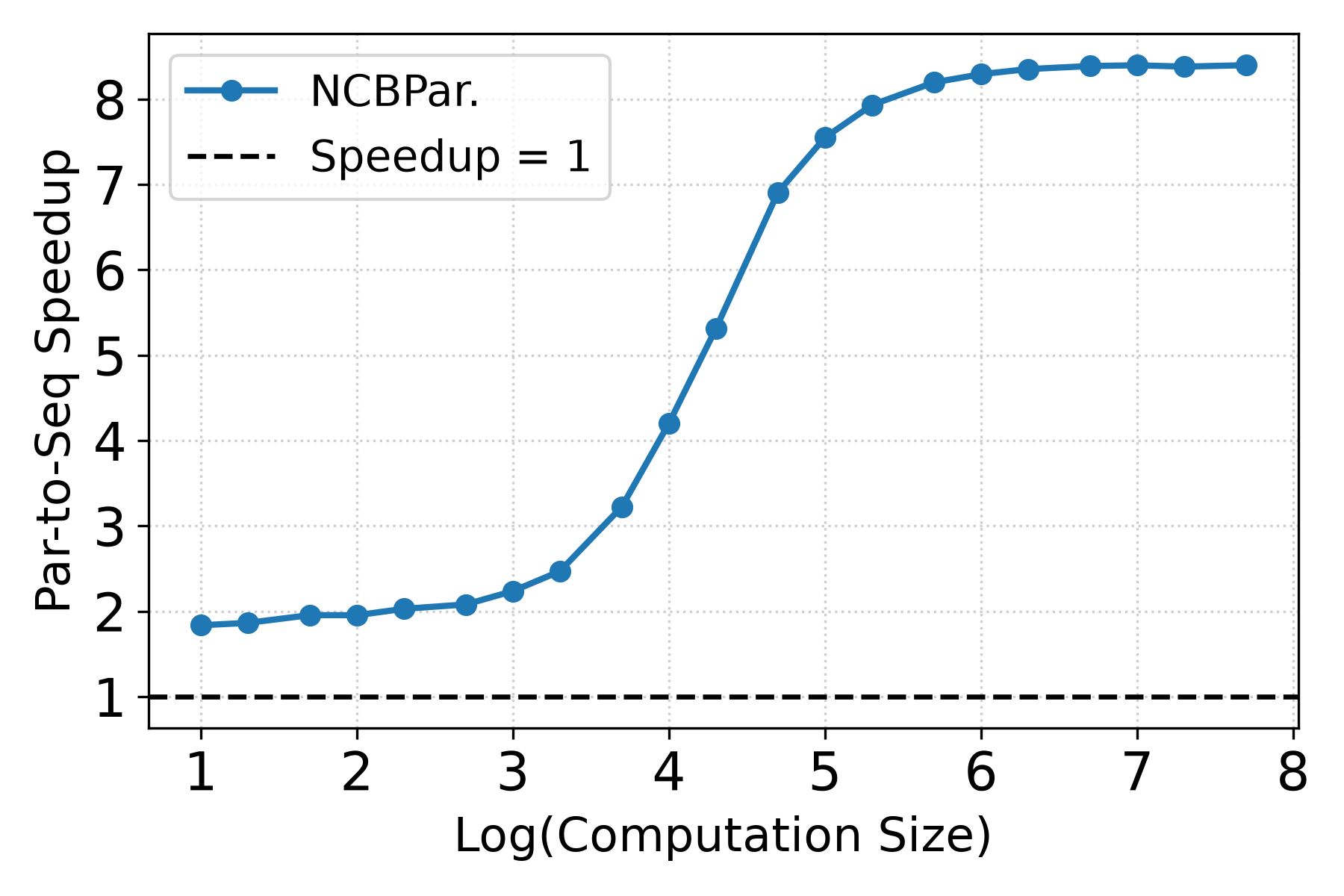}
	\caption{Example adapted from~\citet{ps-dswp}.}
	\label{fig:ps-dswp-2darray}
\end{figure}

{\bf CommSet. (Dependence-based  parallelization; static-only commutativity)} 
CommSet allows users to specify commutativity through pragmas, identifying commutativity conditions that always hold (e.g. when loop index variable is different on each iteration).
Unlike \name{}, CommSet cannot dynamically evaluate commutativity conditions so it cannot parallelize the new benchmarks discussed below.
We evaluated \name{} on CommSet's motivating example (\texttt{commset-md5})
and adapted two of the publicly-available benchmarks (\texttt{commset-potrace} and \texttt{commset-kmeans}) in figure ~\ref{fig:commset-examples}.
(We were unable to obtain the instrumented source code used in the original evaluation, nor could we find the source code for many of the original benchmarks. Additionally, since their implementation operates at the LLVM IR level while ours works at the source level, a meaningful direct performance comparison is not possible. Instead, we include their published results alongside ours to provide context -- the purpose of the comparison is to show that our procedure also automatically parallelizes the case studies.)
\begin{figure}[H]
	\centering
	\begin{subfigure}{0.49\linewidth}
		\centering
		\includegraphics[width=\linewidth]{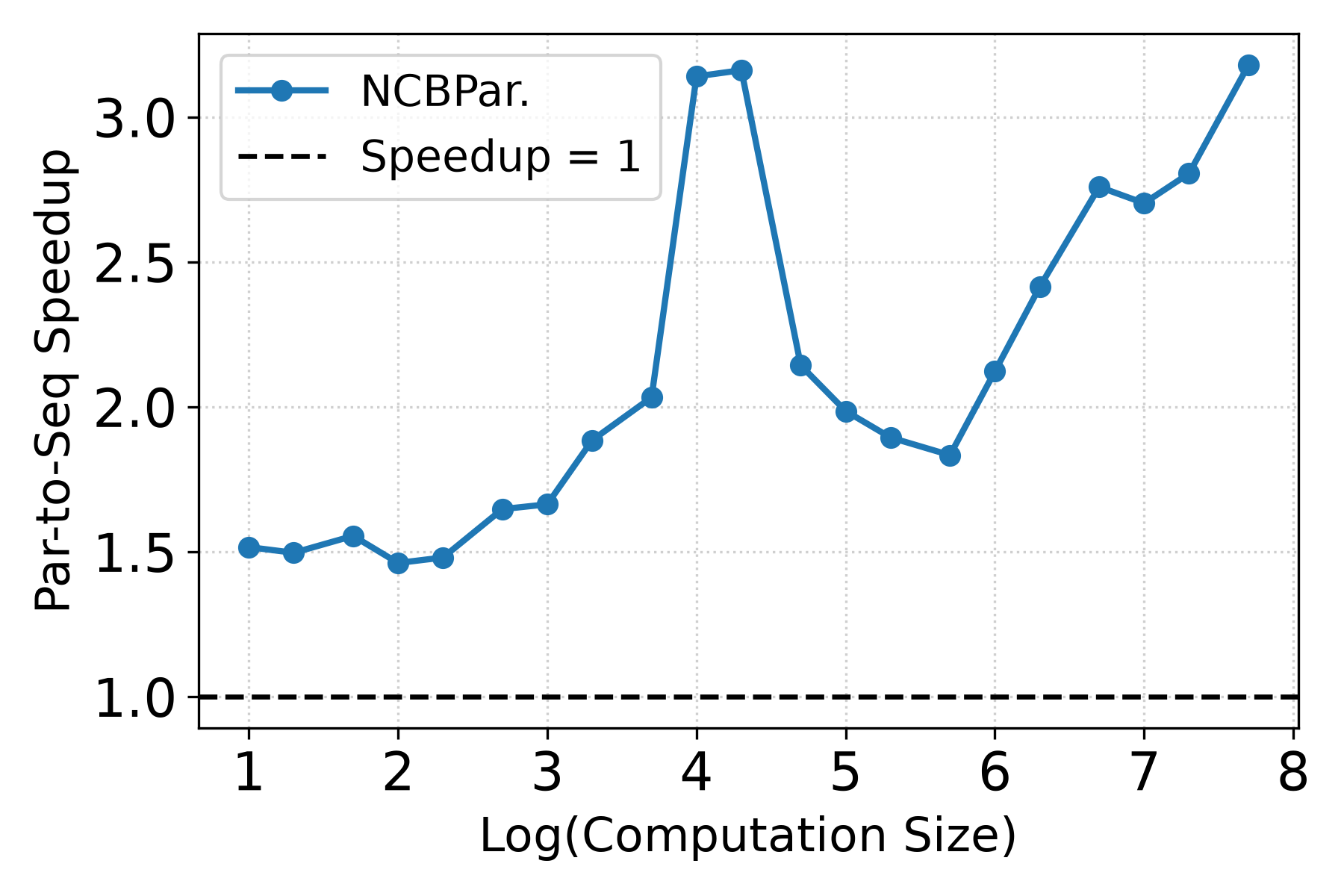}
		\caption{\texttt{commset-md5}}
		\label{fig:commset-md5sum}
	\end{subfigure}
	\begin{subfigure}{0.49\linewidth}
		\centering
		\includegraphics[width=\linewidth]{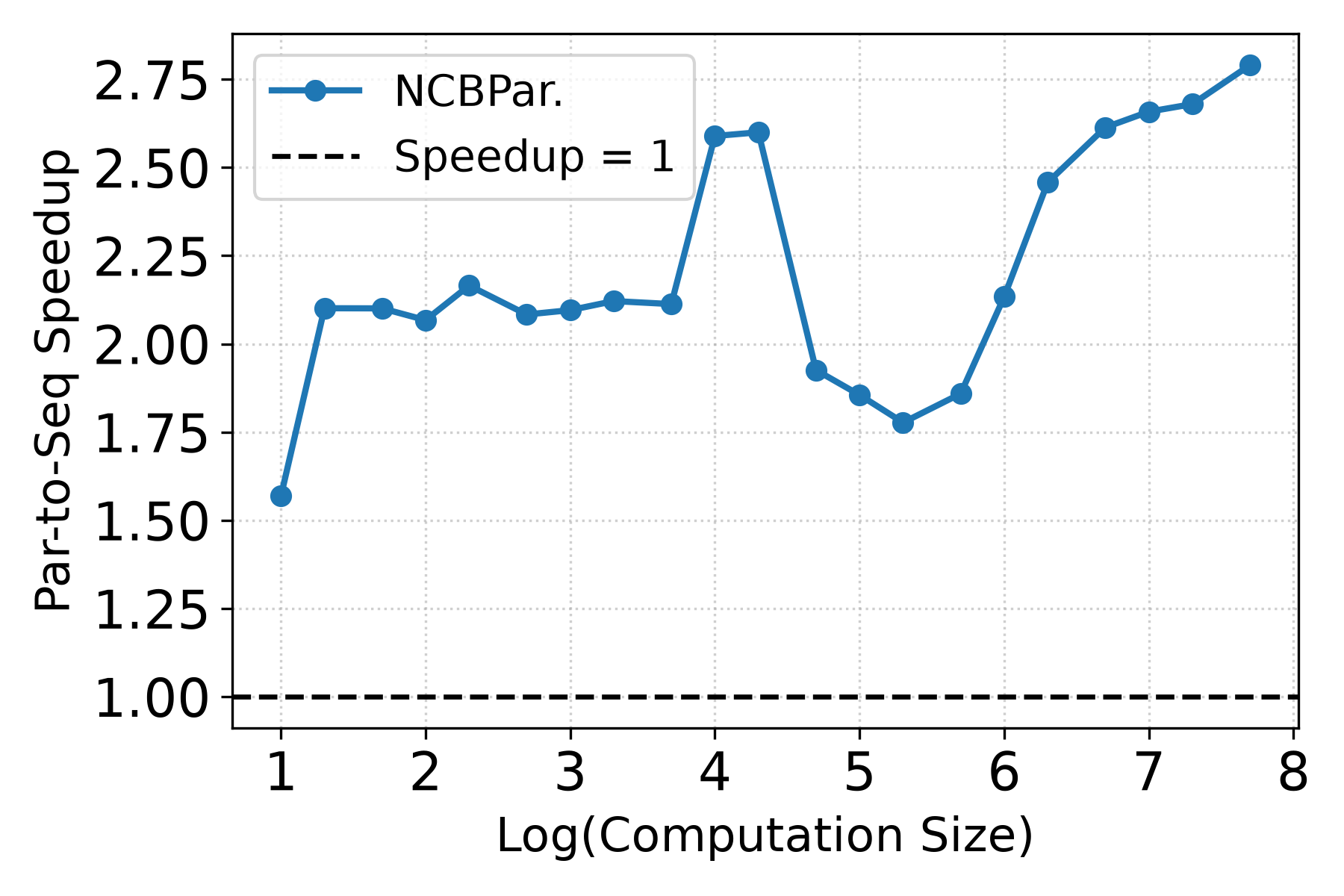}
		\caption{\texttt{commset-potrace}}
		\label{fig:commset-potrace}
	\end{subfigure}

	\begin{subfigure}{0.49\linewidth}
		\centering
		\includegraphics[width=\linewidth]{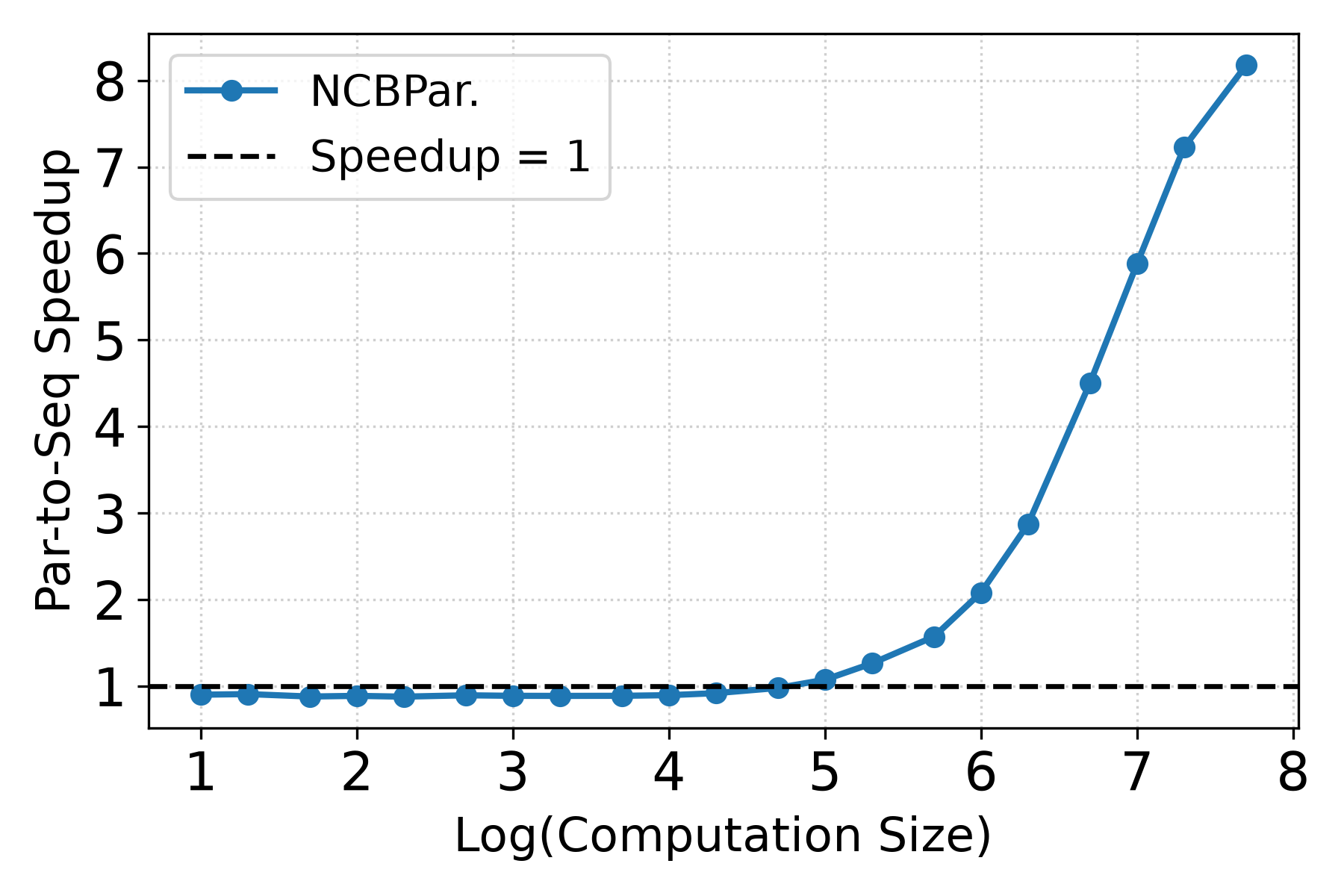}
		\caption{\texttt{commset-kmeans}}
		\label{fig:commset-kmeans}
	\end{subfigure}
	\caption{Examples adapted from~\citet{commset}.}
	\label{fig:commset-examples}
\end{figure}
The commutativity conditions we used were: $(i_1 \neq i_2)$ for \texttt{commset-md5}, $(true)$ for \texttt{commset-kmeans}, and
$(argv[2*i_1+1] \neq argv[2*i_2] \wedge argv[2*i_1] \neq argv[2*i_2+1] \wedge argv[2*i_1+1] \neq argv[2*i_2+1])$
for \texttt{commset-potrace}.
Unlike CommSet, we were also able to verify these commute conditions.

{\bf Veracity (Only adjacent block parallelization; dynamic commutativity).}
Veracity introduced the concept of a \texttt{commute}$(b)\{s_1\}\{s_2\}$ statement, allowing the user to specify the commute condition $b$ of two adjacent blocks $s_1$ and $s_2$. Veracity dynamically evaluates $b$ and if it reduces to
true, executes $s_1$ and $s_2$ in parallel.
Although Veracity allows dynamic commutativity checks, it does not perform a dependence-based evaluation. \name{} by contrast uses NCBs to specify commutativity across arbitrary code regions, performs a dependence-based  parallelization and can dynamically evaluate commute conditions, including ``self-commute'' conditions (e.g. of loop bodies).

Veracity can be considered a primitive version of \name{}, where fragments must be adjacent (the exception is that Veracity allows for nested \texttt{commute} statements). 
We thus compare
\name{} to Veracity using 
the entire suite of benchmarks from \citet{veracity}. We changed \texttt{commute} blocks to \ncb{}s. The commutativity conditions were translated and placed into a global commutativity specification.
We also removed the manually inserted locks from the Veracity benchmarks, relying on \name{} to automatically synthesize the locks needed for synchronization.
All of the benchmarks were run with an increasing computational load while being compared to sequential execution.
Fig.~\ref{fig:veracity-apx} plots the parallel-to-sequential speedup ratio against the logarithm of computational size, with the legend identifying each Veracity benchmark. It confirms that we can reproduce the speedup demonstrated by Veracity's \texttt{commute} blocks.
While \name{} has a different runtime compared to Veracity, it achieves similar scaling behavior.
This shows that \name{} achieves dependence-based  parallelization with broader commutativity support than Veracity, while maintaining comparable performance—demonstrating that increased expressiveness does not introduce significant overhead.

\begin{figure}[H]
	\centering
	\includegraphics[width=0.95\textwidth]{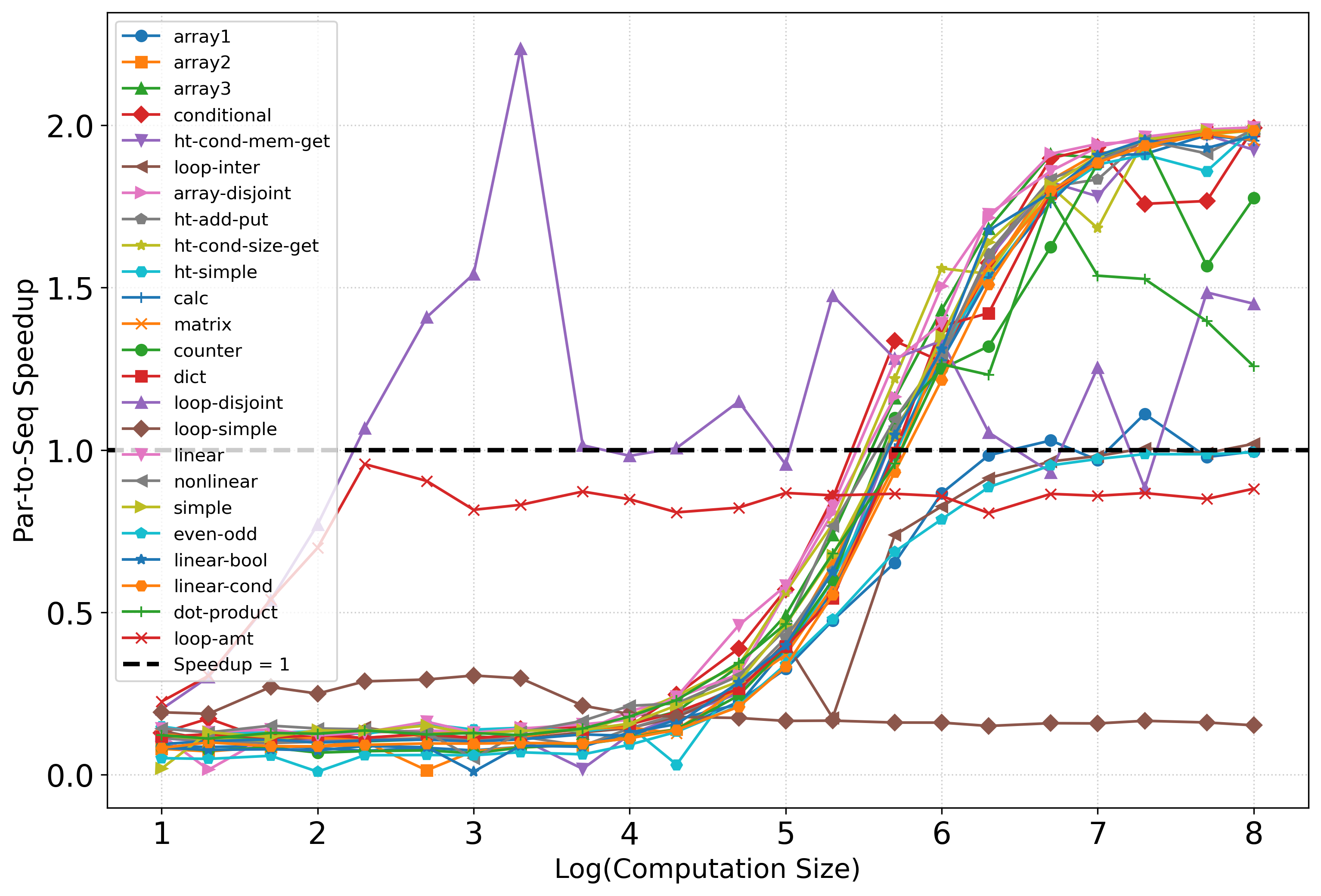}
	\caption{Example adapted from~\citet{veracity}.}
	\label{fig:veracity-apx}
\end{figure}

\input{rocq-examples}

\section{Benchmarks after Parallelization and Lock Synthesis}
\label{apx:emitted-tasks}

Each benchmark is decomposed by the algorithm into a set of tasks.
Fig.~\ref{fig:motivation-tasks} shows the task graph emitted for the \texttt{motivation} benchmark.
Each box is a task, marked \emph{Sequential} or \emph{Doall}, with its \emph{post-synthesis} body:
the \texttt{mutex\_lock}/\texttt{mutex\_unlock} calls (red) have been inserted
automatically by the lock synthesis pass where concurrent tasks access the
same shared variable with at least one write.
Solid edges are dependences, labelled with the variables they carry, and a loop is a self-dependence; orange edges also carry a commutativity condition. Dashed edges create new jobs, and dotted edges are spawns from the initial task.
The task listings for the remaining benchmarks can be found in our public
repository at
\url{https://github.com/veracity-lang/veracity2g/tree/main/reports/tasks},
and their task graphs at
\url{https://github.com/veracity-lang/veracity2g/tree/main/reports/tasks-img}.

\begin{figure}[H]
	\centering
	\includegraphics[width=\textwidth]{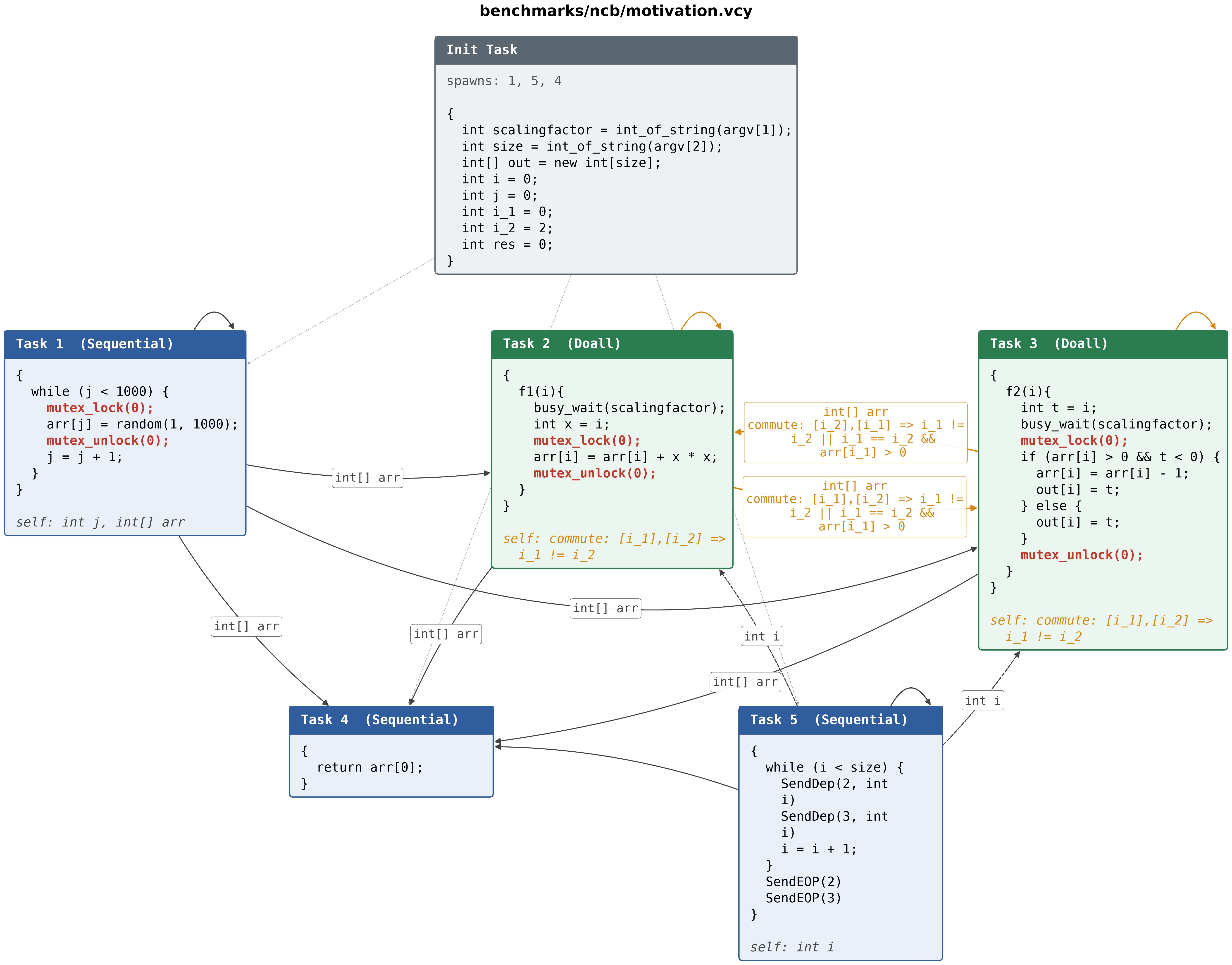}
	\caption{Task graph of the \texttt{motivation} benchmark after parallelization and lock synthesis.}
	\label{fig:motivation-tasks}
\end{figure}

\fi
\end{document}

%% file: varclasses.tex
\vspace{-12pt}
\begin{lstlisting}[style=vcy,language=VCY,basicstyle=\ttfamily\scriptsize,xleftmargin=0pt,xrightmargin=0pt,aboveskip=0pt,belowskip=2pt]
int[] a, b;
for (int i = 0; i < n; i++) {
  int u = h(i);
  |$\nblockA{f}{i,u}$|:{ int x = u*u; a[i] = x; }
  |$\nblockA{g}{i}$|:{ int y = i*2; b[i] = y; } }
\end{lstlisting}
\centering
\begin{tikzpicture}[font=\scriptsize,inner sep=1.5pt,
    par/.style={draw,dashed,fill=white,minimum width=15pt},
    cp/.style={->,>=stealth,thin}]
  \draw[rounded corners,fill=gray!12] (0.15,0.05) rectangle (2.7,1.05);
  \draw[rounded corners,fill=gray!12] (2.9,0.05) rectangle (5.05,1.05);
  \draw[rounded corners] (0,0.5) rectangle (5.3,1.75);
  \node[anchor=west] at (0.05,1.52) {$\GVars{}$};
  \node at (1.05,1.52) {\texttt{a}};
  \node at (1.40,1.52) {\texttt{b}};
  \node at (1.75,1.52) {\texttt{n}};
  \node (i) at (2.45,1.52) {\texttt{i}};
  \node (u) at (3.15,1.52) {\texttt{u}};
  \node[anchor=west] at (0.2,0.27) {$\Varsf$};
  \node[par] (fi) at (0.95,0.78) {\texttt{i}};
  \node[par] (fu) at (1.95,0.78) {\texttt{u}};
  \node at (1.6,0.27) {\texttt{x}};
  \node[anchor=west] at (2.95,0.27) {$\Varsg$};
  \node[par] (gj) at (3.7,0.78) {\texttt{i}};
  \node at (4.35,0.27) {\texttt{y}};
  \draw[cp] (i) -- (fi);
  \draw[cp] (u) -- (fu);
  \draw[cp] (i) -- node[right,pos=0.45,inner sep=1pt] {\tiny copy} (gj);
\end{tikzpicture}
\vspace{-12pt}
\caption{Variable classes. Dashed variables are per-instance input copies of variables (read-only by (W1)).}
\label{fig:scopes}
\vspace{-8pt}

%% file: translation.tex
\subsection{Background: {\sc Servois2} Logical Pre/Post Commutativity}

The tool {\sc Servois2}~\cite{DBLP:conf/atva/ChenFNPBK23} can synthesize or verify commutativity conditions of SMT logical pre/post specifications of methods operating on an object. Subsequent work on Veracity~\cite{veracity}, described a translation $\Tr$ from Veracity programs into those object pre/post specifications, such that the resulting {\sc Servois2} commutativity entails Veracity statement commutativity.

Our approach builds on this, although must be extended to support named commutative blocks and the interpretation over state triples. We now first summarize the Veracity translation and then discuss our extension.
The format of a {\sc Servois2} object is as follows:
\[
\mathcal{O} = \left\{
\begin{array}{lllllllll}
	& \textsf{state} &:& (Var,Type)\; \textsf{list};\;\;\;\;\;& \textsf{methods} &:& \textsf{Meth list};\\ 
	& \textsf{eq}    &:& \textsf{state} \times \textsf{state}; & \textsf{spec} &:& \textsf{Meth} \rightarrow (P,Q) & 
\end{array}
\right\}
\]

\noindent
The above object has a \textsf{state} which is a list of typed variables. Equality is a relation over those state variables. Finally, there is a list of methods, each having a pre/post specification. The {\sc Servois2} tool can then synthesize or verify a provided commutativity condition for a pair of those methods.

Veracity employs an SSA-like translation denoted $Tr(stmt)$ that translates program statements into logical pre/post conditions, in the form of such an object $\mathcal{O}$, for which {\sc Servois2}~\cite{DBLP:conf/atva/ChenFNPBK23} can synthesize or verify commutativity conditions.
Concretely, $\Tr(E,s_1,s_2)$ constructs $\mathcal{O}$ as follows: $\mathcal{O}.\textsf{state}$ contains one typed variable for every program variable visible in context $E$; $\mathcal{O}.\textsf{eq}$ is whole-state equality (all variables must agree); $\mathcal{O}.\textsf{methods}$ are two nullary methods $m_{s_1}$ and $m_{s_2}$; and $\mathcal{O}.\textsf{spec}$ maps each method to a \textsf{true} precondition and an SMT postcondition produced by $\textsf{specOf}$.
The function $\textsf{specOf}(s)$ translates a statement $s$ into an SMT formula.
It first introduces let-bindings $v_0 \leftarrow v$ for every variable accessed in $s$.
Then it recurses over $s$: each assignment $v := e$ introduces a fresh binding $(\texttt{let} \;\; v_{k+1} \;\; e[v_k/v,\ldots] ... )$; conditionals become \texttt{ite} branches (with the remaining statements appended to each branch); and nested \texttt{commute} blocks are treated as sequential composition.
The innermost formula, reached at \textsf{skip}, constrains each $(\texttt{=} \; v_{new} \; v_k)$ for the final index $k$, yielding a relational post-condition over unprimed (pre) and $\mathit{new}$-subscripted (post) variables.
For example, in the Veracity language two adjacent so-called \texttt{commute} blocks 
\texttt{\{ x := x+1; \}\{ if(x>0) y:=y-1; \}} would be translated to:
\[
\mathcal{O} = \left\{
\begin{array}{lllllllll}
	& \textsf{state} &:& \{ (\texttt{x},\mathbb{Z}), (\texttt{y},\mathbb{Z})\}, \;\;\;\;\textsf{methods} : [m_1,m_2], \;\;\;\;\textsf{eq}    : \texttt{x1=x2 \&\& y1=y2}; \\
	& \textsf{spec} &:& m_1 \rightarrow (true, \texttt{x' = (x+1)}), \;\; m_2 \rightarrow (true, \texttt{y' = (x>0 ? y : (y-1))}),  \\
\end{array}
\right\}
\]
Above the post-conditions of $m_1$ and $m_2$ reflect the new values of \texttt{x} or  \texttt{y} in terms of the previous values. 
{\sc Servois2} can then synthesize or verify commute conditions for this object such as \texttt{x>0}.

\subsection{Embedding NCB commutativity in {\sc Servois2}}

We now describe a translation of named commutativity blocks and their commutativity specifications into {\sc Servois2}. We will then prove that commutativity of the {\sc Servois2}-embedded object $\mathcal{O}$ entails soundness of the NCB commutativity specifications.

\begin{definition}[Translation of NCB fragment pairs]
For any two NCBs $\nblockA{f}{i}:\{s_f\}$ and $\nblockA{g}{j}:\{s_g\}$, let $o=Tr(E, s_f, s_g)$ following $\Tr$ defined in Def.~6.2 of~\cite{veracity}, except as follows. Let $\alpha_\texttt{i}$ be a fresh variable for \texttt{i}, and $\beta_\texttt{j}$ for \texttt{j}. 
\begin{itemize*}
\item \underline{$o$.\textsf{state}}: Variables are partitioned into three disjoint sets $Vars_{gl} \uplus Vars_f \uplus Vars_g$, where there is a variable $v \in Vars_{gl}$ for every global variable visible to the scope containing $s_f$ and $s_g$, $\alpha\in Vars_f$ are the $\nblock{f}$-local variables, and $\beta \in Vars_g$ are the $\nblock{g}$-local variables.

\item \underline{$o$.\textsf{eq}}: $o_1$ is equal to $o_2$ provided that they agree on every global variable $v\in Vars_{gl}$.

\item \underline{$o$.\textsf{methods}}: Nullary methods $m_f$ and $m_g$.

\item \underline{$o$.\textsf{spec}}: 
Associating $m_f$ and $m_g$ with true preconditions and post-conditions given by $specOf$ discussed below, i.e., 
$[m_f \mapsto (true,\;specOf(\nblockA{f}{i}: s_f)), \;\;\; m_g \mapsto (true,\;specOf(\nblockA{g}{j}: s_g))]$
\end{itemize*}

\noindent
The $specOf$ follows Fig. 3 of~\cite{veracity}, using translation $Tr(s)$ for statements and $Tr_E(e)$ for expressions. However, for the variables cases in translation of an NCB expression, $Tr_E$ is defined as:
\[\begin{array}{lll}
Tr_E(\texttt{i}) = \alpha_\texttt{i},&
Tr_E(\texttt{j}) = \beta_\texttt{j},&
Tr_E(\texttt{v}_{gl}) = \textsf{v}.
\end{array}\]
NCBs partition variables into global-vs-local and, as seen above,  our translation produces an $o$.\textsf{eq} that requires equality only on global variables, unlike Veracity which required all variables to agree.
\end{definition}

\noindent
For Ex.~\ref{ex:motivation}, the $specOf$/$Tr$ translation produces the following (simplified) post-conditions:

\noindent
\begin{tabular}{ll}
$Tr(\nblockA{f}{i}: s_f)$: & $Tr(\nblockA{g}{i} : s_g)$: \\
\begin{minipage}[t]{0.35\textwidth}
\vspace{-10pt}
\begin{lstlisting}[mathescape=true,basicstyle=\scriptsize\ttfamily,resetmargins=true,xleftmargin=0pt]
(let (x1 (pure_calc alpha_i))
(let (arr1 (store arr alpha_i
      (+ (select arr alpha_i) 
          (* x1 x1))))
(|$\wedge$| (= x' x1) (= arr' arr1))
))
\end{lstlisting}
\end{minipage}
&
\begin{minipage}[t]{0.65\textwidth}
\vspace{-10pt}
\begin{lstlisting}[mathescape=true,basicstyle=\scriptsize\ttfamily,resetmargins=true,xleftmargin=0pt]
(let (t1 (pure_compute beta_i))
(ite (and (> (select arr beta_i) 0) (< t' 0))
  (let (arr1 (store arr beta_i (- (select arr beta_i) 1)))
  (let (out1 (store out beta_i t'))
     (|$\wedge$| (= out' out1) (= arr' arr1))))
... ))
\end{lstlisting}
\end{minipage}
\end{tabular}

\noindent
These SMT expressions are state relations over unprimed and primed variables, as seen on the 5th lines which relate (only) the \emph{post global variables} to the pre variables which may include \emph{fragment-specific input} variables such as \lstinline|alpha_i|. (The SSA-style translation treats pure expressions such as \lstinline|calc| as deterministic but uninterpreted functions.)

Next, in addition to translation of NCB body, we define translation of a commutativity spec.

\begin{definition}[Translation of NCB Commutativity Conditions]
    \label{def:trphi}
    Let $\varphi$ be a commutativity condition declared for two NCBs
    $\nblockA{f}{i}$ and $\nblockA{g}{j}$. The translation $Tr_{phi}$ is defined as follows:
\[\begin{array}{lllllll}
Tr_{phi}(true) &=& \textsf{true} 
& Tr_{ncbe}(c) &=& \textsf{c} \\
Tr_{phi}(\neg \varphi) &=& (\textsf{not} \;\; Tr_{phi}(\varphi))
& Tr_{ncbe}(\texttt{i}) &=& \alpha_\texttt{i} \\
Tr_{phi}(\varphi_1 \vee \varphi_2) &=& (\textsf{or} \;\;\; Tr_{phi}(\varphi_1) \;\;\; Tr_{phi}(\varphi_2)) 
& Tr_{ncbe}(\texttt{j}) &=& \beta_\texttt{j} \\
{Tr_{phi}(\varphi_1 \wedge \varphi_2)} &{=}& {(\textsf{and} \;\;\; Tr_{phi}(\varphi_1) \;\;\; Tr_{phi}(\varphi_2))}
& Tr_{ncbe}(\texttt{v}_{gl}) &=& \textsf{v} \\
Tr_{phi}(ncb\text{-}e_1 \circlelt ncb\text{-}e_2) &=& (\circlelt_{smt} \;\;\; Tr_{ncbe}({ncb\text{-}e_1}) \;\;\; Tr_{ncbe}({ncb\text{-}e_2}))
\end{array}\]
{On compound NCB expressions (operators, conditionals, array indexing and field access), $Tr_{ncbe}$ is $Tr_E$, \eg{} $Tr_{ncbe}(ncb\text{-}e_1[ncb\text{-}e_2]) = (\textsf{select} \;\; Tr_{ncbe}(ncb\text{-}e_1) \;\; Tr_{ncbe}(ncb\text{-}e_2))$.}
\end{definition} 

\noindent
Intuitively, $Tr_{phi}(\varphi)$ translates $\varphi$ into an SMT expression pertaining to the object produced by $\Tr(E,s_f,s_g)$, with variables interpreted over the initial state triple $(\sigma_{gl},\sigma_{\nblock{f}},\sigma_{\nblock{g}}).$

As discussed in Sec.~\ref{sec:implementation}, \tool{} can verify NCB commutativity conditions. The verified commutativity condition is also saved, with partitioned variables, so that the runtime can evaluate the condition correctly. For Ex.~\ref{ex:motivation}, the condition is saved as $\nblockArename{f}{i}, \nblockArename{g}{i} : (\texttt{i}_\nblock{f}\neq \texttt{i}_\nblock{g}) \vee (\texttt{i}_\nblock{f}=\texttt{i}_\nblock{g} \wedge \texttt{arr}[\texttt{i}_\nblock{f}]>0)$.

\subsection{Soundness}
\label{subsec:soundness}

We now show that commutativity of the {\sc Servois2} embedding of NCBs entails commutativity in the source NCB language. We first define an SMT model and show that a model satisfying an embedded commutativity condition is equivalent to the corresponding NCB state satisfying the original commutativity condition.

\begin{definition}[SMT model of a state] For a state $\sigma_{gl},\sigma_{\nblock{f}},\sigma_{\nblock{g}}$, the SMT \emph{model} 
denoted $\mathcal{M}(\sigma_{gl},\sigma_{\nblock{f}},\sigma_{\nblock{g}})$ is the SMT model that 
maps each global variable $v\in Vars_{gl}$ to value $\sigma_{gl}(v)$, 
maps each variable $\alpha{_\texttt{x}}\in Vars_{f}$ to value $\sigma_{\nblock{f}}({\texttt{x}})$, and 
maps each variable $\beta{_\texttt{x}} \in Vars_{g}$ to value $\sigma_{\nblock{g}}({\texttt{x}})$. 
\end{definition}

\begin{lemma}[$Tr_{phi}$ Soundness]
\label{lemma:trphi}
For all $\varphi,\sigma_{gl},\sigma_{\nblock{f}},\sigma_{\nblock{g}}$, 
$
\sem{\varphi}{_\mathbb{B}}(\sigma_{gl},\sigma_{\nblock{f}},\sigma_{\nblock{g}})
\Leftrightarrow
\mathcal{M}(\sigma_{gl},\sigma_{\nblock{f}},\sigma_{\nblock{g}})
\vDash
Tr_{phi}(\varphi)
$.
\ifARXIV
\begin{proof}
By induction on $\varphi$, writing $\sigma$ for $\sigma_{gl},\sigma_\nblock{f},\sigma_\nblock{g}$ and $\mathcal{M}$ for $\mathcal{M}(\sigma)$. We treat \texttt{false} as $\neg\texttt{true}$.
\begin{itemize}
    \item \textbf{Case $\varphi = true$.}
    Immediate, since $Tr_{phi}(true) = \textsf{true}$ and $\sem{true}_\mathbb{B}(\sigma) = true$.
    \item \textbf{Case $\varphi = \neg \varphi'$.}
    $\sem{\neg \varphi'}_\mathbb{B}(\sigma) = \neg \sem{\varphi'}_\mathbb{B}(\sigma)$, so by the induction hypothesis
    $\sem{\neg \varphi'}_\mathbb{B}(\sigma) \iff \mathcal{M} \models (\textsf{not} \; Tr_{phi}(\varphi')) = Tr_{phi}(\neg \varphi')$.
    \item \textbf{Cases $\varphi = \varphi_1 \vee \varphi_2$ and $\varphi = \varphi_1 \wedge \varphi_2$.}
    $\sem{\varphi_1 \vee \varphi_2}_\mathbb{B}(\sigma) = \sem{\varphi_1}_\mathbb{B}(\sigma) \vee \sem{\varphi_2}_\mathbb{B}(\sigma)$; the result follows from the induction hypotheses for $\varphi_1$ and $\varphi_2$ and the semantics of the SMT $\textsf{or}$. Likewise for $\wedge$ and $\textsf{and}$.
    \item \textbf{Case $\varphi = ncb\text{-}e_1 \circlelt ncb\text{-}e_2$.}
    $\sem{\varphi}_\mathbb{B}(\sigma) = \sem{ncb\text{-}e_1}_\mathbb{N}(\sigma) \circlelt \sem{ncb\text{-}e_2}_\mathbb{N}(\sigma)$.
    We show $\sem{ncb\text{-}e}_\mathbb{N}(\sigma) = \mathcal{M}(Tr_{ncbe}(ncb\text{-}e))$ by induction on $ncb\text{-}e$. For a constant this is immediate. For a parameter \texttt{x} of $\nblock{f}$, $\sem{\texttt{x}}_\mathbb{N}(\sigma) = \sigma_\nblock{f}(\texttt{x}) = \mathcal{M}(\alpha_\texttt{x})$ by the definition of $\mathcal{M}$, and likewise for $\nblock{g}$ and for global variables. For a compound expression it follows from the induction hypotheses, since $Tr_E$ translates each operator, conditional, array index and field access to an SMT term with the same meaning, as in~\cite{veracity}. Therefore
    $\sem{\varphi}_\mathbb{B}(\sigma) \iff \mathcal{M} \models (\circlelt_{smt} \; Tr_{ncbe}(ncb\text{-}e_1) \; Tr_{ncbe}(ncb\text{-}e_2)) = Tr_{phi}(\varphi)$.
\end{itemize}
\end{proof}
\fi
\end{lemma}

\begin{theorem}\label{thm:soundness}
For any two {well-formed (Def.~\ref{def:well-formed})} NCBs $\nblockA{f}{a}:\{s_f\}$ and $\nblockA{g}{b}:\{s_g\}$, let $o=Tr(E,s_f, s_g)$. If $Tr_{phi}(\varphi)$ is a sound commutativity condition for $o.m_f$ and $o.m_g$ on object $o$, then  $\varphi$ is a sound commutativity condition for $\nblockA{f}{a}$ and $\nblockA{g}{b}$.
 
\begin{proof}
\ifARXIV
\input{soundness}
\else
The full proof is given \inApx{apx:soundness}. In summary, we correlate the translated SMT transition relations $\sem{Tr(s_f)}_R^\textsf{smt} \circ \sem{Tr(s_g)}_R^\textsf{smt}$ and $\sem{Tr(s_g)}_R^\textsf{smt} \circ \sem{Tr(s_f)}_R^\textsf{smt}$ with the NCB transitions $\angles{s_f,\sigma_{\nblock{f}}\oplus\sigma_{gl}^0}
\leadsto^*
\angles{\sskip,\sigma_{\nblock{f}}'\oplus\sigma_{gl}^1}$, etc. In each of the four cases, we proceed by structural induction over $Tr(s)$, showing that if the SMT global variables agreed with the NCB global variables, then they will still agree after the transitions. This requires accounting for the local-state inputs and the translated local variables.
\fi
\end{proof}

\end{theorem}
\ARXIV{{Independently of {\sc Servois2}, the commutativity specifications of all examples in this paper are proved sound in Rocq (Apx.~\ref{apx:rocq-examples}).}}

%% file: soundness.tex
Unfolding Def.~\ref{Valid NCB Commutativity Condition}, fix $\sigma_{gl}^0, \sigma_\nblock{f}, \sigma_\nblock{g}$ with $\sem{\varphi}_\mathbb{B}(\sigma_{gl}^0,\sigma_\nblock{f},\sigma_\nblock{g}) = true$, and the four NCB executions below. We must show $\sigma_{gl}^2 = \sigma_{gl}^{2'}$.
\[\begin{array}{l|l}
\text{\emph{exec}}_1 =
\angles{s_f,\sigma_{\nblock{f}}\oplus\sigma_{gl}^0}
\leadsto^*
\angles{\sskip,\sigma_{\nblock{f}}'\oplus\sigma_{gl}^1},
& \sem{Tr(s_f)}^\textsf{smt}_R,\\
\text{\emph{exec}}_2 =
\angles{s_g,\sigma_{\nblock{g}}\oplus\sigma_{gl}^1}
\leadsto^*
\angles{\sskip,\sigma_{\nblock{g}}'\oplus\sigma_{gl}^2},
& \sem{Tr(s_g)}^\textsf{smt}_R,\\
\text{\emph{exec}}_3 =
\angles{s_g,\sigma_{\nblock{g}}\oplus\sigma_{gl}^0}
\leadsto^*
\angles{\sskip,\sigma_{\nblock{g}}''\oplus\sigma_{gl}^{1'}},
& \sem{Tr(s_g)}^\textsf{smt}_R,\\
\text{\emph{exec}}_4 =
\angles{s_f,\sigma_{\nblock{f}}\oplus\sigma_{gl}^{1'}}
\leadsto^*
\angles{\sskip,\sigma_{\nblock{f}}''\oplus\sigma_{gl}^{2'}}
& \sem{Tr(s_f)}^\textsf{smt}_R.\\
\end{array}
\]
Written to the right of each execution is the SMT transition relation that it corresponds to. Let $\mathcal{M}_0 = \mathcal{M}(\sigma_{gl}^0,\sigma_\nblock{f},\sigma_\nblock{g})$.

\emph{The condition.} By Lemma~\ref{lemma:trphi}, $\mathcal{M}_0 \vDash Tr_{phi}(\varphi)$.

\emph{Each execution is a model of its relation.} We show: if $\angles{s_f,\sigma_{\nblock{f}}\oplus\sigma_{gl}} \leadsto^* \angles{\sskip,\sigma_{\nblock{f}}'\oplus\sigma_{gl}'}$, then for every $\sigma_\nblock{g}$, the pair of SMT states $(\mathcal{M}(\sigma_{gl},\sigma_\nblock{f},\sigma_\nblock{g}), \mathcal{M}(\sigma_{gl}',\sigma_\nblock{f}',\sigma_\nblock{g}))$ is in $\sem{Tr(s_f)}^\textsf{smt}_R$; symmetrically for $s_g$. This is by structural induction on $s_f$, following the proof of Thm.~6.4 in~\cite{veracity}, which shows that the translation of each statement relates the SMT states that correspond to the states of its executions. The difference lies in the NCB variables: the translation maps the variables of $\nblock{f}$ to fresh $\alpha$ variables and those of $\nblock{g}$ to fresh $\beta$ variables, so $Tr(s_f)$ neither reads nor writes $Vars_g$, just as the execution of $s_f$ neither reads nor writes $\sigma_\nblock{g}$, and the $\beta$ variables keep their values, as $\sigma_\nblock{g}$ does. The global variables are translated as in~\cite{veracity}, and $o.\textsf{state}$ has one for each global variable visible to $s_f$ and $s_g$.

\emph{The two orders.} By the previous step applied to $\text{\emph{exec}}_1$ and then $\text{\emph{exec}}_2$, the SMT states
$\mathcal{M}_0$, $\mathcal{M}(\sigma_{gl}^1,\sigma_\nblock{f}',\sigma_\nblock{g})$, $\mathcal{M}(\sigma_{gl}^2,\sigma_\nblock{f}',\sigma_\nblock{g}')$
form a run of $\sem{Tr(s_f)}_R^\textsf{smt}$ followed by $\sem{Tr(s_g)}_R^\textsf{smt}$. Likewise, by $\text{\emph{exec}}_3$ and $\text{\emph{exec}}_4$, the states
$\mathcal{M}_0$, $\mathcal{M}(\sigma_{gl}^{1'},\sigma_\nblock{f},\sigma_\nblock{g}'')$, $\mathcal{M}(\sigma_{gl}^{2'},\sigma_\nblock{f}'',\sigma_\nblock{g}'')$
form a run of $\sem{Tr(s_g)}_R^\textsf{smt}$ followed by $\sem{Tr(s_f)}_R^\textsf{smt}$. Here $\text{\emph{exec}}_2$ starts from $\sigma_\nblock{g}$ because $\text{\emph{exec}}_1$ does not change the $\beta$ variables, and $\text{\emph{exec}}_4$ from $\sigma_\nblock{f}$ because $\text{\emph{exec}}_3$ does not change the $\alpha$ variables.

\emph{Commutativity in SMT.} Since $Tr_{phi}(\varphi)$ is a sound commutativity condition for $o.m_f$ and $o.m_g$, and $\mathcal{M}_0 \vDash Tr_{phi}(\varphi)$, the final states of the two runs are related by $o.\textsf{eq}$. (The relations are deterministic when $s_f$ and $s_g$ are loop-free; with loops translated through invariants, they may relate a state to several post-states, and we rely on {\sc Servois2}'s check covering every post-state of both compositions.) Hence $\sigma_{gl}^2$ and $\sigma_{gl}^{2'}$ agree on every global variable visible to $s_f$ and $s_g$.

\emph{The remaining variables.} Outside $\GVars{}$, by Lemma~\ref{lem:discard}(ii) applied to each of the four executions (using that the NCBs are well-formed), $\sigma_{gl}^2$ and $\sigma_{gl}^{2'}$ both equal $\sigma_{gl}^0$. The variables of $\GVars{}$ are declared outside both NCB bodies and are visible to both, so they are among those compared by $o.\textsf{eq}$. Hence $\sigma_{gl}^2 = \sigma_{gl}^{2'}$, and $\varphi$ is a sound commutativity condition for $\nblock{f}$ and $\nblock{g}$.

%% file: mainalg.tex
\begin{algorithm}[H]
	\caption{Analysis Phases}
	\label{alg:analysis}
	\begin{algorithmic}
	\Procedure{Analyze}{$Prog$}
		\State $c\text{-}pdg \gets \text{\pdgcAnalysis}(Prog)$ %
		\State $parGraph \gets \text{\mkTaskDepGraph}(c\text{-}pdg)$ %
		\State $\jgTaskGraph \gets \text{\mkJobGenProg}(Prog,parGraph) $%
		\State $\jgTaskGraph \gets \text{\sc LockSynth}(\jgTaskGraph)$
		\State \Return $\jgTaskGraph$ \text{ for execution (Sec.~\ref{sec:execution_model})}%
	\EndProcedure
	\end{algorithmic}
\end{algorithm}

%% file: ex21-runtime.tex
Fig.~\ref{fig:ex21-runtime} shows this concretely for Ex.~\ref{ex:motivation}.

\begin{figure}[t]
\centering
\begin{tikzpicture}[
    job/.style    = {draw, rectangle, rounded corners,
                     minimum width=2.0cm, minimum height=0.55cm,
                     font=\small, thick},
    done/.style   = {job, fill=gray!15, draw=gray!60},
    waiting/.style= {job, fill=blue!10},
    running/.style= {job, fill=green!12, draw=green!60!black},
    tdep/.style   = {-{Latex[length=2mm]}, thick, dashed, gray!60},
    cdep/.style   = {{Latex[length=2mm]}-{Latex[length=2mm]}, thick, blue!60!black},
    bigarr/.style = {-{Latex[length=3mm]}, very thick},
    cfgbox/.style = {draw=gray!55, rounded corners=7pt,
                     inner sep=9pt, semithick, fill=none},
]

\node[done]    (j1a) at (0.0,  0.0)  {$j_1$~(T1)~\textsc{done}};
\node[waiting] (jfa) at (-1.6,-1.8)  {$j_f$~(T2)~\textsc{wait}};
\node[waiting] (jga) at ( 1.6,-1.8)  {$j_g$~(T3)~\textsc{wait}};
\node[below=2pt of jfa, font=\scriptsize, gray] (wsa) {\{$j_g$: comm\}};
\node[below=2pt of jga, font=\scriptsize, gray] (wsg) {\{$j_f$: comm\}};
\node[cfgbox, fit=(j1a)(jfa)(jga)(wsa)(wsg)] (boxA) {};

\node[done]    (j1b) at (9.0,  0.0)  {$j_1$~(T1)~\textsc{done}};
\node[running] (jfb) at (7.75, -1.8)  {$j_f$~(T2)~\textsc{run}};
\node[running] (jgb) at (10.25, -1.8) {$j_g$~(T3)~\textsc{run}};
\node[blue!60!black, font=\large] (par) at (9.0,-1.8) {$\|$};
\node[below=2pt of jfb, font=\scriptsize, text opacity=0] (ghostB) {\{$j_f$: comm\}};
\node[cfgbox, fit=(j1b)(jfb)(jgb)(par)(ghostB)] (boxB) {};

\draw[tdep] (jfa.north) to[out=70,in=225]
    node[left,pos=0.5,font=\scriptsize,gray] {term~$\checkmark$}
    (j1a.south west);
\draw[tdep] (jga.north) to[out=110,in=315]
    node[right,pos=0.5,font=\scriptsize,gray] {term~$\checkmark$}
    (j1a.south east);

\draw[cdep] (jfa.east) -- (jga.west)
    node[midway,above,font=\scriptsize] {$\varphi_f^g$};

\coordinate (mid1) at (5.5, -0.5);
\coordinate (mid2) at (4.5, -2.4);

\draw[bigarr] (boxA.east |- {(0,-0.5)})
    -- node[above,font=\small]           {{\sc DepElim}~(comm)}
       node[below,font=\scriptsize,gray] {$\varphi_f^g{=}\top$~(\texttt{arr}$[0]{>}0$)}
    (mid1);

\draw[very thick] (mid1) .. controls +(0.8, 0) and +(-0.8, 0) .. (mid2);

\draw[bigarr] (mid2)
    -- node[above,font=\small] {{\sc FinishWait}}
    (boxB.west |- {(0,-2.4)});

\end{tikzpicture}
\caption{\label{fig:ex21-runtime}%
  Example of transitions between configurations for Ex.~\ref{ex:motivation} (one loop iteration,
  $\mathtt{arr}[0]{=}5$).
  In the \emph{left configuration}, $j_1$ (Task~1, loop controller) has terminated; $j_f$
  (Task~2, $\nblock{f}$-body) and $j_g$ (Task~3, $\nblock{g}$-body) have
  cleared their data-dep on $j_1$ ($\checkmark$) and each holds only a
  comm dep on the other, guarded by $\varphi_f^g$.
  Since $\mathtt{arr}[0]{=}5{>}0$, the condition $\varphi_f^g{=}\top$ holds;
  {\sc DepElim}~(comm) clears both comm deps and {\sc FinishWait} moves
  both to \textsc{running}. In the \emph{right configuration} $j_f$ and $j_g$ execute  concurrently~($\|$).}
\end{figure}

%% file: correctness.tex
The proof involves three main parts:

\emph{A. Faithfulness.}
As mentioned in the previous section, the task graph produced by \mkJobGenProg{} is \emph{faithful}: inlining all $\newJob$ calls back into a single statement reproduces $s${ exactly, since task bodies keep their NCB labels} (Lemma~\ref{lem:faithful-outline}).
{This is proved by induction on $s$.}

\emph{B. Linearization and job-swap.}
{The jobs of a complete execution in which no job and a job it overtook interleave, run one at a time in the order of their first actions, end in its final shared state; by (S), this covers every execution. Two jobs that run NCBs whose commutativity condition holds end in the same shared state in either order\ARXIV{ (Lemma~\ref{lem:pairswap-outline})}.}

\emph{C. Program order.}
{The jobs of the execution are sorted into program order, using the commutativity conditions and their stability for the jobs that overtook others, and the resulting run is related, through the sequential run of $\jgTaskGraph{}$, to $s$ (Sec.~\ref{sec:correctness-proof}).}

\subsection{Definitions}
\label{sec:correctness-defs}

\begin{definition}[Spawning graph]
\label{def:spawning}
Given a job-generating program $\jgTaskGraph = (\mathit{initTask}, T)$, the
\emph{spawning graph} $G_{\mathsf{spawn}}$ is the directed graph whose vertices are task
identifiers $\{tid \mid (tid,\_,\_) \in T\}$ and whose edges are
\[
  tid_1 \to tid_2
  \;\iff\;
  \text{the body of task } tid_1 \text{ contains a subterm } \newJob(tid_2, \text{---}).
\]
By construction of \mkJobGenProg{} (Sec.~\ref{sec:analysis}), the spawning graph is
a directed acyclic graph (DAG): each $\newJob$ call replaces a strict sub-AST of the
parent's statement, so the nesting of tasks mirrors the nesting of the original AST.
\end{definition}

\begin{definition}[Inlining]
\label{def:inline}
Given a job-generating program $\jgTaskGraph = (\mathit{initTask}, T)$, define
$\mathsf{inline}(s', T)$ for a statement $s'$ by structural recursion:
\[
\begin{array}{lcl}
\mathsf{inline}(s_1 ; s_2,\; T)
  &=& \mathsf{inline}(s_1, T) \;;\; \mathsf{inline}(s_2, T) \\[2pt]
\mathsf{inline}(\texttt{if}(e)\block{s_1}\texttt{else}\block{s_2},\; T)
  &=& \texttt{if}(e)\block{\mathsf{inline}(s_1,T)}\texttt{else}\block{\mathsf{inline}(s_2,T)} \\[2pt]
\mathsf{inline}(\texttt{while}(e)\block{s_1},\; T)
  &=& \texttt{while}(e)\block{\mathsf{inline}(s_1,T)} \\[2pt]
\mathsf{inline}(\nblockA{f}{$\bar{v}$}\block{s_1},\; T)
  &=& \nblockA{f}{$\bar{v}$}\block{\mathsf{inline}(s_1,T)} \\[2pt]
\mathsf{inline}(\newJob(tid, \bar{x}),\; T)
  &=& \mathsf{inline}(s_{tid},\; T)
  \quad\text{where }(tid, s_{tid}, \_) \in T \\[2pt]
\mathsf{inline}(s',\; T)
  &=& s' \quad \text{(\sskip, assignments and stores)}
\end{array}
\]
Then $\mathsf{inline}(\jgTaskGraph) \;=\; \mathsf{inline}(\mathit{initTask}.s,\; T)$.
\ARXIV{\rocqref{JobGen.inlines}}

\medskip\noindent
\emph{Variable names.}
When \mkJobGenProg{} carves a child task out of the parent's AST, the variables in the
child's body that originate from outside the child are passed as arguments to
$\newJob(tid, \bar{x})$ and appear under the \emph{same names} in the child task's
statement.
Therefore the inlining above requires no variable substitution: the child body $s_{tid}$
references exactly the variables $\bar{x}$ that the parent passes.
\end{definition}

\begin{definition}[Faithfulness]
\label{def:faithful}
A job-generating program $\jgTaskGraph$ is \emph{faithful} to the sequential program $s$ if
$\mathsf{inline}(\jgTaskGraph) = s$.
{Task bodies keep their NCB labels, so no reduction is needed to compare the two.}
\end{definition}

\noindent
\textbf{Example (Ex.~\ref{ex:motivation}).}
The job-generating program in Fig.~\ref{fig:jobgenprog} has:
\begin{itemize*}
  \item $\mathit{initTask}.s$ = \lstinline|decls; newJob(T1,[i;size;arr]); newJob(T4,[arr]);|
  \item $T_1.s$ = \lstinline|while(i<size){ newJob(T2,[i]); newJob(T3,[i]); i=i+1; }|
  \item $T_2.s$ = $\nblockA{f}{i,arr}\{\ldots\}$
  \item $T_3.s$ = $\nblockA{g}{i,arr}\{\ldots\}$
  \item $T_4.s$ = \lstinline|return arr[0];|
\end{itemize*}
Applying $\mathsf{inline}$: replace $\newJob(T1,\ldots)$ with $\mathsf{inline}(T_1.s,T)$,
which in turn replaces $\newJob(T2,\ldots)$ and $\newJob(T3,\ldots)$ with the NCB bodies,
yielding:
\[
\texttt{decls; while(i<size)\{} \;\nblockA{f}{i,arr}\!\{\ldots\}\texttt{;}\;\nblockA{g}{i,arr}\!\{\ldots\}\texttt{; i=i+1;\}; return arr[0];}
\]
which is the body of $\mathit{main}$ in Ex.~\ref{ex:motivation}.

\begin{definition}[Execution]
\label{def:exec-history}
An \emph{execution} (or execution history) $H$ of $\jgTaskGraph$ is a finite sequence of transitions
$C_0 \xrightarrow{e_1} C_1 \cdots \xrightarrow{e_n} C_n$ of the rules in Fig.~\ref{fig:rules}, each labeled by the job it concerns:
\[
\begin{array}{ll}
\textsf{local}(j)          & \text{a local step of job } j \text{ ({\sc Local})}, \\
\textsf{term}(j)           & j \text{ finishes ({\sc Terminate})}, \\
\textsf{waitnew}(j)        & j \text{ replaces a task it waits for by that task's current jobs ({\sc WaitNewTask})}, \\
\textsf{depelim}(j, j')    & j \text{ stops waiting for job } j' \text{ ({\sc DepElim})}, \\
\textsf{finish}(j)         & j \text{ moves from waiting to running ({\sc FinishWait})}, \\
\textsf{fork}(j, k)        & j \text{ spawns the new job } k \text{ ({\sc ForkTask})}.
\end{array}
\]
Only $\textsf{local}$ steps change the shared state. $H$ is \emph{complete} if $C_n = (\emptyset, \emptyset, \sigma_{gl}^f)$ for some final shared state $\sigma_{gl}^f$.
The \emph{fork tree} of $H$ has an edge from $j$ to $k$ for each $\textsf{fork}(j, k)$ in $H$.
\ARXIV{\rocqref{Jobs.event, Jobs.trace, Jobs.terminal}}
\end{definition}

\begin{definition}[Program order]
\label{def:program-order}
\emph{Program order} on the jobs of an execution is the pre-order of its fork tree: a job precedes the jobs it spawns, and a job's earlier children, with their descendants, precede its later children.
This is a definition rather than an assumption, since Thm.~\ref{thm:correctness} relates the execution to $s$ directly.
\ARXIV{\rocqref{Spawns.preorder}}
\end{definition}

\subsection{Assumptions}
\label{sec:correctness-assumptions}

{Thm.~\ref{thm:correctness} assumes the following (proof in Sec.~\ref{sec:correctness-proof}).
\begin{itemize}
\item[(V)] Every specification entry is sound (Def.~\ref{Valid NCB Commutativity Condition}) and the entries are stable (Def.~\ref{def:stable}). Soundness is checked by {\sc Servois2}. Stability is checked by \texttt{verify --with-stability}, which asks the solver, for every three NCBs with entries for all three pairs, whether the three conditions can hold in a state from which running the first NCB falsifies the condition between the other two.
\item[(D)] The dependence analysis is sound: {if it relates a job $a$ to a later job $y$, then $a$ is an ancestor of $y$, or the task of $a$ is a predecessor of the task of $y$ (in $pred_{\jgTaskGraph.T}$, waited for at the spawn and, if in $later$, with {\sc WaitNewTask})}, and jobs that it does not relate do not conflict: neither writes a shared variable that the other reads or writes. A job reads the conditions that guard its $\newJob$s, the variables that it passes, and the shared variables of the commutativity conditions between its task and other tasks. Such jobs therefore commute, do not change each other's conditions, and in either order each spawns the same jobs, with the same arguments, in the same order. The mechanization derives these consequences, and includes a decidable check on the task bodies and the specification, proved sound, that implies the absence of such conflicts; (D) then amounts to the analysis relating every two jobs whose tasks fail the check.
{\item[(O)] Spawning is acyclic, and no task spawns the initial task. For each task $t$ and each predecessor $u$ of $t$ not in $later(t)$, one of the following holds. Here a task has a \emph{serial chain} if it has a single spawner, which is the initial task or a task that has a serial chain and is \emph{serial}: it is its own predecessor, without a commutativity condition with itself, so each of its jobs waits for the earlier ones to terminate. (a) $t$ and $u$ have the same single spawner, and $t$ has a serial chain. (b) $u$ spawns $t$, directly or through tasks that each have a single spawner, and $u$ has a serial chain. (c) $t$ spawns $u$ in the same way, the initial task alone spawns $t$, and does so at most once (not inside a loop). The mechanization proves from (D) and (O) that each job waits for, or has overtaken, every earlier job that the analysis relates to it: under (O), a job of a predecessor waited for only at the spawn is created before the jobs it precedes in program order. \name{} checks (O) and falls back to a sequential task graph when it fails; it holds for all our benchmarks.}
\item[(S)] {A job and a job it overtook execute atomically with respect to each other: the execution is equivalent to one, from the same initial configuration, that forks the same jobs, ends in the same shared state, and in which neither of the two takes its first action while the other is between two of its actions. This is what the synchronization layer must provide, and it is orthogonal to the scheduling studied here, which decides which jobs may run concurrently or out of order. Standard protocols such as two-phase locking or optimistic concurrency control provide it: they let the two jobs' steps interleave, but only in ways equivalent to running one entirely before the other. An execution in which the two jobs' steps do not interleave satisfies (S) trivially. All other jobs may run alongside each other: from the wait sets the mechanization derives that a job that starts while another runs is independent of it or its descendant, since otherwise one of them overtook the other, and (A1) keeps a descendant apart from what its ancestor runs after forking. This covers independent jobs, such as the iterations of a DOALL loop, and a spawning loop with the jobs it spawns. Lock synthesis as implemented locks each group of accesses to one shared variable, not each job, so it provides (S) only for blocks whose accesses that conflict with the other block's form one such group, or that touch different array elements, as for most pairs in our benchmarks. \name{} can also enforce (S) directly (\texttt{--atomic-overtaking}): a job holds the locks of the jobs it overtook while it runs. This removes most of the speedup, for example from $6.2\times$ to $1.0\times$ for the running example and from $4.0\times$ to $1.0\times$ for \texttt{vote}, since there the concurrency is precisely that of commuting blocks.}
\ARXIV{\rocqref{Correctness.sync_equiv}}
\item[(T)] Each job terminates when run alone, and so does the sequential run of $\jgTaskGraph{}$.
\end{itemize}
The following two assumptions are added by the step from program order to $s$:
\begin{itemize}
\item[(A1)] After its first $\newJob$, a spawning task reads no variable that a job it can spawn, directly or through descendants, writes, and writes no variable that such a job reads or writes, other than that job's by-value arguments and locals. The code after the first $\newJob$ includes the whole body of any loop that contains a $\newJob$, and an array counts as one variable. The proof runs each job as one block, while $s$ runs the spawning job's code after the $\newJob$ only after the new job's subtree; (A1) ensures that the two orders agree.\ARXIV{ The mechanized condition is \texttt{Footprint.spawn\_ok\_all}.}
\item[(A2)] Code outside NCB bodies neither reads nor writes NCB-local variables, and an NCB body initializes its locals before reading them.
\end{itemize}
(A1) is syntactic and decidable: the mechanization includes a checker, proved sound, which accepts the running example. {\name{} checks (A1) on the generated tasks, mirroring the mechanized checker, and falls back to a sequential task graph when it fails; every benchmark of Sec.~\ref{sec:eval} passes.} Because the reads in (D) include those made when spawning, the jobs of the execution, taken in program order, are proved to be those of the sequential run of $\jgTaskGraph{}$, with the values that each $\newJob$ passes; this need not be assumed. The implementation's dependence analysis counts the reads made when spawning: the conditions that guard a $\newJob$ are statements of the spawning task, with their reads, and a variable that a task passes by value is written by that task or one of its ancestors, so any other task that writes the variable depends on that writer. {It also counts, as reads of an NCB, the shared variables that the commutativity conditions involving it read, so a task that writes one depends on the NCB.} This argument is about the implementation and is not mechanized. The model passes by value only the variables that the new job and its descendants do not write, so no job writes a copy. It also does not pass a variable that a task other than the spawning task and its ancestors writes. Such a task, for example an earlier sibling, may not have run when the copy is taken, so the new job instead waits for it and reads the shared variable. The implementation does the same: it passes only the variables that flow to the new job, or to the jobs below it, from the spawning task and its ancestors. The implementation differs for \texttt{psdswp} and \texttt{2d-array}: it passes the loop variable \texttt{q} by value to a job that modifies it, which is correct there because \texttt{q} is not read afterwards, but these two benchmarks are outside the mechanized theorem.}

\subsection{Rocq Mechanization}
\label{sec:correctness-proof}

The proof of Thm.~\ref{thm:correctness} is mechanized in Rocq; we outline it here. {Part~A (faithfulness) is mechanized in full (\texttt{JobGen.faithful}, Apx.~\ref{apx:rocq-examples}), as is the soundness of the commutativity specifications of all examples (Apx.~\ref{apx:rocq-examples}). Part~B is mechanized as \texttt{linearization}, for serializable executions (each job's local steps contiguous in the trace): running the execution's jobs one at a time, in some order, gives exactly its final shared state. \texttt{Interleave.v} extends it to executions in which no job and a job it overtook interleave, and hence, by (S), to all executions. Part~C is mechanized: a complete serializable execution ends where running its jobs one at a time in program order does (Lemma~\ref{lem:progorder-outline}), which is where the sequential run of $\jgTaskGraph$ ends (Lemma~\ref{lem:seqrun-outline}), which agrees with the final state of $s$ outside NCB-local variables (Lemma~\ref{lem:inline-outline}). This holds under the assumptions above; (A1) and (A2) are added by the last two steps, and (A1) is a mechanized syntactic check.} 

\paragraph{Part A: Faithfulness.}

The job-generating program $\jgTaskGraph$ produced by \mkJobGenProg{} is
\emph{faithful} to the original source $s$:
\begin{lemma}[Faithfulness]
\label{lem:faithful-outline}
If \mkJobGenProg{} carves no AST node into two tasks, then $\mathsf{inline}(\jgTaskGraph) = s$ (Def.~\ref{def:inline}): $\jgTaskGraph$ is faithful to $s$ (Def.~\ref{def:faithful}).
\ARXIV{\rocqref{JobGen.faithful}}
\end{lemma}
\begin{proof}
{By induction on $s$. \mkJobGenProg{} makes $s$ the body of the initial task, with each subtree carved out as a task $V$ replaced by $\newJob(tid_V,\bar{x})$; the body of $V$ is that subtree, with the subtrees carved out of it replaced in the same way. $\mathsf{inline}$ replaces each $\newJob(tid_V,\bar{x})$ by the inlined body of $V$, which is unique because no AST node is carved into two tasks, and which by the induction hypothesis is the subtree. Task bodies keep their NCB labels, so the equality is syntactic.}
\end{proof}

\paragraph{Part B: From concurrent execution to a linearization.}

{An execution is \emph{serializable} if the local steps of each job are contiguous in it.}

{\begin{lemma}[Linearization]\label{lem:linearization-outline}
Running the jobs of a complete serializable execution one at a time, each from its initial state to termination, in the order of their first local steps, from the initial shared state, ends in the final shared state of the execution.
\ARXIV{\rocqref{Linearization.linearization}}
\end{lemma}}
{\begin{proof}
By induction on the execution. Only local steps change the shared state (Def.~\ref{def:exec-history}), and a job's local steps form one block, which runs the job's body from its initial state; since the execution is complete, the job terminates at the end of the block. A local step reads and writes only the job's own local state and the shared state, so the steps of the other jobs before or after the block do not affect it except through the shared state.
\end{proof}}
\noindent
{\texttt{Interleave.v} extends this to executions in which no job and a job it overtook interleave: running the jobs one at a time in the order of their first actions still gives the final state, since a job's later action can be moved before the actions of the independent jobs that started after it (\texttt{interleave\_run}). By (S), this covers all executions.}

{\begin{lemma}[Pairwise job-swap]
\label{lem:pairswap-outline}
Let jobs $j_1$ and $j_2$ run NCBs $\nblock{f}$ and $\nblock{g}$ from their start, let $\varphi$ be a sound commutativity condition for $\nblock{f}$ and $\nblock{g}$ (Def.~\ref{Valid NCB Commutativity Condition}), and let $\varphi$ hold in shared state $\sigma$ on the local states of $j_1$ and $j_2$. Then running $j_1$ and then $j_2$ from $\sigma$, and $j_2$ and then $j_1$, end in shared states that agree on $\GVars{}$, and hence (Lemma~\ref{lem:discard}) are equal.
\ARXIV{\rocqref{Linearization.two_job_comm}}
\end{lemma}}
{\begin{proof}
Running a job that runs an NCB from its start runs the NCB's body on the job's local state merged with the shared state. The two orders thus give the four runs of Def.~\ref{Valid NCB Commutativity Condition}, from $\sigma$ and the local states of $j_1$ and $j_2$, and since $\varphi$ holds there, the final shared states agree on $\GVars{}$. Outside $\GVars{}$ both equal $\sigma$, by Lemma~\ref{lem:discard}(ii).
\end{proof}}

{\begin{lemma}[Overtaking]\label{lem:overtake-outline}
Let a job $w$ run, from shared state $\sigma$ (where its first local step re-checks its conditions), before jobs $o_1, \ldots, o_k$ (in program order), where $w$'s condition for each $o_i$ holds in $\sigma$, the $o_i$ are pairwise unordered in $\sigma$, and the specification is stable (Def.~\ref{def:stable}). Then $w, o_1, \ldots, o_k$ and $o_1, \ldots, o_k, w$ end in the same shared state.
\ARXIV{\rocqref{Overtake.overtake, Overtake.ncb_overtake}}
\end{lemma}
\begin{proof}
By induction on $k$. Swap $w$ with $o_1$ in $\sigma$ (Lemma~\ref{lem:pairswap-outline}). In the state after $o_1$, stability applied to $o_1$, $o_i$, $w$ keeps $w$'s condition for $o_i$, and applied to $o_1$, $o_i$, $o_j$ keeps the $o_i$ pairwise unordered. Mechanized in Rocq as \texttt{overtake} and \texttt{ncb\_overtake}.
\end{proof}}

{\begin{lemma}[Program order]\label{lem:progorder-outline}
Assume the wait sets are complete (every job that precedes a job $y$ in program order, has not started, and is related to $y$ by the dependence analysis, is one $y$ waits for or has overtaken, or does not exist yet while $y$ still waits for a task){, which the mechanization derives from (D) and (O) (\texttt{waits\_complete\_po}, \texttt{created\_in\_order})}, and that independent jobs commute and do not change each other's conditions. Then a complete serializable execution ends in the shared state obtained by running its jobs one at a time in program order.
\ARXIV{\rocqref{Canonical.sort_correct, ProgramOrder.program_order}}
\end{lemma}
\begin{proof}
Take the jobs in the order of their first local steps (the linearization of Part~B), and insert each job behind the jobs that precede it in program order but run after it. At its first step, the re-check gives its conditions for the jobs it overtook and, unless they read only parameters, the pairwise unordered closure; conditions that read only parameters cannot change, and need neither the closure nor stability (\texttt{g\_overtake\_fixed}); completeness gives that the other such jobs are independent of it, and that a job outside the closure is independent of the closure jobs it precedes. The insertion generalizes Lemma~\ref{lem:overtake-outline}. Mechanized in Rocq as \texttt{sort\_correct} and \texttt{program\_order}; \texttt{Counterexample2.v} shows that without the closure the lemma fails.
\end{proof}}

\paragraph{Part C: Program order and final assembly.}

{Part~C consists of the following two lemmas, after Lemma~\ref{lem:progorder-outline}. The \emph{sequential run} of $\jgTaskGraph$ runs one job at a time: a job runs up to its next $\newJob$, the new job runs together with its descendants, and then the job continues. Each job works on its own copy of the variables passed to it, merged with the shared state.}

{\begin{lemma}[Sequential run]\label{lem:seqrun-outline}
Assume (A1). The sequential run of a job ends where running the job and then its descendants in pre-order, each job as one block, does. Under (D), (O) and (T), a complete serializable execution from a single job is such a run, and therefore ends where the sequential run of its initial job does.
\ARXIV{\rocqref{Sequential.dfs_blocks, Footprint.dfs_free_dfs, Spawns.dfs_pre_run, SpawnExec.trace_pre_run, SpawnExec.exec_dfs_sp}}
\end{lemma}
\begin{proof}
By induction on the sequential run. At a $\newJob$, the run is the job's code up to the $\newJob$, then the new job's subtree, then the rest of the job. (A1) moves the rest of the job before the subtree, and the two pieces of the job form its block.
For the execution, take the jobs in the order of their first actions, a local step or a $\newJob$, each run as one block that records the jobs it spawns. Sort them into program order as in Lemma~\ref{lem:progorder-outline}, on states that also hold the jobs spawned so far by each job. Independent jobs spawn the same jobs in either order by (D), and jobs swapped under a condition run NCBs, which spawn nothing. By the definition of program order, the sorted order, with the jobs that have nothing to do put back, is the pre-order of the fork tree, so the sorted run is a pre-order run of the initial job; such runs are unique.
\end{proof}}

{\begin{lemma}[Inlined program]\label{lem:inline-outline}
Assume (A2). The sequential run of $\jgTaskGraph$ from $\sigma_{gl}^0$ ends in a shared state that agrees, outside the NCB-local variables, with the final state of $\mathsf{inline}(\jgTaskGraph)$, and hence (Part~A) of $s$, from $\sigma_{gl}^0$.
\ARXIV{\rocqref{Inline.dfs_sim, Inline.build_args_ro, Correctness.main_theorem}}
\end{lemma}
\begin{proof}
A simulation: each step of a job's code outside NCBs is matched by one step of the inlined program, in states that agree outside the NCB-local variables. At a $\newJob$, the inlined program runs the new job's inlined body. The variables passed by value are not written by the new job or its descendants, so the copies stay equal to the shared values. (A2) makes an NCB body end in agreeing states.
\end{proof}}

\subsection{Main Theorem}

{\noindent\textbf{Theorem~\ref{thm:correctness} (restated).}
\emph{\correctnessstatement{the assumptions (V), (D), (O), (S), (T), (A1) and (A2) of Sec.~\ref{sec:correctness-assumptions}}}
\ARXIV{\rocqref{Correctness.main_theorem_sync}}}

{\begin{proof}
By (S), the execution is equivalent to one in which no job and a job it overtook interleave, which forks the same jobs and ends in the same shared state; it suffices to prove the theorem for that execution. It ends where running its jobs one at a time in program order does (Lemma~\ref{lem:progorder-outline}, extended by \texttt{Interleave.v} to executions in which no job and a job it overtook interleave). By Lemma~\ref{lem:seqrun-outline}, this is where the sequential run of $\jgTaskGraph$ ends, which by Lemma~\ref{lem:inline-outline} agrees, outside the NCB-local variables, with the final state of $\mathsf{inline}(\jgTaskGraph)$ and hence, by Lemma~\ref{lem:faithful-outline}, of $s$.
\end{proof}}

{\noindent\textbf{Remark.}
The theorem compares shared states, and only outside the NCB-local variables. A job discards its local state when it terminates, and an NCB's locals belong to the job that runs it, whereas the sequential run of $s$ keeps both in its final state. (A2) ensures that no code outside the NCB bodies observes the difference.}

%% file: rocq-examples.tex
\lstdefinelanguage{Rocq}{
  morekeywords={Definition,Fixpoint,Inductive,Record,Theorem,Lemma,Example,
    Corollary,Proof,Qed,Variables,Section,End,Notation,match,with,end,fun,
    forall,exists,let,in,if,then,else,Type,Prop},
  morecomment=[s]{(*}{*)},
  morestring=[b]",
  sensitive=true}
\lstdefinestyle{rocq}{language=Rocq,basicstyle=\ttfamily\scriptsize,
  keywordstyle=\bfseries,commentstyle=\itshape,breaklines=true,keepspaces=true,
  showstringspaces=false,mathescape=true,escapechar={},xleftmargin=0pt,aboveskip=2pt,belowskip=4pt}

\section{{Rocq mechanization of the examples}}
\label{apx:rocq-examples}
{This appendix shows the exact Rocq text, extracted from the sources, for every example of the paper. For each commutativity specification entry we prove \texttt{valid\_entry}: the pair is well-formed (W1, W2, scoping), the condition only mentions parameters and shared variables, and the condition is sound (Def.~\ref{Valid NCB Commutativity Condition}). Proofs are by the tactic \texttt{solve\_entry}, which computes the four runs of Def.~\ref{Valid NCB Commutativity Condition} symbolically and compares the final shared states location by location. The language has integer values, arrays (\texttt{a !! e} reads \code{a[e]}, \texttt{a <[ e ]> ::= e'} stores), and calls of pure functions such as \code{calc}, which are section variables, so each result holds for all such functions. The paper's \texttt{\&\&}, \texttt{||}, \texttt{\%} are written \texttt{/\textbackslash}, \texttt{\textbackslash/}, \texttt{mod}. In a condition, \texttt{Fp x} and \texttt{Gp x} are the positional aliases of the parameter \code{x} of the first and second NCB of the entry (so the paper's \code{j} or \code{i'} is \texttt{Gp "i"}), and \texttt{Gl x} is a shared variable. Apart from functional extensionality (used only for Fig.~\ref{fig:psdswp}), the results below use no axioms.}

\subsection{{Core definitions}}

\noindent{An NCB declaration $\nblock{f}\langle\bar{x}\rangle:\{s_\nblock{f}\}$, with the scoping facts read off the program text: parameters, the variables $In_\nblock{f}$ (empty in the paper body), and body locals. The frame is held by the fragment-local state.}
\begin{lstlisting}[style=rocq]
Record ncb : Type := {
  n_label  : ncb_label;
  n_params : list var;
  n_inputs : list var;
  n_locals : list var;
  n_body   : cmd
}.

Definition ncb_cmd (n : ncb) : cmd := CNCB (n_label n) (n_params n) (n_body n).

(** The frame: the variables held by the fragment-local state while the
    body runs.  Everything else lives in the shared state. *)
Definition frame (n : ncb) : list var := n_params n ++ n_inputs n ++ n_locals n.
\end{lstlisting}

\noindent{$\sigma_l \oplus \sigma_{gl}$ (Sec.~\ref{sec:prelim}): the local state wins on the frame. Locations are scalar variables and array cells; an array is local or shared as a whole.}
\begin{lstlisting}[style=rocq]
Definition merge (D : list var) (l g : state) : state :=
  fun v => if memb (base v) D then l v else g v.
\end{lstlisting}

\noindent{$\langle s, \sigma_l \oplus \sigma_{gl}\rangle \leadsto^* \langle \sskip, \sigma_l' \oplus \sigma_{gl}'\rangle$, discarding $\sigma_l'$.}
\begin{lstlisting}[style=rocq]
Definition run_ncb (n : ncb) ($\sigma$_l $\sigma$_gl $\sigma$_gl' : state) : Prop :=
  exists s',
    $\langle$ ncb_cmd n , merge (frame n) $\sigma$_l $\sigma$_gl $\rangle$ $\leadsto$* $\langle$ SKIP , s' $\rangle$ /\
    $\sigma$_gl' = merge (frame n) $\sigma$_gl s'.
\end{lstlisting}

\noindent{Well-formedness (Def.~\ref{def:well-formed}).}
\begin{lstlisting}[style=rocq]
(** (W1) parameters are read-only: neither body assigns a parameter of the pair. *)
Definition W1 (f g : ncb) : Prop :=
  forall x, In x (writes (n_body f) ++ writes (n_body g)) ->
            ~ In x (n_params f ++ n_params g).

(** (W2) no other non-shared writes: every variable the body assigns is
    declared in the body or belongs to Vars_gl. *)
Definition W2 (G : is_global) (n : ncb) : Prop :=
  forall x, In x (writes (n_body n)) -> G x = true \/ In x (n_locals n).

(** Scoping facts that hold by construction in Def. var-assoc: In_f lies
    outside Vars_gl, and a body-declared variable is neither a parameter
    nor an input. *)
Definition scoped (G : is_global) (n : ncb) : Prop :=
  (forall x, In x (n_inputs n) -> G x = false) /\
  (forall x, In x (n_locals n) -> ~ In x (n_params n ++ n_inputs n)).

Definition wf_pair (G : is_global) (f g : ncb) : Prop :=
  W1 f g /\ W2 G f /\ W2 G g /\ scoped G f /\ scoped G g.
\end{lstlisting}

\noindent{Variables of a condition, associated with $gl$, $\nblock{f}$, or $\nblock{g}$ (Def.~\ref{def:var-assoc}). In the examples, \texttt{Gl x}, \texttt{Fp x}, \texttt{Gp x} abbreviate \texttt{CEVar (VGl x)}, \texttt{CEVar (VF x)}, \texttt{CEVar (VG x)}.}
\begin{lstlisting}[style=rocq]
Inductive cvar : Type :=
  | VGl : var -> cvar
  | VF  : var -> cvar
  | VG  : var -> cvar.
\end{lstlisting}

\noindent{Def.~\ref{Valid NCB Commutativity Condition}.}
\begin{lstlisting}[style=rocq]
Definition sound_comm_cond (G : is_global) ($\varphi$ : cond) (f g : ncb) : Prop :=
  forall $\sigma$_gl0 $\sigma$_gl1 $\sigma$_gl1' $\sigma$_gl2 $\sigma$_gl2' $\sigma$_f $\sigma$_g,
    run_ncb f $\sigma$_f $\sigma$_gl0  $\sigma$_gl1  ->
    run_ncb g $\sigma$_g $\sigma$_gl1  $\sigma$_gl2  ->
    run_ncb g $\sigma$_g $\sigma$_gl0  $\sigma$_gl1' ->
    run_ncb f $\sigma$_f $\sigma$_gl1' $\sigma$_gl2' ->
    cond_eval $\sigma$_gl0 $\sigma$_f $\sigma$_g $\varphi$ = true ->
    globally_eq G $\sigma$_gl2 $\sigma$_gl2'.
\end{lstlisting}

\noindent{A valid specification entry: well-formed, well-associated, and sound.}
\begin{lstlisting}[style=rocq]
Definition valid_entry (G : is_global) ($\varphi$ : cond) (f g : ncb) : Prop :=
  wf_pair G f g /\ cond_ok G f g $\varphi$ /\ sound_comm_cond G $\varphi$ f g.

(** Vars_gl given as a list of names. *)
Definition globals (xs : list var) : is_global := fun x => memb x xs.
\end{lstlisting}

\subsection{{Examples}}

\smallskip\noindent{Introduction (Sec.~\ref{sec:intro}).}
\begin{lstlisting}[style=rocq]
(* Sec. 1:
     commutativity { f, g<i> : (y == 0 || x != i);
                     g<i>, g<i'> : (i != i'); }
     int main(int[] arr, int size, int x) {
       f:{ arr[x] = arr[x] / 2; }
       int y = x;
       for (int i = 0; i < size; i++) { g<i>:{ arr[i] = arr[i] + y; } } } *)
Definition G_intro := globals ["arr"; "size"; "x"; "y"; "i"].

Definition intro_f : ncb := {|
  n_label := "f"; n_params := []; n_inputs := []; n_locals := [];
  n_body := "arr" <[ "x" ]> ::= "arr" !! "x" / 2 |}.

Definition intro_g : ncb := {|
  n_label := "g"; n_params := ["i"]; n_inputs := []; n_locals := [];
  n_body := "arr" <[ "i" ]> ::= "arr" !! "i" + "y" |}.

Theorem intro_fg : valid_entry G_intro (Gl "y" == 0 \/ Gl "x" != Gp "i") intro_f intro_g.
Proof. solve_entry. Qed.

Theorem intro_gg : valid_entry G_intro (Fp "i" != Gp "i") intro_g intro_g.
Proof. solve_entry. Qed.
\end{lstlisting}

\smallskip\noindent{Sec.~\ref{sec:overview}, first example.}
\begin{lstlisting}[style=rocq]
(* Sec. 2, first example:
     commutativity { f<i>, f<j> : (i != j); }
     int main(int[] arr, int size) {
       for (int i = 0; i < size; i++) { f<i>:{ arr[i] = (arr[i] + 1) / 2; } } } *)
Definition G_ov1 := globals ["arr"; "size"; "i"].

Definition ov1_f : ncb := {|
  n_label := "f"; n_params := ["i"]; n_inputs := []; n_locals := [];
  n_body := "arr" <[ "i" ]> ::= ("arr" !! "i" + 1) / 2 |}.

Theorem ov1_ff : valid_entry G_ov1 (Fp "i" != Gp "i") ov1_f ov1_f.
Proof. solve_entry. Qed.
\end{lstlisting}

\smallskip\noindent{Sec.~\ref{sec:overview}, second example.}
\begin{lstlisting}[style=rocq]
(* Sec. 2, second example:
     commutativity { f, g<i> : (x <= y || x != i); }
     int main(int[] arr, int size, x, y) {
       f:{ if (x > y) arr[x] = y; }
       x = y;
       for (int i = 0; i < size; i++) { g<i>:{ arr[i] = (arr[i] + 1); } } } *)
Definition G_ov2 := globals ["arr"; "size"; "x"; "y"; "i"].

Definition ov2_f : ncb := {|
  n_label := "f"; n_params := []; n_inputs := []; n_locals := [];
  n_body := IF "x" > "y" THEN "arr" <[ "x" ]> ::= "y" ELSE SKIP END |}.

Definition ov2_g : ncb := {|
  n_label := "g"; n_params := ["i"]; n_inputs := []; n_locals := [];
  n_body := "arr" <[ "i" ]> ::= "arr" !! "i" + 1 |}.

Theorem ov2_fg : valid_entry G_ov2 (Gl "x" <= Gl "y" \/ Gl "x" != Gp "i") ov2_f ov2_g.
Proof. solve_entry. Qed.
\end{lstlisting}

\smallskip\noindent{Sec.~\ref{sec:overview}, multiple loops.}
\begin{lstlisting}[style=rocq]
(* Sec. 2, multiple loops:
     commutativity { f<i>, g<j> : (j % 2 != 0);
                     g<i>, g<j> : (i != j);
                     f<i>, f<j> : (i != j); }
     int main(int[] arr, int size) {
       for (int i = 0; i < size; i++) {
         f<i>:{ if (i % 2 == 0) arr[i] = 0; else arr[i] = arr[i] + 1; } }
       for (int i = 0; i < size; i++) { g<i>:{ arr[i] = arr[i] - 1; } } } *)
Definition G_ml := globals ["arr"; "size"; "i"].

Definition ml_f : ncb := {|
  n_label := "f"; n_params := ["i"]; n_inputs := []; n_locals := [];
  n_body := IF "i" mod 2 == 0 THEN "arr" <[ "i" ]> ::= 0
                              ELSE "arr" <[ "i" ]> ::= "arr" !! "i" + 1 END |}.

Definition ml_g : ncb := {|
  n_label := "g"; n_params := ["i"]; n_inputs := []; n_locals := [];
  n_body := "arr" <[ "i" ]> ::= "arr" !! "i" - 1 |}.

Theorem ml_fg : valid_entry G_ml (Gp "i" mod 2 != 0) ml_f ml_g.
Proof. solve_entry. Qed.

Theorem ml_gg : valid_entry G_ml (Fp "i" != Gp "i") ml_g ml_g.
Proof. solve_entry. Qed.

Theorem ml_ff : valid_entry G_ml (Fp "i" != Gp "i") ml_f ml_f.
Proof. solve_entry. Qed.
\end{lstlisting}

\smallskip\noindent{Ex.~\ref{ex:motivation}.}
\begin{lstlisting}[style=rocq]
(* Ex. motivation (Sec. 2):
     commutativity { f<i>, g<j> : i != j || (i == j && arr[i] > 0);
                     f<i>, f<j> : i != j;
                     g<i>, g<j> : i != j; }
     int main(int[] arr, int size) {
       int[] out = new int[size]; int i = 0;
       while (i < size) {
         f<i>:{ int x = calc(i); arr[i] = arr[i] + (x*x); }
         g<i>:{ int t = compute(i);
                if (arr[i] > 0 && t < 0) { arr[i] = arr[i] - 1; out[i] = calc2(t); }
                else { out[i] = calc3(t); } }
         i = i + 1; } } *)
Definition G_mot := globals ["arr"; "size"; "out"; "i"].

Definition mot_f : ncb := {|
  n_label := "f"; n_params := ["i"]; n_inputs := []; n_locals := ["x"];
  n_body := "x" ::= AFun calc "i" ;;
            "arr" <[ "i" ]> ::= "arr" !! "i" + "x" * "x" |}.

Definition mot_g : ncb := {|
  n_label := "g"; n_params := ["i"]; n_inputs := []; n_locals := ["t"];
  n_body := "t" ::= AFun compute "i" ;;
            IF "arr" !! "i" > 0 /\ "t" < 0
            THEN "arr" <[ "i" ]> ::= "arr" !! "i" - 1 ;;
                 "out" <[ "i" ]> ::= AFun calc2 "t"
            ELSE "out" <[ "i" ]> ::= AFun calc3 "t" END |}.

Theorem mot_fg :
  valid_entry G_mot (Fp "i" != Gp "i" \/ (Fp "i" == Gp "i" /\ "arr" !! (Fp "i") > 0))
              mot_f mot_g.
Proof. solve_entry. Qed.

Theorem mot_ff : valid_entry G_mot (Fp "i" != Gp "i") mot_f mot_f.
Proof. solve_entry. Qed.

Theorem mot_gg : valid_entry G_mot (Fp "i" != Gp "i") mot_g mot_g.
Proof. solve_entry. Qed.
\end{lstlisting}

\smallskip\noindent{Sec.~\ref{sec:ncb}, NCB syntax example, and why the earlier version violated (W1).}
\begin{lstlisting}[style=rocq]
(* Sec. 4, NCB syntax example ("a" is a global array):
     int main(int x, int y) {
       f<x>:{ a[x] = a[x] + 1; }
       if (x < y) g<x,y>:{ a[y] = x; } }
   with the entry  f<x>, g<x',y'> : (x != y'). *)
Definition G_s4 := globals ["a"; "x"; "y"].

Definition s4_f : ncb := {|
  n_label := "f"; n_params := ["x"]; n_inputs := []; n_locals := [];
  n_body := "a" <[ "x" ]> ::= "a" !! "x" + 1 |}.

Definition s4_g : ncb := {|
  n_label := "g"; n_params := ["x"; "y"]; n_inputs := []; n_locals := [];
  n_body := "a" <[ "y" ]> ::= "x" |}.

Theorem s4_fg : valid_entry G_s4 (Fp "x" != Gp "y") s4_f s4_g.
Proof. solve_entry. Qed.

(* The earlier version f<x>:{ x = x + 1; }, g<x,y>:{ x = y; } assigns its
   parameters, so it violates (W1). *)
Definition s4_old_f : ncb := {|
  n_label := "f"; n_params := ["x"]; n_inputs := []; n_locals := [];
  n_body := "x" ::= "x" + 1 |}.

Definition s4_old_g : ncb := {|
  n_label := "g"; n_params := ["x"; "y"]; n_inputs := []; n_locals := [];
  n_body := "x" ::= "y" |}.

Example s4_old_not_W1 : ~ W1 s4_old_f s4_old_g.
Proof. intros H. apply (H "x"); simpl; auto. Qed.
\end{lstlisting}

\smallskip\noindent{Fig.~\ref{fig:scopes}.}
\begin{lstlisting}[style=rocq]
(* Fig. scopes (In_f = In_g = empty; u is passed as a parameter):
     int[] a, b;
     for (int i = 0; i < n; i++) {
       int u = h(i);
       f<i,u>:{ int x = u*u; a[i] = x; }
       g<i>:{ int y = i*2; b[i] = y; } }
   with the entry  f<i,u>, g<j> : i != j. *)
Definition G_sc := globals ["a"; "b"; "n"; "i"; "u"].

Definition sc_f : ncb := {|
  n_label := "f"; n_params := ["i"; "u"]; n_inputs := []; n_locals := ["x"];
  n_body := "x" ::= "u" * "u" ;; "a" <[ "i" ]> ::= "x" |}.

Definition sc_g : ncb := {|
  n_label := "g"; n_params := ["i"]; n_inputs := []; n_locals := ["y"];
  n_body := "y" ::= "i" * 2 ;; "b" <[ "i" ]> ::= "y" |}.

Theorem sc_fg : valid_entry G_sc (Fp "i" != Gp "i") sc_f sc_g.
Proof. solve_entry. Qed.

(* The two bodies write different arrays, so they even commute unconditionally. *)
Theorem sc_fg_true : valid_entry G_sc CTrue sc_f sc_g.
Proof. solve_entry. Qed.
\end{lstlisting}

\smallskip\noindent{Fig.~\ref{fig:scopes-gen}, with $In_\nblock{f}$ and $In_\nblock{g}$.}
\begin{lstlisting}[style=rocq]
(* Fig. scopes-gen (Apx. in-vars), with In_f = {u} and In_g = {u, t}:
     int[] a, b;
     for (int i = 0; i < n; i++) {
       int u = h(i);
       f<i>:{ int x = u*u; a[i] = x; }
       int t = u + 1;
       g<i>:{ int y = t*2; b[i] = y; } }
   with the entry  f<i>, g<j> : i != j.  Vars_gl excludes u and t. *)
Definition G_scg := globals ["a"; "b"; "n"; "i"].

Definition scg_f : ncb := {|
  n_label := "f"; n_params := ["i"]; n_inputs := ["u"]; n_locals := ["x"];
  n_body := "x" ::= "u" * "u" ;; "a" <[ "i" ]> ::= "x" |}.

Definition scg_g : ncb := {|
  n_label := "g"; n_params := ["i"]; n_inputs := ["u"; "t"]; n_locals := ["y"];
  n_body := "y" ::= "t" * 2 ;; "b" <[ "i" ]> ::= "y" |}.

Theorem scg_fg : valid_entry G_scg (Fp "i" != Gp "i") scg_f scg_g.
Proof. solve_entry. Qed.
\end{lstlisting}

\smallskip\noindent{Why (W2) is needed (Apx.~\ref{apx:in-vars}).}
\begin{lstlisting}[style=rocq]
(* Apx. in-vars, why (W2) is needed:
     { int k; g:{ k = arr[0]; } use(k); }   with   f:{ arr[0] = arr[0] + 1; }
   In the general setting k is in In_g, not in Vars_gl.  Def. Valid then
   accepts the condition true, although the final k depends on the order;
   (W2) rejects g. *)
Definition k_f : ncb := {|
  n_label := "f"; n_params := []; n_inputs := []; n_locals := [];
  n_body := "arr" <[ 0 ]> ::= "arr" !! 0 + 1 |}.

(* k is neither in Vars_gl nor declared in g's body, so g's write to k
   lands in the shared state but is not compared by Def. Valid. *)
Definition k_g : ncb := {|
  n_label := "g"; n_params := []; n_inputs := []; n_locals := [];
  n_body := "k" ::= "arr" !! 0 |}.

Definition G_k := globals ["arr"].

Theorem k_accepted : sound_comm_cond G_k CTrue k_f k_g.
Proof. solve_comm. Qed.

Lemma k_not_W2 : ~ W2 G_k k_g.
Proof. intros H. destruct (H "k") as [Hg | Hl]; simpl in *; auto; discriminate. Qed.

Example k_order_matters :
  exists s1 s2 s1' s2',
    run_ncb k_f empty_state empty_state s1 /\ run_ncb k_g empty_state s1 s2 /\
    run_ncb k_g empty_state empty_state s1' /\ run_ncb k_f empty_state s1' s2' /\
    s2 "k" <> s2' "k".
Proof.
  do 4 eexists.
  split; [apply run_ncb_exec; reflexivity |].
  split; [apply run_ncb_exec; reflexivity |].
  split; [apply run_ncb_exec; reflexivity |].
  split; [apply run_ncb_exec; reflexivity |].
  vm_compute. intros E. discriminate E.
Qed.

(* In the setting of the paper body (In_f = In_g = empty), k counts as a
   shared variable, and Def. Valid correctly rejects true. *)
Definition G_k_body := globals ["arr"; "k"].

Theorem k_rejected : ~ sound_comm_cond G_k_body CTrue k_f k_g.
Proof.
  intros H.
  specialize (H _ _ _ _ _ empty_state empty_state
    (run_ncb_exec k_f empty_state empty_state eq_refl)
    (run_ncb_exec k_g empty_state _ eq_refl)
    (run_ncb_exec k_g empty_state empty_state eq_refl)
    (run_ncb_exec k_f empty_state _ eq_refl) eq_refl (LVar "k") eq_refl).
  vm_compute in H. discriminate.
Qed.
\end{lstlisting}

\smallskip\noindent{Fig.~\ref{fig:psdswp} (Sec.~\ref{sec:execution_model}). The mechanization shows that the entry is \emph{not} sound as written, because \code{q} is declared outside $\nblock{f}$ and written by both instances. Declaring \code{q} inside $\nblock{f}$ makes the instances commute for all rows.}
\begin{lstlisting}[style=rocq]
(* Apx. model, example illustrating waiting in topological order:
     commutativity { f<rowi>, f<rowj> : rowi != rowj; }
     while (p < pSize) {
       id = pIds[p];
       if (!visited[id]) {
         visited[id] = true; row = p; int q = 0;
         f<row>:{ while (q < 99 && !pqFlag[row][q]) {
                    q = q+1; if (q < 100) sum = sum + pValue[row]; } } }
       p = p + 1; }
   The 2-D array pqFlag[row][q] is pqFlag[row*100 + q]. *)
Definition pq_guard : bexp := "q" < 99 /\ "pqFlag" !! ("row" * 100 + "q") == 0.

Definition pq_body : cmd :=
  "q" ::= "q" + 1 ;;
  IF "q" < 100 THEN "sum" ::= "sum" + "pValue" !! "row" ELSE SKIP END.

Definition pq_loop : cmd := WHILE pq_guard DO pq_body END.

(* As written, q is declared outside f, so (In_f = empty) it is shared. *)
Definition G_pq :=
  globals ["visited"; "pqFlag"; "pValue"; "pSize"; "pIds"; "p"; "id"; "row"; "q"; "sum"].

Definition pq_f : ncb := {|
  n_label := "f"; n_params := ["row"]; n_inputs := []; n_locals := [];
  n_body := pq_loop |}.

Lemma pq_wf : wf_pair G_pq pq_f pq_f.
Proof. apply wf_pairb_sound. reflexivity. Qed.

(* The entry is not sound: the second instance continues from the first
   one's final q.  Rows 1 and 2, with flags pqFlag[1][1] = pqFlag[2][0] =
   pqFlag[2][2] = 1 and all other flags 0, end with sum = 110 in one
   order and sum = 10 in the other. *)
Definition pq_s0 : state :=
  (LElem "pqFlag" 101 !-> 1 ; LElem "pqFlag" 200 !-> 1 ; LElem "pqFlag" 202 !-> 1 ;
   LElem "pValue" 1 !-> 10 ; LElem "pValue" 2 !-> 100 ; empty_state).

Definition pq_r (sl sg : state) : state :=
  merge (frame pq_f) sg (run_res 1000 (n_body pq_f) (merge (frame pq_f) sl sg)).

Definition pq_sf : state := ("row" !-> 1 ; empty_state).
Definition pq_sg : state := ("row" !-> 2 ; empty_state).

Theorem pq_unsound : ~ sound_comm_cond G_pq (Fp "row" != Gp "row") pq_f pq_f.
Proof.
  intros H.
  pose proof (run_ncb_run 1000 pq_f pq_sf pq_s0 ltac:(vm_compute; reflexivity)) as R1.
  pose proof (run_ncb_run 1000 pq_f pq_sg (pq_r pq_sf pq_s0)
                          ltac:(vm_compute; reflexivity)) as R2.
  pose proof (run_ncb_run 1000 pq_f pq_sg pq_s0 ltac:(vm_compute; reflexivity)) as R3.
  pose proof (run_ncb_run 1000 pq_f pq_sf (pq_r pq_sg pq_s0)
                          ltac:(vm_compute; reflexivity)) as R4.
  change (run_ncb pq_f pq_sf pq_s0 (pq_r pq_sf pq_s0)) in R1.
  change (run_ncb pq_f pq_sg pq_s0 (pq_r pq_sg pq_s0)) in R3.
  specialize (H _ _ _ _ _ _ _ R1 R2 R3 R4 eq_refl (LVar "sum") eq_refl).
  vm_compute in H. discriminate H.
Qed.

(* Fix: declare q inside f (int q = 0; as the first statement of f).
   Then each instance adds k * pValue[row] to sum, where k depends only on
   row and pqFlag, and the instances commute for all rows. *)
Definition pq_fix_f : ncb := {|
  n_label := "f"; n_params := ["row"]; n_inputs := []; n_locals := ["q"];
  n_body := "q" ::= 0 ;; pq_loop |}.

Definition G_pq_fix :=
  globals ["visited"; "pqFlag"; "pValue"; "pSize"; "pIds"; "p"; "id"; "row"; "sum"].

(* Number of loop iterations from q, with fuel n. *)
Fixpoint pq_count (n : nat) (flag : Z -> Z) (row q : Z) : Z :=
  match n with
  | O => 0
  | S m => if (q <? 99) && (flag (row * 100 + q) =? 0)
           then 1 + pq_count m flag row (q + 1) else 0
  end.

Lemma pq_loop_spec : forall n (s s' : state),
  (Z.to_nat (99 - s "q") < n)%nat ->
  multi_step pq_loop s SKIP s' ->
  s' "sum" = s "sum" + pq_count n (fun i => s (LElem "pqFlag" i)) (s "row") (s "q")
                       * s (LElem "pValue" (s "row")) /\
  (forall l, l <> LVar "q" -> l <> LVar "sum" -> s' l = s l).
Proof.
  induction n as [| m IH]; intros s s' Hn R; [lia |].
  apply while_inv in R as [[Hb ->] | [Hb [s1 [R1 R2]]]].
  - simpl. simpl in Hb. rewrite Hb. split; [lia | reflexivity].
  - apply run_loop_free in R1; [| reflexivity]. subst s1.
    simpl in Hb. apply andb_true_iff in Hb as [Hq Hf].
    apply Z.ltb_lt in Hq.
    assert (Hlt : (s "q" + 1 <? 100) = true) by (apply Z.ltb_lt; lia).
    simpl in R2. rewrite update_same, Hlt in R2.
    apply IH in R2 as [Hsum Hrest].
    2: { unfold update; cbn -[Z.sub Z.add Z.to_nat]. lia. }
    unfold update in Hsum, Hrest; simpl in Hsum, Hrest.
    split.
    + rewrite Hsum. cbn [pq_count]. rewrite Hf, (proj2 (Z.ltb_lt _ _) Hq).
      cbn [andb]. ring.
    + intros l Hl1 Hl2. rewrite Hrest by assumption.
      apply loc_eqb_neq in Hl1, Hl2. rewrite Hl1, Hl2. reflexivity.
Qed.

Definition pq_K (sl sg : state) : Z :=
  pq_count 100 (fun i => sg (LElem "pqFlag" i)) (sl "row") 0.

Lemma pq_fix_run : forall sl sg sg', run_ncb pq_fix_f sl sg sg' ->
  sg' "sum" = sg "sum" + pq_K sl sg * sg (LElem "pValue" (sl "row")) /\
  (forall l, l <> LVar "sum" -> sg' l = sg l).
Proof.
  intros sl sg sg' [s' [R ->]]. apply ncb_reduces in R. simpl in R.
  apply seq_inv in R as [s1 [R1 R2]]. apply asgn_reduces_to in R1. subst s1.
  apply pq_loop_spec with (n := 100%nat) in R2 as [Hsum Hrest];
    [| unfold update; simpl; lia].
  split.
  - unfold merge at 1. simpl. rewrite Hsum. reflexivity.
  - intros l Hl. unfold merge at 1.
    destruct (memb (base l) (frame pq_fix_f)) eqn:E; [reflexivity |].
    rewrite Hrest; [| intros ->; discriminate E | exact Hl].
    unfold update, merge. rewrite E.
    destruct (loc_eqb l (LVar "q")) eqn:Eq; [| reflexivity].
    apply loc_eqb_eq in Eq. subst. discriminate E.
Qed.

Lemma pq_K_frame : forall sl sa sb,
  (forall i, sa (LElem "pqFlag" i) = sb (LElem "pqFlag" i)) -> pq_K sl sa = pq_K sl sb.
Proof.
  intros sl sa sb H. unfold pq_K. f_equal. apply functional_extensionality. exact H.
Qed.

Theorem pq_fix_true : valid_entry G_pq_fix CTrue pq_fix_f pq_fix_f.
Proof.
  split; [apply wf_pairb_sound; reflexivity |]. split; [exact I |].
  intros s0 s1 s1' s2 s2' sf sg R1 R2 R3 R4 _.
  apply pq_fix_run in R1 as [S1 F1], R2 as [S2 F2], R3 as [S3 F3], R4 as [S4 F4].
  intros l _. destruct (loc_eqb l (LVar "sum")) eqn:E.
  - apply loc_eqb_eq in E. subst l.
    rewrite S2, S1, S4, S3.
    rewrite (pq_K_frame sg s1 s0) by (intros i; apply F1; discriminate).
    rewrite (pq_K_frame sf s1' s0) by (intros i; apply F3; discriminate).
    rewrite (F1 (LElem "pValue" (sg "row"))) by discriminate.
    rewrite (F3 (LElem "pValue" (sf "row"))) by discriminate.
    ring.
  - apply loc_eqb_neq in E. rewrite F2, F1, F4, F3 by exact E. reflexivity.
Qed.

Theorem pq_fix : valid_entry G_pq_fix (Fp "row" != Gp "row") pq_fix_f pq_fix_f.
Proof.
  destruct pq_fix_true as [Hwf [_ Hs]].
  split; [exact Hwf |]. split; [apply cond_okb_sound; reflexivity |].
  apply sound_true_weaken. exact Hs.
Qed.
\end{lstlisting}

\subsection{{ReconstructAST and faithfulness}}

{We mechanize \textproc{ReconstructAST} and \mkJobGenProg{} (Sec.~\ref{sec:analysis}) and prove Lemma~\ref{lem:faithful-outline}: inlining the job-generating program gives back the original program. The parallelization graph is given as an annotation of the AST, assigning each statement to a vertex; task ids are vertex ids. The graph is well-formed when no vertex is carved out twice. Because NCB wrappers stay in the task bodies, inlining yields the original program exactly. In the execution model (\texttt{Jobs.v}), {\sc ForkTask} executes the generated $\newJob$ statements.}

\smallskip\noindent{The AST annotated with parallelization-graph vertices, and \textsc{GetDataDependences}.}
\begin{lstlisting}[style=rocq]
Notation vertex := nat.

(* The program AST, each node annotated with its parallelization-graph vertex. *)
Inductive tcmd : Type :=
  | TSkip  : tcmd
  | TAsgn  : vertex -> var -> aexp -> tcmd
  | TStore : vertex -> var -> aexp -> aexp -> tcmd
  | TSeq   : vertex -> tcmd -> tcmd -> tcmd
  | TIf    : vertex -> bexp -> tcmd -> tcmd -> tcmd
  | TWhile : vertex -> bexp -> tcmd -> tcmd
  | TNCB   : vertex -> ncb_label -> list var -> tcmd -> tcmd.

(* The original program. *)
Fixpoint erase (c : tcmd) : cmd :=
  match c with
  | TSkip => SKIP
  | TAsgn _ x a => x ::= a
  | TStore _ a i e => CStore a i e
  | TSeq _ c1 c2 => erase c1 ;; erase c2
  | TIf _ b c1 c2 => CIf b (erase c1) (erase c2)
  | TWhile _ b c1 => CWhile b (erase c1)
  | TNCB _ l ps c1 => CNCB l ps (erase c1)
  end.

(* GetDataDependences: the arguments passed by value to a new job, given
   the carved subtree (with its descendants).  For an NCB these are its
   parameters; otherwise the scalar variables it reads (arrays are shared
   memory and are not copied).  Two kinds of variable are not passed:
   - a variable that the subtree writes: it stays shared, so the writes are
     seen by the code that runs after the subtree, as in the sequential
     program.  Hence no job, nor any of its descendants, writes a variable
     it received by value.
   - a variable that a task other than the spawner and its ancestors writes
     ([fw], see [owrites]).  A copy is taken when the job is forked, and
     such a task (e.g. an earlier sibling) may not have run yet; the job
     instead waits for it and reads the shared variable.  Ancestors are
     harmless: what the spawner does not write, it received by value
     itself.  As in the implementation, a job is thus passed only the
     variables that flow to it from its spawner (SendDep). *)
Fixpoint reads_a (a : aexp) : list var :=
  match a with
  | ANum _ => []
  | AVar x => [x]
  | AArr _ e | AFun _ e => reads_a e
  | APlus a1 a2 | AMinus a1 a2 | AMult a1 a2 | ADiv a1 a2 | AMod a1 a2 =>
      reads_a a1 ++ reads_a a2
  end.

Fixpoint reads_b (b : bexp) : list var :=
  match b with
  | BTrue | BFalse => []
  | BEq a1 a2 | BLe a1 a2 | BLt a1 a2 => reads_a a1 ++ reads_a a2
  | BNot b1 => reads_b b1
  | BAnd b1 b2 | BOr b1 b2 => reads_b b1 ++ reads_b b2
  end.

Fixpoint reads (c : cmd) : list var :=
  match c with
  | CSkip => []
  | CAsgn _ a => reads_a a
  | CStore _ i e => reads_a i ++ reads_a e
  | CSeq c1 c2 => reads c1 ++ reads c2
  | CIf b c1 c2 => reads_b b ++ reads c1 ++ reads c2
  | CWhile b c1 => reads_b b ++ reads c1
  | CNCB _ ps _ => ps
  | CNewJob _ args => args
  end.

(* Remove duplicates, keeping first occurrences. *)
Fixpoint dedup (xs : list var) : list var :=
  match xs with
  | [] => []
  | x :: xs' => x :: filter (fun y => negb (String.eqb y x)) (dedup xs')
  end.

Definition deps (fw : list var) (c : cmd) : list var :=
  dedup (filter (fun x => negb (memb x (writes c)) && negb (memb x fw)) (reads c)).

(* The variables written by the nodes of the vertices not in [keep]. *)
Fixpoint owrites (keep : list vertex) (c : tcmd) : list var :=
  match c with
  | TSkip => []
  | TAsgn w x _ => if existsb (Nat.eqb w) keep then [] else [x]
  | TStore w a _ _ => if existsb (Nat.eqb w) keep then [] else [a]
  | TSeq _ c1 c2 | TIf _ _ c1 c2 => owrites keep c1 ++ owrites keep c2
  | TWhile _ _ c1 => owrites keep c1
  | TNCB w _ _ c1 => if existsb (Nat.eqb w) keep then [] else writes (erase c1)
  end.
\end{lstlisting}

\smallskip\noindent{\textproc{ReconstructAST} and \mkJobGenProg.}
\begin{lstlisting}[style=rocq]
Section Recon.

(* The whole program, for the variables that other tasks write. *)
Variable root : tcmd.

(* [recon anc v c = (body, tasks)]: the code of [c] as seen from a task of
   vertex [v], whose ancestors (with [v] itself) are [anc].  A node of
   another vertex [w] is carved out: it becomes newJob(w, deps), where deps
   are computed on the original subtree [sub], and the task (w, its
   reconstruction) is added. *)
Definition carve (anc : list vertex) (v w : vertex) (node sub : cmd)
                 (ts : list (vertex * cmd)) : cmd * list (vertex * cmd) :=
  if Nat.eqb w v then (node, ts)
  else (CNewJob w (deps (owrites anc root) sub), ts ++ [(w, node)]).

(* The ancestors seen from a node of vertex [w]. *)
Definition down (anc : list vertex) (v w : vertex) : list vertex :=
  if Nat.eqb w v then anc else w :: anc.

Fixpoint recon (anc : list vertex) (v : vertex) (c : tcmd)
  : cmd * list (vertex * cmd) :=
  match c with
  | TSkip => (SKIP, [])
  | TAsgn w x a => carve anc v w (x ::= a) (x ::= a) []
  | TStore w a i e => carve anc v w (CStore a i e) (CStore a i e) []
  | TSeq w c1 c2 =>
      let (b1, t1) := recon (down anc v w) w c1 in
      let (b2, t2) := recon (down anc v w) w c2 in
      carve anc v w (b1 ;; b2) (erase c1 ;; erase c2) (t1 ++ t2)
  | TIf w b c1 c2 =>
      let (b1, t1) := recon (down anc v w) w c1 in
      let (b2, t2) := recon (down anc v w) w c2 in
      carve anc v w (CIf b b1 b2) (CIf b (erase c1) (erase c2)) (t1 ++ t2)
  | TWhile w b c1 =>
      let (b1, t1) := recon (down anc v w) w c1 in
      carve anc v w (CWhile b b1) (CWhile b (erase c1)) t1
  | TNCB w l ps c1 =>
      carve anc v w (CNCB l ps (erase c1)) (CNCB l ps (erase c1)) []
  end.

End Recon.

(* buildJobGenProg: the Init task's body and the other tasks. *)
Definition build (c : tcmd) : cmd * list (vertex * cmd) := recon c [0%nat] 0 c.

(* The tasks for the execution model (Jobs.v), given the spawning DAG's
   predecessor relation. *)
Definition to_tasks (preds : vertex -> list vertex) (locals : vertex -> list var)
                    (c : tcmd) : task_list :=
  map (fun p => {| t_id := fst p; t_body := snd p; t_preds := preds (fst p);
                   t_locals := locals (fst p) |})
      ((0%nat, fst (build c)) :: snd (build c)).
\end{lstlisting}

\smallskip\noindent{Inlining $\newJob$ statements.}
\begin{lstlisting}[style=rocq]
Fixpoint lookup (T : list (vertex * cmd)) (w : vertex) : option cmd :=
  match T with
  | [] => None
  | (w', b) :: T' => if Nat.eqb w' w then Some b else lookup T' w
  end.

(* [inlines T b d]: inlining every newJob(w, _) in [b] by the body of task
   [w] in [T], recursively, yields [d]. *)
Inductive inlines (T : list (vertex * cmd)) : cmd -> cmd -> Prop :=
  | IN_Skip : inlines T SKIP SKIP
  | IN_Asgn : forall x a, inlines T (x ::= a) (x ::= a)
  | IN_Store : forall a i e, inlines T (CStore a i e) (CStore a i e)
  | IN_Seq : forall c1 c2 d1 d2,
      inlines T c1 d1 -> inlines T c2 d2 -> inlines T (c1 ;; c2) (d1 ;; d2)
  | IN_If : forall b c1 c2 d1 d2,
      inlines T c1 d1 -> inlines T c2 d2 -> inlines T (CIf b c1 c2) (CIf b d1 d2)
  | IN_While : forall b c d, inlines T c d -> inlines T (CWhile b c) (CWhile b d)
  | IN_NCB : forall l ps c d, inlines T c d -> inlines T (CNCB l ps c) (CNCB l ps d)
  | IN_NewJob : forall w args b d,
      lookup T w = Some b -> inlines T b d -> inlines T (CNewJob w args) d.
\end{lstlisting}

\smallskip\noindent{Faithfulness (Lemma~\ref{lem:faithful-outline}). The lemma \texttt{recon\_inlines} generalizes the statement to any vertex and any task table containing the generated tasks.}
\begin{lstlisting}[style=rocq]
(* Faithfulness: if no vertex is carved out twice, inlining the
   job-generating program gives back the original program. *)
Theorem faithful : forall c,
  NoDup (map fst (snd (build c))) ->
  inlines (snd (build c)) (fst (build c)) (erase c).
Proof.
  intros c Hnd. apply recon_inlines. intros x y Hin. apply lookup_In; assumption.
Qed.
\end{lstlisting}

\smallskip\noindent{The running example: \mkJobGenProg{} produces the tasks of Fig.~\ref{fig:jobgenprog}.}
\begin{lstlisting}[style=rocq]
(* Ex. motivation, annotated with the parallelization graph: the
   declarations are in the Init vertex 0, the loop (with i = i + 1) in
   vertex 1, and the NCBs f and g in vertices 2 and 3.  The NCB bodies
   are those of Examples.v. *)
Fixpoint tag (v : vertex) (c : cmd) : tcmd :=
  match c with
  | CAsgn x a => TAsgn v x a
  | CStore a i e => TStore v a i e
  | CSeq c1 c2 => TSeq v (tag v c1) (tag v c2)
  | CIf b c1 c2 => TIf v b (tag v c1) (tag v c2)
  | CWhile b c1 => TWhile v b (tag v c1)
  | CNCB l ps c1 => TNCB v l ps (tag v c1)
  | _ => TSkip
  end.

Definition mot_ast : tcmd :=
  TSeq 0 (TAsgn 0 "i" 0%Z)
    (TWhile 1 ("i" < "size")
       (TSeq 1 (TNCB 2 "f" ["i"] (tag 2 (n_body (mot_f calc))))
       (TSeq 1 (TNCB 3 "g" ["i"] (tag 3 (n_body (mot_g compute calc2 calc3))))
               (TAsgn 1 "i" ("i" + 1)%ncb)))).

(* The generated tasks, as in Fig. jobgenprog: Init spawns Task1 (the
   loop), which spawns one job of Task2 (f) and one of Task3 (g) per
   iteration, passing i.  The loop writes i, so Init does not pass it: the
   loop's i stays shared, and only the NCBs get a copy. *)
Example mot_build :
  build mot_ast =
  ("i" ::= 0%Z ;; CNewJob 1 ["size"],
   [ (2%nat, ncb_cmd (mot_f calc));
     (3%nat, ncb_cmd (mot_g compute calc2 calc3));
     (1%nat, WHILE "i" < "size" DO
               CNewJob 2 ["i"] ;; CNewJob 3 ["i"] ;; "i" ::= "i" + 1
             END) ]).
Proof. reflexivity. Qed.

Theorem mot_faithful :
  inlines (snd (build mot_ast)) (fst (build mot_ast)) (erase mot_ast).
Proof. apply faithful. apply nodupb_sound. reflexivity. Qed.
\end{lstlisting}

\smallskip\noindent{The AST reconstruction example of Sec.~\ref{sec:analysis}.}
\begin{lstlisting}[style=rocq]
(* Fig. reconstruct-ex: a vertex owns n3..n6, and the subtree n7 belongs
   to another vertex, so n7 becomes a newJob in Task_i.  It reads y and z
   but writes z, so only y is passed by value. *)
Definition fig_ast : tcmd :=
  TSeq 1 (TIf 1 ("x" < 0%Z) (TAsgn 1 "y" 1%Z) (TAsgn 1 "y" 2%Z))
         (TSeq 2 (TAsgn 2 "z" ("y" + 1)%ncb) (TAsgn 2 "w" "z")).

Example fig_recon :
  recon fig_ast [1%nat] 1 fig_ast =
  (IF "x" < 0%Z THEN "y" ::= 1%Z ELSE "y" ::= 2%Z END ;; CNewJob 2 ["y"],
   [(2%nat, "z" ::= "y" + 1 ;; "w" ::= "z")]).
Proof. reflexivity. Qed.
\end{lstlisting}